\documentclass[12pt]{article}
\usepackage{amsmath}
\numberwithin{equation}{section}
\usepackage{graphicx}
\usepackage{bm}
\usepackage{slashed}
\usepackage{color} 
\newcommand{\la}{\langle}
\newcommand{\ra}{\rangle}
\newcommand{\muT}{\frac{\mu}{T}}

\newcommand{\pt}{{\bf p_T}}
\usepackage{color} 
\begin{document}
\setcounter{page}{0}
\pagenumbering{roman}
\setcounter{page}{0}
\pagenumbering{arabic}
\pagestyle{myheadings}
\markright{}

\title{A short course on Relativistic Heavy Ion Collisions}
  
 \author{A. K. Chaudhuri\\ 
 Theoretical Physics Division\\
 Variable Energy Cyclotron Centre\\
 1-AF, Bidhan Nagar,Kolkata - 700 064, India\\
Email:akc@vecc.gov.in} 
\maketitle 
\begin{abstract}
Some  ideas/concepts in relativistic heavy ion collisions are discussed. 
To a large extent, the discussions are non-comprehensive and non-rigorous. 
It is intended for fresh graduate students of Homi Bhabha National Institute,
Kolkata Centre, 
who are intending to pursue career in theoretical /experimental high energy nuclear physics. 
Comments and criticisms will be appreciated. 
\end{abstract}  
 
 \newpage
 
{\bf{
\noindent Contents:\\ \\
1. Introduction\\ \\
2. Conceptual basis for QGP formation\\\\
3. Kinematics of HI collisions\\\\
4. QGP and hadronic resonance gas in the ideal gas limit\\\\
5. Quantum chromodynamics: theory of strong interaction\\\\
6. Color Glass Condensate\\\\
7. Relativistic kinetic Theory\\\\
8. Hydrodynamic model for heavy ion collisions\\\\
9. Signals of Quark-Gluon-Plasma\\\\
10. Summary
}}
\newpage


\section{Introduction}\label{sec1}

\begin{figure}[h]
 \center
  \includegraphics[scale=0.5]{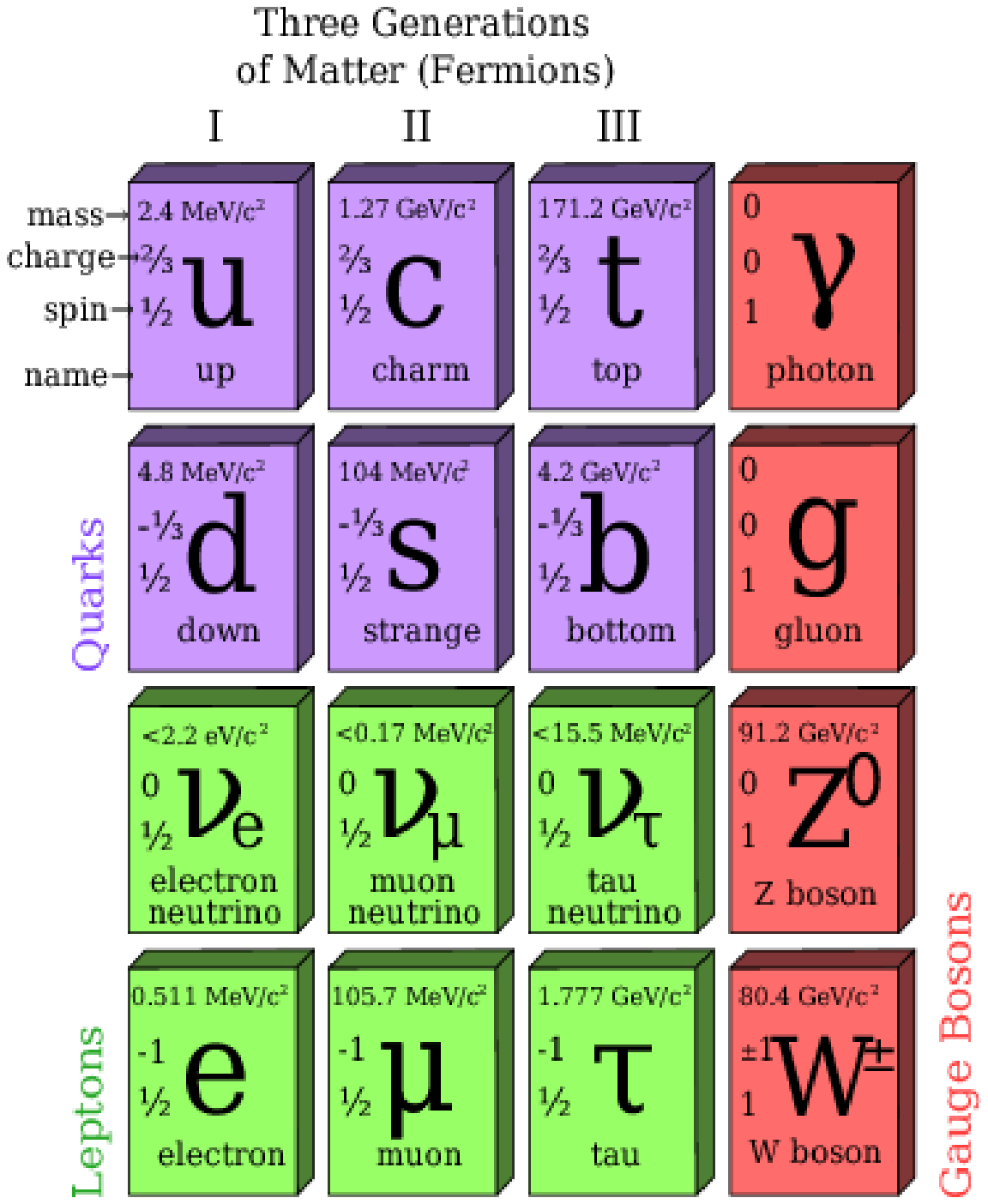}
\caption{\label{F1}Three generation of matter particles in the standard model. The mediator particles (Gauge bosons) are also shown.}\end{figure}

Surprisingly, our diverse universe consists of a handful of 'elementary' or 'fundamental' particles. In Fig.\ref{F1}, I have listed the presently known elementary particles. These elementary particles can be classified as (i) matter particles, the fermions and (ii) mediator particles, the bosons. The handful of fundamental particles, can interact only in four definite manner, (i) strong interaction (ii) electromagnetic interaction (iii) weak interaction and (iv) gravitational interaction. In table.\ref{table1}, I have listed the mediators of the interactions, also shown the relative strength of the interactions. All these particles interact gravitationally.

Study of strong interaction is generally called nuclear physics. Historically, nuclear physics started with Rutherford's discovery of 'Nucleus'
  in his celebrated gold foil experiment (1909). The term 'Nucleus' was coined by  Robert Brown, the botanist, in 1831,    describing the cell structure (alternatively, by Michael Faraday in 1844), from the latin word 'Nux' which means 'nut'. The result of gold foil experiment was so bizarre at that time that Rutherford commented like this, "It was almost as if you fire a 15 inch shell into a piece of tissue paper and it came back and hit you". The concept of Atomic Nucleus was completed with James Chadwick's discovery of 'Neutron' in 1932. Indeed, one can say that proper Nuclear Physics started in 1932 after the discovery of neutron. 
  
  For a long time 'Atomic Nucleus' supposed to be composed of protons (a term possibly coined by Rutherford for hydrogen nucleus) and neutrons and they are supposed to interact strongly.
In the mean time there was much progress in the understanding of electromagnetic (EM) interaction. It was recognised that EM interaction arises due to exchange of photons between two charged particles. In analogy to EM interaction,  in 1934 Hideki Yukawa put forward the hypothesis that strong interaction between nucleons originate from exchange of mesons. At that time mesons were not known. He made this bold conjecture to obtain a theory analogous to electromagnetic interaction, where a photon mediates the force. He was only 27 years old then. 
In 1937 pions were discovered and in 1949 Yukawa was awarded the Noble prize in Physics. However, in later years, with the advent of particle accelerators,  experimentalists discovered hundreds of particles (mesons and baryons) many of which can be thought to be mediators of the strong interaction. People then tried to characterize those particles, study their internal symmetry [internal symmetry refers to the fact that one generally find a family of particles called multiplet, all with same or nearly same mass. Each multiplet can be looked upon as a realisation of some internal symmetry].   I will not go into detail, suffice to say that  Murray Gell-Mann and George Zweig (1964) found that all these particles, including protons and neutrons, consists of only a few building blocks which he termed as quarks. Murray Gellman picked the word 'quark'
from the sentence 'Three quarks for Muster Mark' in James Joyce book, 'Finnegans Wake'. Simplest version of the quark model faces problem. Some baryons e.g. $\Omega^-$ or $\Delta^{++}$  then composes of identical quarks and violate Pauli's exclusion principle. To eliminate the contradiction,    the concept of color was introduced. Color is a new quantum number. Only three colors required to be hypothesised. Murray Gell-Mann was born in September 1929. When he postulates quarks, he was   35 years old. He got Nobel prize in the year 1969.  
One can borrow G. H. Hardy's (known for discovering Ramanujan) words and say, 'creative physics is young man's game'.
Take for example: Newton, at the age 23-24   gave the law of Gravitation, discovered 
Fluxions (calculus), Einstein discovered relativity at the age of 25-26. 
Wolfgang Pauli formulated his exclusion principle when he is 25 years old.

\begin{table}[t] \label{table1}
\caption{Four fundamental forces, their relative strength and their mediators are listed.}
	\centering
		\begin{tabular}{|l|l|l|l|l|}
		\hline
interaction  & theory & Mediators & relative & interaction\\ 
             &        &           & strength & range (m)\\	\hline	\hline
 strong      & QCD            & Gluon & $10^{38}$ & $10^{-15}$\\ \hline
electromagnetic & QED          & Photon & $10^{36}$ & infinity \\ \hline
weak &electroweak & W, Z         & $10^{25}$ & $10^{-18}$\\ \hline
 gravitational  & general & graviton & 1 & infinity\\
                &relativity           &         &   &  \\ \hline
   		\end{tabular}
	\textsl{}
\end{table}

Traditionally, nuclear physics is the study of nuclear matter at zero temperature and at densities of the order of the atomic nuclei, nucleon density, $\rho\sim 0.17 fm^{-3}$ or energy density $\epsilon \sim 0.16$ $GeV/fm^3$. Advent of accelerators has extended the study to hundreds MeV of temperature and energy densities several order of magnitude higher. At such high density/temperature, individual hadrons loss their identity and the matter is best described in terms of the constituents of the matter, e.g. quarks and gluons, commonly called Quark-Gluon-Plasma (QGP). Historically, T. D. Lee, in collaboration with G. C. Wick first speculated about an abnormal nuclear
state, where mucleon mass is zero or near zero in an extended volume and non-zero out side the volume \cite{Lee:1974ma}\cite{Lee:1974kn}.
They also suggested that an effective way to search for these
new objects is through high-energy heavy ion collisions. In this short lecture course, I will try to discuss some aspects of the matter at such high density and temperature. For a general introduction to the subject, see  , \cite{BRAHMSwhitepaper}\cite{PHOBOSwhitepaper}\cite{PHENIXwhitepaper}\cite{STARwhitepaper}\cite{Wong}\cite{Csernai}\cite{QM}.


\begin{figure}[t]
  \vspace{1.0cm} 
 \center
  \includegraphics[scale=0.5]{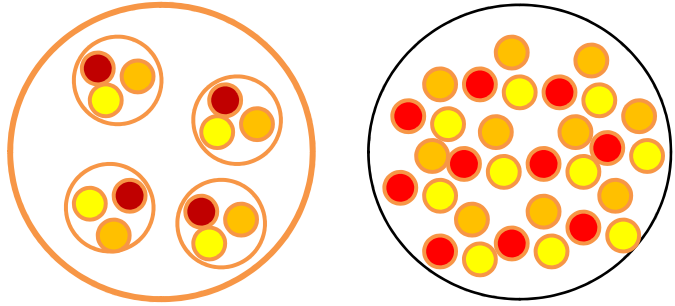}  
\caption{Left panel shows a nucleus at normal density. The right panel shows the same at high density.}
 \label{F2}
\end{figure}

\section{Conceptual basis for QGP formation}\label{sec2}

For composite hadrons, with finite spatial extension, concept of hadronic matter appears to lose its meaning at sufficiently high density. Once we have a system of mutually interpenetrating hadrons, each quark will find in its vicinity, at a distance less  than the hadron radius, a number of quarks. The situation is shown schematically in Fig.\ref{F2}. At low density, a particular quark in a hadron knows in partner quarks. However, at high density, when the hadrons starts to interpenetrate each other,
a particular quark will not able to identify the quark which was its partner at lower density. Similar phenomena can happen at high temperature. As the temperature of a nuclear matter is increased, more and more low mass hadrons (mostly pions) will be created. The system again will be dense enough and  hadrons will starts to   interpenetrate.  
The system where, hadrons interpenetrate is best considered as a Quark matter, rather than made of hadrons. It is customary to call the quark matter as Quark-Gluon-Plasma (QGP). We define QGP as  a thermalised, or near to thermalised state of quarks and gluons, where quarks and gluons are free to move over a nuclear volume rather than a nucleonic volume. Model calculations indicate that beyond a critical energy density $\epsilon_{cr}\sim$ 1 $GeV/fm^3$, or temperature $T_{cr} \sim$ 200 MeV, matter can exist only as QGP.

QGP is the deconfined state of strongly interacting mater. Since at low density or low temperature quarks are confined within the hadrons and at high density or at high temperature,  quarks are deconfined, one can talk about a confinement-deconfinement phase transition. I will discuss it later, but it turns out that the confinement-deconfinement transition is not a phase transition in thermodynamic sense (in thermodynamic phase transition, free energy or its derivative have singularity at the transition point), rather it is a smooth cross-over, from confinement to deconfinement or vice-versa.
The mechanism of deconfinement is provided by the screening of the color charge. It is analogous to the Mott transition in atomic physics. In dense matter, the long range coulomb potential, which binds ions and electrons into electrically neutral atom, is partially screened due to presence of other charges,
the potential become much more short range,

\begin{equation}
V(r)=e^2_0/r \rightarrow e^2_0/r \times exp(-r/r_D)
\end{equation}

\noindent here $r$ is the distance of the probe from the test charge $e_0$. $r_D$ is the Debye screening radius and is inversely proportional to density,

\begin{equation}
r_D\sim n^{-1/3}
\end{equation}

At sufficiently high density, $r_D$ can be smaller than the atomic radius. 
A given electron can no longer feel the binding force of its ion, alternatively, at such density, coulomb potential can no longer bind electron and ion into a neutral atom. The insulating matter becomes a conducting matter. This is the  Mott transition. We expect deconfinement to be the quantum chromodynamic analog of   Mott transition. Due to screening of color potential, quarks can not be bound into a hadron. 
Now one may wonder about the very different nature of QCD and QED forces.
 Interaction potential in QED and QCD can be expressed as,

\begin{eqnarray}
QED&:& V(r)\sim -e^2/r \\
QCD&:& V(r)\sim -\alpha/r +\sigma r
\end{eqnarray}

While in QED,  potential decreases continuously with increasing distance, in QCD, at large distance, it increases with distance.
However, screening is a phenomenon at high density, or at short distance. The difference in QED and QCD at large distance is of no consequence then. More over, due to asymptotic freedom, in QCD interaction strength decreases at short distances, thereby enhancing the deconfinement.

It may be noted that in insulating solid, at $T>0$, conductivity is not exactly zero, it is exponentially small,

\begin{equation}
\sigma_E \sim e^{-\Delta E/T}
\end{equation}

\noindent where $\Delta E$ is the ionisation potential. Above the Mott transition temperature, $\sigma_E$ is non-zero because Debye screening has globally dissolved coulomb binding between ion and electrons, but below the Mott transition temperature, ionisation can produce locally free electrons, making $\sigma_E$ small but non-zero. Corresponding phenomenon in QCD is the creation of quark-antiquark pairs in the form of a hadron. If we try to remove a quark from a hadron, the confining potential will rise with the distance of separation, until it reaches the value $m_H$, the lowest $q\bar{ q}$ state. At this point, an additional hadron will form, whose anti-quark neutralises the quark we were trying to separate. This is the mechanism of quark fragmentation.

\subsection{Why QGP is important to study?}

\begin{figure}[t]
 \center
 \resizebox{0.45\textwidth}{!}
 {\includegraphics{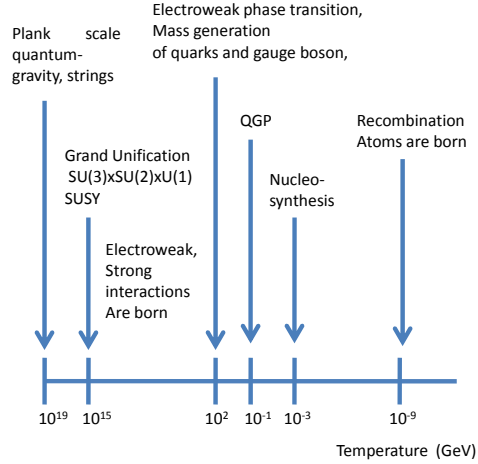}}%
\caption{Schematic representation of temporal evolution of universe in big bang theory.}
\label{F3}
\end{figure}

QGP surely existed in very early universe. In Fig.\ref{F3}, different stages of evolution of universe, in the Big bang model, are shown. 

(i) At the earliest time, temperatures are of the order of $T\sim 10^{19} GeV$, it is the Plank scale temperature. At this stage, quantum gravity is important. Despite an enormous effort by string theorists, little is  understood about this era. 

(ii) We have better understanding of the  later stage of evolution, say, around temperature $T\sim 10^{16}$ GeV. It is the Grand unification scale.
Strong, and electroweak interactions are unified at this scale. The universe at this scale may also be supersymmetric (for each fermion a boson exists and vice-versa). 

(iii) As the universe further expands and cools, strong and electroweak
interactions are separated. At much lower temperature $T\sim$ 100 GeV,
electroweak symmetry breaking takes place. Baryon asymmetry may be produced here. Universe exists as QGP, deconfined state of quarks and gluons 

(iv) Somewhere around $T\sim$ 100 MeV, deconfinement-confinement transition occur, hadrons are formed. Relativistic Heavy Ion collider (RHIC) at Brookhaven National Laboratory (BNL), and Large Hadron Collider (LHC) at CERN, are designed to study matter around this temperature.

(v) at temperature $T\sim$ 1 MeV, nucleosynthesis starts and light elements are formed. This temperature range is well studied in nuclear physics experiments. For example at our centre (Variable Energy Cyclotron Centre, Kolkata), nuclear collisions produces matter around this temperature.

(vi) at temperature $T\sim$ 1 ev, universe changes from ionised gas to a gas of neutral atoms and structures begin to form.

QGP may also exist at the core of a neutron star. Neutron stars are remnants of gravitational collapse of massive stars. They are small objects, radius $\sim$ 10 Km, but very dense, central density $\sim$ 10 normal nuclear matter density.
At such high density hadrons loss their identity and matter is likely to be in the form of QGP. One important difference between QGP at the early universe and that in neutron stars is the temperature. While in early universe, QGP is at temperature $T\sim$100 MeV, at the core of the neutron star it is cold  QGP, $T\sim$ 0 MeV. Hot and dense matter with energy density exceeding 1 $GeV/fm^3$ may also occur in supernova explosions, collisions between neutron stars or between black holes. 

\section{Kinematics of HI collisions}\label{sec3}

Our knowledge of universe is gained through experiments. Horizon of human mind and of science is increased by solving puzzles posed by new and newer experiments.  It is thus appropriate that we discuss kinematics of heavy ion collisions, which is very relevant for experimentalists.

Throughout the note, I have used natural units,

\begin{equation} \nonumber
\text{(i)} \hbar = c = k_B= 1, \text{(ii) Metric:} g^{\mu\nu} =diag(1, -1, -1, -1).
\end{equation}

When we calculate some observable, the missing $\hbar$, $c$ and $k_B$ must be put into the equation taking into account the appropriate dimension of the observable.
We also use the Einstein's summation convention, repeated indices are summed over (unless otherwise stated). Thus,

\begin{equation} \nonumber
\partial_\mu J^\mu\equiv\partial_t J^t+\partial_x J^x+\partial_y J^y+\partial_z J^z
\end{equation}

 \begin{figure}[t]
 \center
 \resizebox{0.45\textwidth}{!}
 {\includegraphics{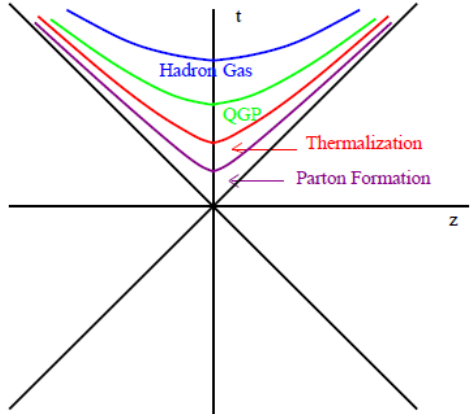}}
\caption{A space-time diagram for the evolution of matter produced in relativistic heavy ion collisions.} \label{F4}
\end{figure}

\subsection{Space-time picture}

Fig.\ref{F4} depicts the collision of two nuclei in (t,z) plane. Two Lorentz contracted nuclei approaching each other with velocity of light and collide at (t=0,z=0). In the collision process a fireball is created. The fireball expands in space-time going through various processes till the created particles freeze-out. 
In relativistic mechanics, neither $\Delta t$ nor $\Delta x$ are invariant distance.
Invariant distance is $\Delta \tau^2 = \Delta t^2 -\Delta x^2$.
Appropriate coordinates in a relativistic collision is  then proper time and space-time rapidity,

\begin{eqnarray}
\text{proper time}:\tau&=&\sqrt{t^2-z^2} \label{eq2.4} \\ 
\text{space-time rapidity:} \eta_s&=&\frac{1}{2}\ln \frac{t+z}{t-z} \label{eq2.5}
\end{eqnarray}

Region of space-time for which $\tau^2=t^2-z^2 >0$ is called time like region,
$\tau^2=t^2-z^2 <0$ is called space-like region. 
$t=z$ line is called lightlike (only light or massless particles can travel along this line). Space-like region is inaccessible to a physical particle,
it need to travel faster than light. For a massive particle, with speed $v<c$,
only accessible region is the time-like region.
Particle production then occurs only in the time like region. Space-time rapidity ($\eta_s$) is properly defined in the time like region only.
$\eta_s$ is positive and negative infinity along the beam direction, $t=\pm z$. $\eta_s$ is not defined in space-like region .

\subsection{Lorentz transformation}
In relativistic nucleus-nucleus collisions it is convenient to use kinematic variables which take simple form under Lorentz transformation for the change
of frame of reference. For completeness, we briefly discuss Lorentz transformation. 

If $x^\mu$ is the coordinate in one frame of reference, then in any other frame of reference the coordinates ${x^\prime}^\mu$ must satisfy,

\begin{equation}\label{eq2.1}
g_{\mu\nu} {dx^\prime}^\mu {dx^\prime}^\mu=g_{\mu\nu} dx^\mu dx^\nu 
\end{equation}

\noindent or equivalently,

\begin{equation} \label{eq2.2}
g_{\mu\nu} \frac{ {dx^\prime}^\mu}{dx^\rho} \frac{ {dx^\prime}^\mu}{dx^\sigma},
=g_{\rho\sigma} 
\end{equation}

The transformation has the special property that speed of light is same in the two frame of reference, a light wave travels at the speed $|d\vec{x}/dt|=1$. 
The transformation $x^\mu \rightarrow {x^\prime}^\mu= \Lambda^\mu_\nu x^\nu+a^\mu$, $a^\mu$ being an arbitrary constant, satisfying Eq.\ref{eq2.2}, i.e,

\begin{equation} \label{eq2.3}
g_{\mu\nu} \Lambda^\mu_\rho \Lambda^\mu_\sigma=g_{\rho\sigma} 
\end{equation}

\noindent is called a Poincar\'{e} transformation.  Lorentz transformation is the special case of Poincar\'{e} transformation when $a^\mu=0$. The matrix $\Lambda^\mu_\nu$ form a group called Lorentz group.

A general Lorentz transformation consists of rotation and translation. Lorentz transformation without rotation is called Lorentz boost. As an example, consider the Lorenz boost along the $z$ direction by velocity $\beta$. The transformation can be written as,

\begin{equation} \label{eq3.6}
\begin{pmatrix} t^\prime \\ z^\prime \end{pmatrix}=\begin{pmatrix}\gamma & -\beta\gamma \\ -\beta\gamma & \gamma \end{pmatrix}
\begin{pmatrix} t \\ z \end{pmatrix}
\end{equation}

\noindent where, $\gamma=1/\sqrt{1-\beta^2}$ is the Lorentz factor.

\subsection{Mandelstam variables}

\begin{figure}[h]
 \center
 \resizebox{0.35\textwidth}{!}
 {\includegraphics{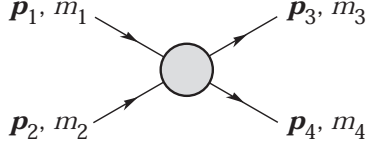}}
\caption{Pictorial diagram of $A +B\rightarrow C+D$ process. } \label{F5}
\end{figure}

In Fig.\ref{F5}, a two body collision process is shown. Two particles of momenta $p_1$ and $p_2$ and masses $m_1$ and $m_2$ scatter to particles   of momenta $p_3$ and $p_4$ and masses $m_3$ and $m_4$. The Lorentz-invariant Mandelstam variables are defined as,

\begin{eqnarray}
s&=&(p_1+p_2)^2=(p_3+p_4)^2 \nonumber \\
 &=& m_1^2+m_2^2+2E_1E_2-2 \bf{p_1}.\bf{p_2}\\
t&=&=(p_1-p_3)^2=(p_2-p_4)^2  \nonumber\\
 &=&m_1^2+m_3^2-2E_1E_3+2\bf{p_1}.\bf{p_3}\\
u&=&=(p_1-p_4)^2=(p_2-p_3)^2 \nonumber\\
&=&m_1^2+m_4^2-2E_1E_4+2\bf{p_1}.\bf{p_4} 
\end{eqnarray}

They satisfies the constrain,

\begin{equation}
s+t+u=m_1^2+m_2^2+m_3^2+m_4^2
\end{equation}

\subsection{Rapidity variable:}

In relativistic energy, rapidity variable, defined as,

\begin{eqnarray} \label{eq2.6}
y&=&\frac{1}{2}\ln \frac{E+p_z}{E-p_z}\\
&=&\frac{1}{2}\ln \frac{1+p_z/E}{1-p_z/E}=\tanh^{-1}\left (\frac{p_z}{E}\right )=\tanh^{-1}(\beta_L)
\end{eqnarray}

\noindent
is more appropriate than the longitudinal velocity ($\beta_L=p_z/E$).  
Rapidity has the advantage that they are additive under a longitudinal boost. A particle with rapidity $y$ in a given inertial frame has rapidity $y+dy$
in a frame which moves relative to the first frame  with rapidity $dy$  in the $-z$ direction. One can see this from the addition formula of relativistic velocity $\beta_1$ and $\beta_2$. The resultant velocity,

\begin{equation}
\beta=\frac{\beta_1+\beta_2}{1+\beta_1\beta_2}
\end{equation}

\noindent  is also the addition formula for hyperbolic tangents,

\begin{equation}
\tanh(y_1+y_2)=\frac{\tanh(y_1)+\tanh(y_2)}{1+\tanh(y_1)\tanh(y_2)}
\end{equation}

The underlying reason is that Lorentz boost can be thought of 
as a hyperbolic rotation of the coordinates in Minkowski space.  In terms of rapidity variable, velocity and Lorentz
factor can be written as,

\begin{eqnarray*}
\beta&=&\tanh(y)\\
\gamma&=&\cosh(y),
\end{eqnarray*}

\noindent and the transformation in Eq.\ref{eq3.6} can be rewritten as,

\begin{equation}
\begin{pmatrix} t^\prime \\ z^\prime \end{pmatrix}=\begin{pmatrix}\cosh(y) & -\sinh(y) \\ -\sinh(y) & \cosh(y) \end{pmatrix}
\begin{pmatrix} t \\ z \end{pmatrix}
\end{equation}

\noindent which is a hyperbolic rotation.

Rapidity is the relativistic analog of non-relativistic velocity. In the non-relativistic limit, $p<<m$ and Eq.\ref{eq2.6} can be written as,

\begin{eqnarray} 
y&=&\frac{1}{2}\ln \frac{\sqrt{p^2+m^2}+mv_z}{\sqrt{p^2+m^2}-mv_z}=\frac{1}{2}\ln \frac{m+mv_z}{m-mv_z} \nonumber\\
&=&\frac{1}{2}[\ln(1+v_z)-\ln(1-v_z)] \approx v_z \label{eq2.6a}
\end{eqnarray}

In terms of the rapidity variables, particle 4-momenta can be parameterised as,

\begin{equation} \label{eq2.7}
p^\mu=(E, p_x,p_y,p_z)=(m_T \cosh y, p_x,p_y,m_T \sinh y)
\end{equation}

with transverse mass ($m_T$),

\begin{equation}\label{eq2.8}
 m_T=\sqrt{m^2+p_T^2}=\sqrt{m^2+p_x^2+p_y^2}
\end{equation}

\subsection{Pseudo-rapidity Variable:}

For a particle emitted at an angle $\theta$ with respect to the beam axis, rapidity variable is,

\begin{eqnarray}
y&=&\frac{1}{2}\ln \frac{E+p_z}{E-p_z} \nonumber \\
&=&\frac{1}{2}\ln \frac{\sqrt{m^2+p^2}+p\cos \theta}{\sqrt{m^2+p^2}-p\cos\theta} \label{eq2.9}
\end{eqnarray}

At very high energy, $p>> m$,the mass can be neglected,

\begin{eqnarray}
y&=&\frac{1}{2}\ln \frac{p+p\cos \theta}{p-p\cos\theta} \nonumber \\
&=&-\ln \tan \theta/2\equiv \eta \label{eq2.10}
\end{eqnarray}

$\eta$ is called pseudorapidity. Only angle $\theta$ determine the pseudorapidity. It is a convenient  parameter for experimentalists when details of the particle, e.g. mass, momentum etc. are not known, but only the angle of emission is known (for example in emulsion experiments). 


\subsection{Light cone momentum:} For a particle with 4-momentum $p(p_0,\pt,p_z)$, forward and
backward light cone variables are defined as,

\begin{eqnarray}
p_+&=&p_0+p_z \label{eq2.11}\\
p_-&=&p_0-p_z \label{eq2.12}
\end{eqnarray}

It is apparent that for a particle traveling along the beam axis, forward light cone momentum is higher than for a particle traveling opposite to the beam axis. An important property of the light cone is that in case of a boost, light cone momentum is multiplied by a constant factor. It can be seen as follow, write the momentum in terms of rapidity variable, $p^\mu=(m_T \cosh y, p_x,p_y, m_T\sinh y)$,

\begin{eqnarray}
p_+&=&m_T e^y \label{eq2.13}\\ 
p_-&=&m_T e^{-y} \label{eq2.14}
\end{eqnarray}

\subsection{Invariant distribution:} Let us show $\frac{d^3p}{E}$ is Lorentz invariant. The differential of Lorentz boost in longitudinal direction is,

\begin{eqnarray}
dp_z^*&=&\gamma (dp_z-\beta dE)=\gamma  (dp_z-\beta \frac{p_z dp_z}{E}), \nonumber\\  
&=&\frac{dp_z}{E} \gamma (E-\beta p_z)=\frac{dp_z}{E} E^* \label{eq2.15}
\end{eqnarray}

\noindent where we have used, $E^2=m^2+p_T^2+p_z^2 \Rightarrow EdE=p_zdp_z$.
Then $dp_z/E$ is Lorentz invariant. Since $p_T$ is Lorentz invariant, $d^3p/E$ is also Lorentz invariant.

The Lorentz invariant differential yield is,

\begin{equation}\label{eq2.16}
E\frac{d^3N}{d^3p}=E\frac{d^3N}{d^2p_Tdp_z}=\frac{d^3N}{d^2p_Tdy}
\end{equation}

\noindent where the relation $dp_z/E=dy$ is used. Some times experimental results are given in terms of pseudorapidity. The transformation from $(y,p_T)$ to $(\eta,p_T)$ is the following,

\begin{equation} \label{eq2.17}
\frac{dN}{d\eta dp_T}=\sqrt{1-\frac{m^2}{m_T^2 \cosh^2y}} \frac{dN}{dydp_T}
\end{equation}

\subsection{Luminosity:} The luminosity is an important parameter in collider experiments. The reaction rate in a collider is given by,

\begin{equation} \label{eq2.18}
R=\sigma L
\end{equation}

\noindent where, $\sigma$ is the interaction cross section and $L$ is the luminosity (in $cm^{-2}s^{-1}$), defined as,

\begin{equation}\label{eq2.19}
L=f n \frac{N_1N_2}{A}
\end{equation}

\noindent where,\\
$f\equiv$revolution frequency ,\\ $N_1,N_2 \equiv$ number of particles in each bunch,\\
$n\equiv$number of bunches in one beam in the storage ring,\\ $A\equiv$cross-sectional area of the beams.

\subsection{Collision centrality}

Nucleus is an extended object. Accordingly, depending upon the impact parameter of the collision, several types of collision can be defined,  e.g. central collision when two nuclei collide head on, peripheral collision when only glancing interaction occur between the two nuclei. System created in a central  collision can be qualitatively as well as quantitatively different from the system created in a peripheral collision. Different aspects of reaction dynamics can be understood if heavy ion collisions are studied as a function of impact parameter. Impact parameter of a collision can not be measured experimentally. However, one can 
have one to one correspondence between impact parameter of the collision and some experimental observable. e.g. particle multiplicity, transverse energy ($E_T=\sum_i E_iSin\theta_i$) etc.  For example, one can safely assume that multiplicity or transverse energy is a monotonic function of the impact parameter. High multiplicity or transverse energy  events are from central collisions and low multiplicity or low transverse energy events are from peripheral collisions. One can then group the collisions according to multiplicity or transverse energy.

\begin{figure}[t]
\center
\resizebox{0.45\textwidth}{!}
 {\includegraphics{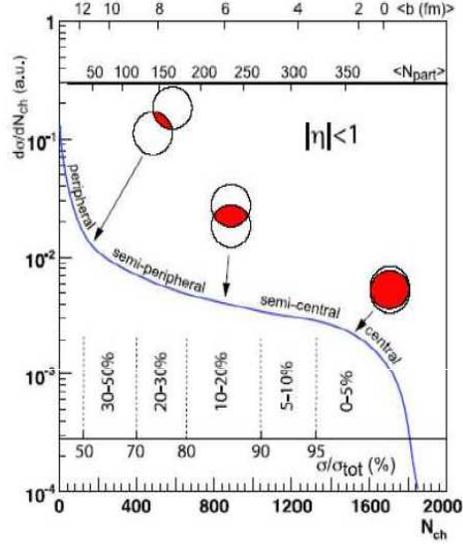}}
\caption{\label{F6}Schematic representation of  multiplicity distribution
in minimum bias nucleus-nucleus collision. }
\end{figure}

It can be done quantitatively. Define a  minimum bias collision where all possible collisions are allowed. In Fig.\ref{F6} charged particles multiplicity ($N_{ch}$) in a minimum bias collision is shown schematically. 
Minimum bias yield can be cut into successive intervals starting from maximum value of multiplicity.  First 5\% of the high $N_{ch}$ events corresponds to top 5\% or 0-5\% collision centrality. Similarly, first 10\% of the high $N_{ch}$ corresponds to 0-10\% centrality. The overlap region
between 0-5\% and 0-10\% corresponds to 5-10\% centrality and so on.  
Similarly, centrality class can be defined by measuring the transverse energy.

Instead of impact parameter, one often defines centrality in terms of number of participating nucleons (the nucleons that undergo at least one inelastic collision) or in terms of binary nucleon collision number. These measures have one to one relationship with impact parameter and can be calculated in a Glauber model.

\subsubsection{Optical Glauber model}
 
Glauber model views AA collisions in terms of the individual interactions of constituent nucleons. It is assumed that at sufficient high energy, nucleons carry enough momentum and are undeflected as the nuclei
pass through each other. It is also assumed that the nucleons move independently in the nucleus and size is large compared to  NN interaction range. The hypothesis of independent linear trajectories of nucleons made it possible to obtain simple analytical expression for nuclear cross section, number of binary collisions, participant nucleons etc. Details of Glauber modeling of heavy ion collisions can be found in  \cite{Miller:2007ri}. Below, salient features of the model are described.

\begin{figure}[t]
 \center
 \resizebox{0.55\textwidth}{!}
 {\includegraphics{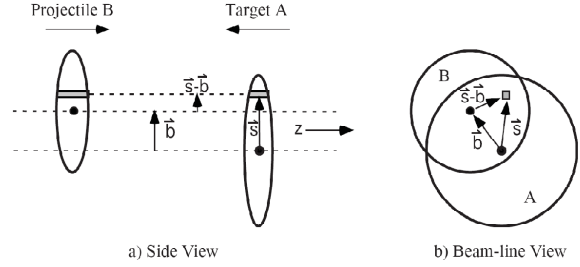}}
\caption{Nucleus-nucleus collisions as viewed in a Glauber model.}
\label{F7}
\end{figure}

In Fig.\ref{F7} collisions of two heavy nuclei at impact parameter ${\bf b}$ is shown. Consider the two flux tubes, (i) located at a displacement  ${\bf s}$ from the centre of target nucleus and (ii) located at a displacement ${\bf s-b}$ from the centre of the projectile nucleus.
During the collision, these two flux tube overlap. Now, for most of the nuclei,  density distribution can be conveniently parameterised by a three parameter Fermi function,

\begin{equation} \label{eq2.20}
\rho(r)=\rho_0\frac{1+w(r/R)^2}{1+exp(\frac{r-R}{a})}
\end{equation}

\noindent where $\rho_0$ is the nucleon density, $R$ the radius, $a$ the skin thickness. $w$ measure the deviation from a spherical shape. In table.\ref{t2}, for selected nuclei, these parameters are listed.

\begin{table}[h]
\caption{ \label{t2} Parameters $R$, $a$ and $w$ of a three parameter Fermi distribution for selected nuclei are given.}
	\centering
		\begin{tabular}{|c|c|c|c|}
		\hline
Nucleus & R (fm) & a(fm) & w (fm)\\	\hline	\hline
$^{16}O$ & 2.608 & 0.513& -0.51 \\ \hline
$^{62}Cu$ & 4.2 & 0.596 & 0.0 \\ \hline
$^{197}Au$& 6.38& 0.535 & 0.0\\	 \hline
$^{208}Pb$& 6.62& 0.594 & 0.0 \\ \hline
$^{238}U$& 6.81& 0.6 & 0.0 \\ \hline
  		\end{tabular}
	\textsl{}
\end{table}

$\rho(r)$ in Eq.\ref{eq2.20}, normalised to unity, can be interpreted as the probability to find a given nucleon at a position $r(=x,y,z)$. Then,

\begin{equation}
T_A({\bf s})=\int dz \rho_A({\bf s},z),\\
\end{equation}

\noindent is the probability that a given nucleon in the nucleus A (say projectile) is at a transverse distance $s$. Similarly, $T_B({\bf s-b})=\int dz \rho({\bf s-b},z)$ is the probability that
a given nucleon in the target nucleus B is at a transverse distance ${\bf s-b}$.
Then $T_A({\bf s}) T_B({\bf s-b})$ is the joint probability that in an impact parameter ${\bf b}$ collision, two nucleons in target and projectile are in the overlap region. One then define a overlap function,
at impact parameter b,

\begin{equation}
T_{AB}({\bf b})=\int d^2s T_A({\bf s}) T_B({\bf s-b})
\end{equation}

Overlap function is in unit of inverse area. We can interpret it as the effective area with which a specific nucleon in A interact with a given nucleon at B. If $\sigma_{NN}$ is the inelastic cross section, then probability of an inelastic interaction is  
$\sigma_{NN} T_{AB}({\bf b})$. Now there can be $AB$ interactions between nucleus A and B. Probability that at an impact parameter {\bf b} there is n interaction is,

\begin{equation}
P(n,{\bf b})= \left (
\begin{matrix}AB \\ n
\end{matrix} \right ) [\sigma_{NN} T_{AB}({\bf b})]^n [1-\sigma_{NN} T_{AB}({\bf b})]^{AB-n}
\end{equation}

The first term is the number of combinations for finding $n$ collisions out of $AB$ collisions, the 2nd term is the probability for having $n$ collisions and the 3rd term is the probability that $AB-n$ collisions do not occur.

The total probability of an interaction between A and B is

\begin{equation}
\frac{d\sigma}{db^2}=\sum_{n=1}^{AB} P(n,{\bf b})=1-[1-\sigma_{NN}T_{AB}({\bf b})]^{AB}
\end{equation}

Total inelastic cross-section is,

\begin{eqnarray}
\sigma_{inel}&=&\int_0^\infty 2\pi b db (1-[1-\sigma_{NN}T_{AB}(b)]^{AB})  \nonumber \\
&\approx& \int_0^\infty 2\pi b db (1-exp(-\sigma_{NN}T_{AB}(b))
\end{eqnarray}
 
Total number of binary collisions is,

\begin{equation} \label{eq4.9}
N_{coll}(b)=\sum n P(n,b)=AB T_{AB}(b) \sigma_{NN}
\end{equation}
 
The number of nucleons in  projectile and target that interacts is called participant nucleons or the wounded nucleons. One obtains,

\begin{eqnarray}
N_{part}(b)&=&A\int d^2sT_A(s)(1-[1-\sigma_{NN}T_{B}(b-s)]^{B}) \nonumber \\
&+&
B\int d^2s T_B(b-s)(1-[1-\sigma_{NN}T_{A}(s)]^{A}) \label{eq4.10}
\end{eqnarray}

 \begin{figure}[t]
 \center
 \resizebox{0.45\textwidth}{!}
 {\includegraphics{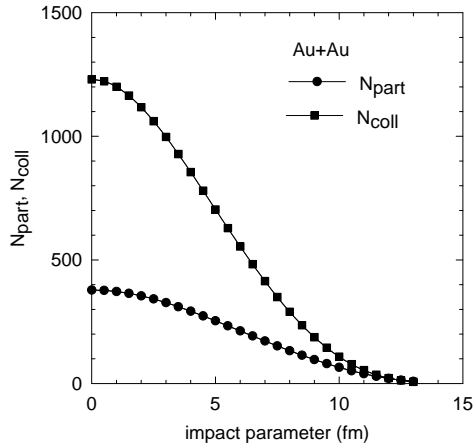}}
\caption{\label{F8}A optical Glauber model calculation for the impact parameter dependence of number of participant nucleons ($N_{part}$) and number of binary collisions ($N_{coll}$) in $\sqrt{s}_{NN}$=200 GeV Au+Au collision. Inelastic nucleon-nucleon cross section is $\sigma_{NN}$=42 mb.}
\end{figure}

Glauber model calculation of binary collision number or participant number is energy dependent through the inelastic NN cross section $\sigma_{NN}$. It is common to take, $\sigma_{NN}\approx$ 30 mb at Super Proton Synchrotron (SPS; $\sqrt{s}_{NN}\approx$20 GeV), 40 mb at Relativistic Heavy Ion Collider (RHIC; $\sqrt{s}_{NN}\approx$200 GeV) and 70 mb at Large Hadron Collider (LHC; $\sqrt{s}_{NN}\geq$1000 GeV).
For demonstration purpose, in Fig.\ref{F8}, I have shown a Glauber model calculation for $N_{part}$ and $N_{coll}$ as a function of impact parameter in $\sqrt{s}_{NN}$=200 GeV Au+Au collision. One understands that there is a one-to-one correspondence between impact parameter $b$ and participant number or collision number.

\subsubsection{Monte-Carlo Glauber model}

In Monte-Carlo Glauber model, individual nucleons are stochastically distributed 
event-by-event and collision properties are calculated averaging over many events. Optical Glauber model and  Monte-Carlo Glauber model give very close results for average quantities like binary collision number or participant numbers. However, in the quantities where fluctuations are important, e.g. participant eccentricity, the results are different. 
Monte-Carlo Glauber model calculations proceed as follows: (i) nucleons in the colliding nuclei are distributed randomly following the probability distribution $\rho(r)$, (ii) an impact parameter is selected randomly from a distribution $dN/db \propto b$, (iii) assuming the nuclei are moving in the straight line, two nuclei are collided,   (iv) if the transverse separation between two colliding nucleons are less than the 'ball diameter'
$D=\sqrt{\sigma_{NN}/\pi}$, they are tagged as interacted, and a register, keeping the coordinates of the colliding nucleons is updated. More details about the model can be found in \cite{Miller:2007ri},\cite{Alver:2008aq}.


\section{QGP and hadronic resonance gas in the ideal gas limit}\label{sec4}

The basic difference between quarks and gluons inside a hadron and quarks and gluons in QGP as existed in early universe or in neutron star or as produced in high energy nuclear collisions, is that as opposed to the former, the later can be treated as a macroscopic system.  
A macroscopic system is generally characterised by some state variables, e.g. number density ($n$), pressure ($p$), energy density ($\varepsilon$), temperature ($T$) etc. Dynamics of the system is then obtained in terms of these state variables.
In kinetic theory this programme is realised by means of a statistical
description, in terms of 'one-particle distribution function' and its transport
equation. Later, I will discuss some aspects of relativistic kinetic theory. Here, following \cite{groot}, I present
some simple calculations for the number density, energy density and pressure of a macroscopic system of particles of mass $m$, chemical potential $\mu$, and at temperature $T$, when particles follow  Maxwell-Boltzmann, Bose and Fermi-Dirac distributions.  

\subsection{Maxwell-Boltzmann distribution}

Maxwell-Boltzmann distribution function is,

\begin{equation}
f(p)=\frac{1}{(2\pi)^3} exp\left (\frac{\mu-E}{T}\right )
\end{equation}

The distribution is of fundamental importance. Bose as well as Fermi-Dirac distributions can always be written as an infinite sum of Boltzmann distribution,

\begin{eqnarray}
f(p)&=&\frac{1}{(2\pi)^3}\frac{1} {e^{(E-\mu)/T} \pm 1} \nonumber\\
&=&\frac{1}{(2\pi)^3} \sum_{i=1}^\infty (\mp)^{n+1} e^{-n(E-\mu)/T}
\end{eqnarray}

The $\pm$ corresponds to Fermi and Bose distribution respectively.

For Boltzmann distribution, the number density, energy density and pressure can be obtained as,

\begin{eqnarray}
n&=& \frac{1}{(2\pi)^3} \int d^3p  exp\left (\frac{\mu-E}{T} \right) \label{eqs.2}\\
\varepsilon&=& \frac{1}{(2\pi)^3} \int d^3p E exp\left (\frac{\mu-E}{T} \right)\\ 
p&=& \frac{1}{(2\pi)^3} \int d^3p \frac{1}{3}\frac{|\vec{p}|^2}{E} exp\left (\frac{\mu-E}{T} \right)
\end{eqnarray} 
 
Let us introduce the dimensionless variables, $z$ and $\tau$

\begin{eqnarray}
z&=&\frac{m}{T} ; \tau=\frac{E}{T}=\frac{\sqrt{|\vec{p}|^2+m^2}}{T},\\
|\vec{p}|&=&T\sqrt{\tau^2-z^2},|\vec{p}|d|\vec{p}|=T^2 \tau d\tau,\\
|\vec{p}|^2 d|\vec{p}|&=&T^3 \tau \sqrt{\tau^2-z^2} d\tau,
\end{eqnarray}

In terms of $\tau$ and $z$, the number density can be written as,

\begin{equation}\label{eq7.7}
n=4\pi\frac{T^3} {(2\pi)^3} e^{\mu/T} \int_z^\infty
d\tau (\tau^2-z^2)^{1/2} \tau e^{-\tau}
\end{equation}
 
Closed form expression can be given for $n$ in terms of the modified Bessel function of the second kind \cite{Abramowitz},

\begin{equation} \label{eq7.8}
K_n(z)=\frac{2^n n!}{(2n)!} \frac{1}{z^n} \int_z^\infty d\tau (\tau^2-z^2)^{n-1/2}  e^{-\tau}
\end{equation} 
 
$K_n(z)$ has another representation which can be obtained from the Eq.\ref{eq7.8} by partial integration, 
 
\begin{equation}\label{eq7.9}
K_n(z)=\frac{2^{n-1} (n-1)!}{(2n-2)!} \frac{1}{z^n} \int_z^\infty\tau (\tau^2-z^2)^{n-3/2} \tau e^{-\tau}
\end{equation}  
 
Modified Bessel function has a nice recurrence relation. If $K_0$ and $K_1$ are known, all the others can be easily obtained. For completeness, the recurrence relation is noted below,

\begin{equation}
K_{n+1}(z)=\frac{2nK_n(z)}{z}+K_{n-1}(z)
\end{equation}

From Eq.\ref{eq7.9} one easily obtain,
 
\begin{equation}
K_2(z)= \frac{1}{z^2} \int_z^\infty\tau (\tau^2-z^2)^{1/2} \tau e^{-\tau},
\end{equation}   
 
\noindent and the number density in Eq.\ref{eq7.7} can be written in a closed form, 
 
\begin{eqnarray} \label{eq4.13}
n&=&\frac{T^3} {2\pi^2}  z^2 K_2(z)=\frac{T^3}{2\pi^2} \left (\frac{m}{T}\right )^2K_2\left (\frac{m}{T}\right )e^{\frac{\mu}{T}}
\end{eqnarray}

Similarly, the energy density can be obtained as, 
 
\begin{eqnarray}
\varepsilon&=&\frac{T^4}{2\pi^2} e^{\mu/T} \int d\tau (\tau^2-z^2)^{1/2} \tau^2 e^{-\tau} \nonumber \\
&=&\frac{T^4}{2\pi^2} e^{\mu/T}  \int d\tau (\tau^2-z^2)^{1/2} (\{\tau^2-z^2\}+z^2) e^{-\tau} \nonumber  
\end{eqnarray}  

Now, from Eq.\ref{eq7.8}

\begin{eqnarray}
zK_1(z)&=&\int_z^\infty d\tau (\tau^2-z^2)^{1/2}e^{-\tau} \\
3z^2 K_2(z)&=&\int_z^\infty d\tau (\tau^2-z^2)^{3/2}e^{-\tau} 
\end{eqnarray} 

and  the final expression for energy density is,
 
\begin{eqnarray}
\varepsilon&=&\frac{T^4}{2\pi^2} e^{\mu/T}[3z^2K_2(z)+z^3 K_1(z)] \nonumber\\
&=&\frac{T^4}{2\pi^2} \left [3\left (\frac{m}{T}\right )^2K_2\left (\frac{m}{T}\right )+\left (\frac{m}{T}\right )^3 K_1\left (\frac{m}{T}\right )\right ]e^{\muT} 
\end{eqnarray}

The expression for pressure is similarly obtained, 

\begin{eqnarray}
p&=&\frac{T^4}{6\pi^2}e^{\muT} \int_z^\infty d\tau (\tau^2-z^2)^{3/2}     e^{-\tau} \nonumber \\
&=&\frac{T^4}{2\pi^2} \left (\frac{m}{T}\right )^2 K_2\left (\frac{m}{T}\right ) e^{\muT}
\end{eqnarray}
 
The expressions for $n$, $\varepsilon$ and $p$ are simplified in the massless limit, $z=m/T \rightarrow 0$, when one can used the asymptotic relation for the modified Bessel function,

\begin{equation}
\lim_{z\rightarrow 0} z^nK_n(z)=2^{n-1} (n-1)!
\end{equation}


\begin{eqnarray}
\mbox{Number density:}n&=&\frac{T^3}{2\pi^2} \left (\frac{m}{T}\right )^2K_2\left (\frac{m}{T}\right )e^{\frac{\mu}{T}}\nonumber\\
_{m/T \rightarrow 0} &=&\frac{T^3}{\pi^2}e^{\muT}
\end{eqnarray} 

\begin{eqnarray}
\mbox{energy density:} \varepsilon&=&\frac{T^4}{2\pi^2} \left [3\left (\frac{m}{T}\right )^2K_2\left (\frac{m}{T}\right )+\left (\frac{m}{T}\right )^3 K_1\left (\frac{m}{T}\right )\right ]e^{\muT} \nonumber \\
_{m/T \rightarrow 0}&=&\frac{3T^4}{\pi^2} e^{\mu/T}
 \end{eqnarray}

\begin{eqnarray}
\mbox{pressure:} p&=&\frac{T^4}{2\pi^2} \left (\frac{m}{T}\right )^2 K_2\left (\frac{m}{T}\right ) e^{\muT} \nonumber\\
_{m/T \rightarrow 0}&=& \frac{T^4}{\pi^2}e^{\muT}
\end{eqnarray}

One do notice that for massless gas, $p=\frac{1}{3}\varepsilon$ relation is obtained. Above equations implicitly assumed that the degeneracy factor $g=1$. If the particle has degeneracy $g$, the expressions for $n$, $\varepsilon$ and $p$ has to be multiplied by the same. 

\subsection{Bose distribution}

We write the Bose distributions  as an infinite sum 
of Boltzmann distributions  
  
\begin{eqnarray}
f(p)&=&\frac{1}{(2\pi)^3}\frac{1} {e^{(E-\mu)/T} - 1} \nonumber\\
&=&\frac{1}{(2\pi)^3} \sum_{n=1}^\infty (+)^{n+1} e^{-n(E-\mu)/T}
\end{eqnarray}

To obtain close expressions, we will need Riemann zeta function.
Riemann zeta function is a function of complex
variable $s=(x+iy)$ and expressed as the infinite series,

\begin{equation}
\zeta(s)=\sum_{n=1}^\infty \frac{1}{n^s}=\frac{1}{1^s}+\frac{1}{2^s}+\frac{1}{3^s}+
\frac{1}{4^s}+...
\end{equation}

one can compute,

\begin{eqnarray*}
\zeta(0)&=&-1/2 \\
\zeta(1)&=&\infty \\
\zeta(2)&=&\pi^2/6 \approx  1.645 \\
\zeta(3)&=&1.202 \\
\zeta(4)&=&\pi^4/90=1.0823
\end{eqnarray*}

One also note an important relation, between Riemann zeta function and Dirichlet eta function,

\begin{equation}
\eta(s)=\sum_{n=1}^\infty \frac{(-1)^{n-1}}{n^{s}}=(1-2^{1-s}) \zeta(s),
\end{equation}

\noindent giving,

\begin{eqnarray*}
\eta(0)&=&(-1)\zeta(0)=1/2\\
\eta(2)&=&\frac{1}{2}\zeta(2)=\frac{1}{2} \frac{\pi^2}{6}\\
\eta(3)&=&\frac{3}{4}\zeta(3)=\frac{3}{4} 1.202\\
\eta(4)&=&\frac{7}{8} \zeta(4)=\frac{7}{8}\frac{\pi^4}{90}
\end{eqnarray*}

Riemann zeta function or more precisely, Riemann hypothesis played and continue to play  an important part in the development of mathematical theory. Riemann zeta function have trivial and non-trivial zeros. It has zeros at the negative even integers. Riemann hypothesis states that non-trivial zeros of zeta function has real part $\frac{1}{2}$, i.e. non-trivial zeros lie on the line $\frac{1}{2}+ i t$, $t$ being a real number. The hypothesis is one of the most challenging problems in mathematics, and is not proved until now. Once Hilbert was asked about what would be in his mind if he is resurrected 1000 years later. He answered that he will inquire if Riemann hypothesis is proved. 

Let us now calculate the number density of  a Bose gas,

\begin{eqnarray}
n&=& \frac{1}{(2\pi)^3} \int d^3p  \left [exp\left (\frac{\mu-E}{T} \right) -1 \right ]^{-1}\nonumber \\
 &=& \frac{1}{(2\pi)^3} \int d^3p \sum_{n=1}^\infty (+1)^{n+1} exp\left (\frac{n\mu-nE}{T} \right)
\end{eqnarray} 

If we define a temperature $T^\prime=T/n$ then above expression can be written as,

\begin{eqnarray}
n&=& \sum_{n=1}^\infty \left [\frac{1}{(2\pi)^3} \int d^3p  exp\left (\frac{\mu-E}{T^\prime} \right) \right ] 
\end{eqnarray} 

The bracketed quantity is just the Eq.\ref{eqs.2}, which has been evaluated in Eq.\ref{eq4.13}. We immediately get,

\begin{eqnarray} 
n &=&\sum_{n=1}^\infty \frac{T^3}{2\pi^2}\frac{1}{n^3} \left (\frac{nm}{T}\right )^2 K_2\left (\frac{nm}{T}\right ) e^{\frac{n\mu}{T}} \label{eq5.30}
\end{eqnarray}

Similarly, one obtain for energy density and pressure,

\begin{eqnarray}
\varepsilon&=& \sum_{n=1}^\infty \frac{1}{n^4}\frac{T^4}{2\pi^2}  \left [3(\frac{nm}{T})^2K_2(\frac{nm}{T})+(\frac{nm}{T})^3 K_1(\frac{nm}{T})\right ] e^{n\mu/T} \label{eq5.31} \\
p&=&\sum_{n=1}^\infty \frac{T^4}{2\pi^2} \frac{1}{n^4}(\frac{nm}{T})^2K_2(nm/T) e^{n\mu/T}\label{eq5.32} 
\end{eqnarray}

in the limit ${m\rightarrow 0,\mu \rightarrow 0}$

\begin{eqnarray}
n&=& \frac{T^3}{\pi^2}\sum_{n=1}^\infty \frac{1}{n^3}=\frac{T^3}{\pi^2}\zeta(3)= 1.202\frac{T^3}{\pi^2} \label{eqs.33}\\
\varepsilon&=&\sum_{n=1}^\infty \frac{3T^4}{\pi^2} \frac{1}{n^4}= 
\frac{3T^4}{\pi^2}\zeta(4)=\frac{\pi^2}{30}T^4 \label{eqs.34}\\
p&=&\sum_{n=1}^\infty \frac{T^4}{\pi^2} \frac{1}{n^4}= 
\frac{T^4}{\pi^2}\zeta(4)=\frac{\pi^2}{90}T^4 \label{eqs.35}
\end{eqnarray}

\subsection{Fermi distribution}

In analogy to Bose particles described in the previous section, 
for Fermion, number density, energy density, pressure can be written as,

\begin{eqnarray} 
n &=&\sum_{n=1}^\infty (-1)^{n+1} \frac{T^3}{2\pi^2}\frac{1}{n^3} \left (\frac{nm}{T}\right )^2 K_2\left (\frac{nm}{T}\right ) e^{\frac{n\mu}{T}} \label{eq5.36}\\
\varepsilon&=& \sum_{n=1}^\infty (-1)^{n+1} \frac{1}{n^4}\frac{T^4}{2\pi^2} [3\left (\frac{nm}{T}\right )^2K_2\left (\frac{nm}{T}\right ) \nonumber \\
&+&\left (\frac{nm}{T}\right )^3 K_1\left (\frac{nm}{T}\right )]e^{\frac{n\mu}{T}} \label{eq5.37} \\
p&=&\sum_{n=1}^\infty (-1)^{n+1} \frac{T^4}{2\pi^2} \frac{1}{n^4}\left (\frac{nm}{T}\right )^2K_2\left (\frac{nm}{T}\right ) e^{\frac{n\mu}{T}}\label{eq5.38}
\end{eqnarray} 

In the limit $m\rightarrow 0$, $\mu\rightarrow 0$,

\begin{eqnarray} \label{eqs.37}
n &=&\frac{T^3}{\pi^2} \sum_{n=1}^\infty (-1)^{n+1} \frac{1}{n^3}=\frac{T^3}{\pi^2} \frac{3}{4}\zeta(3)\\
\varepsilon&=& \frac{3T^4}{\pi^2} \sum_{n=1}^\infty (-1)^{n+1}\frac{1}{n^4}= \frac{7}{8}\frac{\pi^2}{30} T^4  \\
p&=& \frac{T^4}{\pi^2} \sum_{n=1}^\infty (-1)^{n+1} \frac{1}{n^4}=\frac{7}{8}\frac{\pi^2}{90}T^4 
\end{eqnarray} 

One notes that in the massless limit, energy density, pressure in Bose and Fermi distribution differ by the factor $\frac{7}{8}$ only. 

\begin{table}[h]
\caption{ \label{t3} Summary of properties of quarks}
	\centering
		\begin{tabular}{|l|l|l|l|l|}
		\hline
quark  & symbol & Charge & constituent mass & current mass\\		
 flavor &  & Q/e & (MeV) & (MeV)\\ \hline
 up & u & 2/3 & $\sim$ 350 & 1.7-3.1 \\	
 down & d &-1/3 & $\sim$ 350 & 4.1-5.7 \\
 strange & s & -1/3 & $\sim$ 550 & $100^{+30}_{-20}$\\	
 charm & c &2/3 & $\sim$ 1800 & $1.29^{+0.5}_{-0.11}\times 10^3$\\
 bottom & b & -1/3 & $\sim$ $4\times 10^3$ & $4.19^{+0.18}_{-0.06}\times 10^3$\\	
top & t & 2/3 & $\sim$ $170\times 10^3$ & $172.9\pm 1.08\times 10^3$\\	\hline  
  		\end{tabular}
	\textsl{}
\end{table}

\subsection{Number density, energy density and pressure in QGP}

At high temperature, QCD coupling is weak and to a good approximation, quarks and gluons can be treated as interaction free particles. Gluons are massless boson, and Eqs.\ref{eqs.33},\ref{eqs.34} and \ref{eqs.35} derived for massless bosons are applicable. However, they have to be multiplied by the degeneracy factor $g_g$. For gluons, there are 8 colors and two helicity state and degeneracy factor is,

\begin{equation}
g_{gluon}={\mbox color} \times {\mbox spin}=8\times 2
\end{equation}

Quarks are fermions with three color and two spin state. Also, for each quark, there
is an anti-quark. Quarks comes in different (six in total) flavors. However, mass of all the quarks flavors are not the same. In table\ref{t3}, I have listed the constituent and current quark mass of the six known flavors. Current quark mass is the relevant mass here, it enters into the QCD Lagrangian. Constituent quark masses are used in modeling hadrons. In a sense they are dressed current quarks. As seen in table.\ref{t3}, $u$ and $d$ quarks current mass is approximately same and can be assumed to be degenerate.  If mass of $N_f$ flavors are assumed to be same,  the degeneracy factor for quarks can be obtained as,

 \begin{eqnarray}
g_{quark}&=&{\mbox particle-antiparticle}\times{\mbox spin}\times{\mbox color} \times {\mbox flavor} \nonumber\\
& =&2\times 2\times 3\times N_f
\end{eqnarray}

Considering that difference in distribution introduce an additional factor $\frac{7}{8}$ in quark energy density/pressure, one can define a effective degeneracy factor for QGP,

  \begin{eqnarray}
g_{QGP}&=&g_{gluon}+\frac{7}{8}g_{quark} \nonumber \\
&=&16+\frac{21}{2}N_f
\end{eqnarray}

Now, in 1974, a group of physicist at MIT gave a model for hadron structure. The model become very popular and is known as MIT bag model \cite{Chodos:1974je}.
In the model, the quarks are forced by a fixed external pressure to move only inside a fixed spatial region (bag). Inside the bag, they are quasi-free. Appropriate boundary conditions are imposed such that no quark can leave the bag. MIT bag model predict fairly accurate hadron masses. Color confinement is built in the model. However, chiral symmetry is explicitly broken at the bag surface. A remedy was suggested in cloudy bag model \cite{Theberge:1980ye}. 

Equation of state (equation of state is a relation between the state variables, pressure, energy density and number density) of QGP can be approximated by the Bag model. As in the bag model, in high temperature QGP, quarks are approximately free and even though it is a deconfined medium, it is confined in a limited region (albeit, confinement region is of nuclear size rather than of hadronic size).
If $B$ is the 'external bag pressure', the expressions derived earlier can  be augmented with the bag pressure to obtain  energy density and pressure as,
 
\begin{eqnarray}
\varepsilon&=&g_{QGP}\frac{\pi^2}{30}T^4+B\\ 
p&=&g_{QGP}\frac{\pi^2}{90}T^4-B\\
n&\approx&g_{QGP} \frac{3}{4\pi^2}T^3
\end{eqnarray}

In MIT bag model for hadrons, bag pressure $B^{1/4}\sim$200 MeV. However, in the QGP equation of state, bag pressure is obtained by the consideration that
QGP is a transient state and below a critical or (pseudo) critical temperature $T_c$, QGP transform into a hadronic matter or Hadron Resonance Gas. If the transformation is a first order phase transition, the Bag constant is obtained by 
demanding that at the transition temperature $T_c$, pressure of the two phases are equal,

\begin{equation}
p_{QGP}(T_c)=p_{HRG}(T_c)
\end{equation} 

It will be discussed later, but explicit simulations of QCD on lattice indicate that for baryon free ($\mu_B=0$) matter, the transformation of QGP to HRG is not a phase transition in the thermodynamic sense, rather it is a smooth cross-over. In that case, thermodynamic variables in two phases can be joined smoothly to obtain the Bag pressure.

\subsection{Hadronic resonance gas}  

QGP is a transient state. If formed in heavy ion collisions, it will cool back  
to hadronic matter at low temperature.
At sufficiently low temperature, thermodynamics of a strongly interacting matter is dominated by pions. As the temperature increase, larger and larger fraction of available energy goes into excitation of more and more heavier resonances. 
For temperature $T\geq$ 150 MeV, heavy states dominate the energy density. However, densities of heavy particles are still small, $\rho_i \sim e^{-M_i/T}$.
There mutual interaction, being proportional to $\rho_i\rho_j\sim e^{-(M_i+M_j)/T}$, are suppressed. One can use Virial expansion to obtain an effective interaction. Virial expansion together with experimental phase shifts were used by Prakash and Venugopal to study thermodynamics of low temperature hadronic matter \cite{Venugopalan:1992hy}. It was shown that interplay of attractive interactions (characterised by positive phase shifts) and repulsive interactions (characterised by negative phase shifts) is such that effectively, theory is interaction free. One can then consider interaction free resonances constitute the hadronic matter at low temperature.

The   expressions for energy density, pressure and number density for hadronic resonance gas, comprising $N$ hadrons, at temperature $T$ and   chemical potential $\mu$ can be obtained by summing over the same for individual components of HRG,

\begin{eqnarray}
\varepsilon(T,\mu) &=& \sum_{i=1}^N \varepsilon_i(T,\mu_i)\\
P(T,\mu) &=& \sum_{i=1}^N P_i(T,\mu_i)\\
n(T,\mu) &=& \sum_{i=1}^N n_i(T,\mu_i) 
\end{eqnarray}

The chemical potential $\mu_i$ is,
  
\begin{equation}
\mu_i=B_i \mu + S_i\mu_s
\end{equation}

\noindent where $B_i=0,\pm 1, \pm 2 ...$ and   $S_i=0,\pm 1, \pm 2 ...$ are the baryon and strangeness quantum number of the $i$th hadron. 

Earlier, I have derived the expressions for $n_i$, $\varepsilon_i$ and $p_i$, for particles obeying  Fermi distribution (Eqs.\ref{eq5.36},\ref{eq5.37}, \ref{eq5.38}) and for particles obeying Bose distribution (Eqs.\ref{eq5.30},\ref{eq5.31}, \ref{eq5.32}). They can be used in the above equations.
However, in deriving those expressions it was implicitly assumed that particles are point particles. The expressions can be corrected to account for finite size of hadrons. The correction is called 'excluded volume correction'. If $v_i$ is the volume of the $i$th hadron, then available volume is,

\begin{equation}
V^\prime=V(1-\sum_{i=1}^N v_i n_i)
\end{equation}

One can estimate the excluded volume per particle as $1/2$ of spherical volume of radius $2r_h$,

\begin{equation}
v_i=v=\frac{16\pi}{3}r_h^3
\end{equation}

Several procedures are in vogue to include the finite volume effect\cite{Rischke:1991ke},\cite{Yen:1997rv} \cite{Cleymans},\cite{Kapusta:1982qd},\cite{De:2010zg},\cite{Gorenstein:2012sj}. 
For example, in \cite{Rischke:1991ke}, \cite{Cleymans} excluded volume effect is taken into account by reducing all the thermodynamic quantities including pressure by the reduction factor $r=[1+\sum_j v_j n_j^{id}(T,\mu_j)]^{-1}$. How ever the procedure is not thermodynamically consistent. Kapusta and Olive    \cite{Kapusta:1982qd} advocated the following procedure, which is supposed to be 'thermodynamically' consistent. Finite or excluded volume corrected pressure, energy density, temperature and entropy density are,

\begin{eqnarray}
P_{xv}&=&\frac{P_{pt}(T^*)}{1-\frac{P_{pt}(T^*)}{4v}}\\
T_{xv}&=&\frac{T^*}{1-\frac{P_{pt}(T^*)}{4v}}\\
\varepsilon_{xv}&=&\frac{\varepsilon_{pt}(T^*)}{1+\frac{\varepsilon_{pt}(T^*)}{4v}}
\end{eqnarray}

\noindent where $T^*$ is the temperature of the system having point particles. B is the bag pressure, $B^{1/4}$=340 MeV.

In \cite{Yen:1997rv} the 'excluded volume model' pressure is expressed in terms of the ideal (point particle) gas pressure as,

\begin{equation} \label{eq8.13q}
P(T,\mu)=P^{id}(T,\tilde{\mu}), \tilde{\mu}=\mu -v P(T,\mu) 
\end{equation}

For a given excluded volume $v$, Eq.\ref{eq8.13q} can be solved to obtained pressure at a given temperature and chemical potential. Particle number density, energy density can be obtained as,

 \begin{eqnarray} \label{eq8.14}
\varepsilon(T,\mu)&=&\frac{\varepsilon^{id}(T,\tilde{\mu})}{1+v n^{id}(T,\tilde{\mu}) } \\
n(T,\mu)&=&\frac{n^{id}(T,\tilde{\mu})}{1+v n^{id}(T,\tilde{\mu}) }
\end{eqnarray}

See \cite{De:2010zg},\cite{Gorenstein:2012sj}  for more details on excluded volume correction in HRG.

\section{Quantum chromodynamics: theory of strong interaction}\label{sec5}

Modern theory of strong interaction is Quantum Chromodynamics (QCD).
Formally, QCD can be defined as a field theoretical scheme for describing strong interaction. QCD is built on three major   concepts, (i) colored quarks, (ii) interaction between colored quarks results from exchange of spin 1 colored gluon fields and (iii) local gauge symmetry.

(i) Quarks: Quarks are fundamental constituents of matter. Quarks have various intrinsic properties, including electric charge, color charge, spin, and mass. Quarks can come in three colors (e.g. red, green and blue). In table.\ref{t3},   properties of the presently known quarks are listed.
One notes that quarks posses fractional charges. Fractional charges are not observed in isolation. Millikan's oil drop type experiments give negative result for fractional charges.  The experimental fact that quarks (fractional charges) are not observed in isolation, was accommodated in the theory by postulating 'color confinement'. Due to   color confinement, quarks are never found in isolation.   Quarks combine to form physically observable, 'color neutral', particles; mesons (pion, kaon etc.) and hadrons (protons, neutrons etc.)  From table.\ref{t3}, one can identify protons as composite of (uud) and neutrons
as composite of (ddu).  It may be mentioned here that the mechanism of color confinement is not properly understood as yet. QCD Lagrangian is highly singular at small momentum (large distance limit). Numerical simulation of QCD on lattice does indicate confinement.  

(ii) Gluons: Gluons are the mediators of the strong interaction. They are mass less bosons (spin 1).
Indeed, role of photons in QED is played by gluons in QCD. But unlike photons, 
which are not self-interacting, gluons are.
There are eight types of gluons. This can be understood if we note that quarks (anti-quarks) can carry three color charges. They can be combined in 9 different ways, 1 (singlet) colorless state  and 8 (octet) colored states ($3\times \bar{3}=1+8$). Gluons can not occur in a singlet state (color singlet states can not interact with colored states). Hence there can only be 8 types of gluons.

(iii) Gauge theory: QCD is a gauge theory, i.e. Lagrangian is invariant under a   continuous group of local transformations. The Gauge group corresponding to QCD is SU(3). Below, I briefly discuss Gauge theory and SU(3) symmetry group. More detailed exposition can be found in text books,  e.g.\cite{peskin}\cite{Greiner}.

\subsection{Gauge theory in brief}

QCD is based on the principle of local gauge symmetry of color interaction.
Here, I briefly describe the procedure to obtain local gauge symmetric Lagrangian. 

Consider a complex scalar field $\phi(x)$, with Lagrangian density,

\begin{equation}
 \mathcal{L}_0(\phi(x),\partial^\mu\phi(x))=\partial_\mu \phi^* \partial^\mu \phi - V(\phi \phi^*)
\end{equation}

The Lagrangian is invariant under a constant phase change,
\begin{eqnarray}
\phi(x) &\rightarrow & U \phi(x); \hspace{.5cm} U=e^{-i \alpha}
\end{eqnarray}

\noindent
where $\alpha$ is an arbitrary real constant. This transformation is called 'global gauge transformation'. The theory is said to be invariant under 
global gauge transformation under the group $U(1)$. Note $U=e^{-i\alpha}$ is a unitary matrix in one dimension, $U U^{\dagger}=1$. The transformation,

\begin{equation}
\phi^\prime \rightarrow e^{-i\alpha} \phi,
\end{equation}

\noindent is a global gauge transformation under U(1). 

 
If the complex field is written as, 

\begin{eqnarray}
\phi&=&\frac{1}{\sqrt{2}}(\phi_1 + i \phi_2) \\
\phi^*&=&\frac{1}{\sqrt{2}}(\phi_1 - i \phi_2) 
\end{eqnarray}

the transformation: $\phi\rightarrow e^{-i\alpha} \phi$, $\phi^*\rightarrow e^{i\alpha} \phi^*$ gives,

 \begin{eqnarray}
\phi^\prime&=   &\frac{1}{\sqrt{2}}(\phi^\prime_1 + i \phi^\prime_2) =e^{-i\alpha} (\phi_1 + i \phi_2)\\
\phi^{*\prime}&=&\frac{1}{\sqrt{2}}(\phi^\prime_1 - i \phi^\prime_2) =e^{ i\alpha} (\phi_1 - i \phi_2)),
\end{eqnarray}
 
\noindent which is equivalent to,
 \begin{eqnarray}
\phi^\prime&=& (\phi_1 \cos\alpha +  \phi_2 \sin\alpha)\\
\phi^{*\prime}&=&(-\phi_1 \sin\alpha +  \phi_2 \cos\alpha)
\end{eqnarray}

The transformation $\phi(x) \rightarrow e^{-i\alpha}\phi(x)$ can be thought of as a rotation in some internal space by an angle $\alpha$.
Thus U(1) group is isomorphic to O(2), the group of rotation in two dimensions.
(In group theory, two groups are called isomorphic when there is one to one
correspondence between the group elements. Isomorphic groups have the same properties and need not be distinguished).

 In a global gauge transformation, $\phi(x)$ must be rotated by the same angle $\alpha$ in all space-time points. This is contrary to the spirit of relativity, according to which signal speed is limited by the velocity of light.
Then without violating causality, $\phi$ in all the spatial positions can not be rotated by the same angle at the same time. This inconsistency is corrected in
local gauge transformation, where freedom is given to chose the phase locally, the phase angle $\alpha$ become space-time dependent, 
 
\begin{equation}
\phi(x) \rightarrow U(x)\phi(x)=e^{-i\alpha(x)}\phi(x)
\end{equation}

Under such a transformation, 

\begin{equation}
\partial^\mu \phi(x) \rightarrow U(x)\partial^\mu \phi(x) +\underline{ \phi(x) \partial^\mu U(x) }
\end{equation}

\noindent
and Lagrangian is not invariant under the gauge transformation. The 'underlined' term  must be compensated. This can be done by introducing
a gauge field $A^\mu(x)$, which under the local gauge transformation transform  as.

\begin{equation}
A^\mu(x) \rightarrow A^\mu(x) + \frac{1}{e} \partial^\mu \alpha(x),
\end{equation}

\noindent and replacing the partial derivative ($\partial_\mu$) to \underline{covariant} derivative ($D_\mu$)  defined as,

\begin{equation} \label{eq4.16}
D^\mu \phi(x) = [\partial^\mu + i e A^\mu(x) ]\phi(x)
\end{equation}

While, Lagrangian is now invariant under local gauge transformation, it is not the same Lagrangian as before. A gauge field $A^\mu(x)$ is now present as an external field. To obtain a closed system, we need to add a kinetic energy term, to be constructed from $A_\mu$ and its derivatives. The only term which is invariant under the gauge transformation is,

\begin{equation}
F^{\mu\nu}=\partial^\mu A^\nu - \partial^\nu A^\mu
\end{equation}

Thus we arrive at a Lagrangian density for a closed dynamical system, invariant under local $U(1)$ gauge transformation,

\begin{equation} \label{eq6.17}
\mathcal{L}=-\frac{1}{4} F^{\mu\nu}F_{\mu\nu} + D_\mu \phi^* D^\mu \phi - V(\phi \phi^*)
\end{equation}

Lagrangian in Eq.\ref{eq6.17} is essentially for QED, which is a local gauge theory with U(1) group symmetry. Symmetry group for QCD on the other hand is SU(3). In contrast to U(1), which is an abelian group (i.e. group elements commute), SU(3) is non-abelian (group elements do not commute). Non-abelian nature of SU(3) group introduces additional complications.  

\subsection{Brief introduction to SU(3)} 

Special unitary group 
$SU(N)$ is a Lie group isomorphic to that of all $N\times N$ special unitary matrices,

\begin{eqnarray}
det U= 1\\
U^{\dagger}U=1
\end{eqnarray} 

In general $N\times N$ complex matrices has $2N^2$ arbitrary real parameters. The condition $U^{\dagger}U=1$ imposes $N^2$ condition and $det U=1$ one condition. Hence $SU(N)$ has $N^2-1$ arbitrary parameters. Correspondingly $SU(N)$ has $N^2-1$ generators, $L_\alpha$, obeying,

\begin{equation} \label{2eq3}
[L_\alpha, L_\beta]=i \sum_{\gamma=1}^{\gamma=N} f_{\alpha\beta\gamma}L_\gamma 
\end{equation}

$f_{\alpha\beta\gamma}$ are the 'antisymmetric' structure constants (changes sign
for interchange of consecutive indices, $f_{\alpha\beta\gamma}$=$-f_{\alpha\gamma\beta}$=$f_{\gamma\alpha\beta}$. One immediately notes that $SU(3)$ is non-abelian, generators or the group element do not commute (in an abelian' group, structure constants are zero and generator and the group elements commute). For reference purpose, structure constants for SU(3) are noted in table.\ref{t4}.

Naturally $SU(3)$ has 8 generators, 
$L_\alpha=\frac{1}{2}\lambda_\alpha, (\alpha=1,2...8)$.
$\lambda_\alpha$ are ($3\times 3$) Gell-Mann matrices, they act on the (color) basis states

\begin{equation}
x= \left( \begin{array}{c}
x_1  \\
x_2  \\
x_3  \end{array} \right)\ 
\end{equation}

For the sake of completeness, I have listed the 8 Gell-Mann matrices. 

\begin{equation} \nonumber
\lambda_1=  \left( \begin{array}{ccc}
0 & 1 & 0 \\
1 & 0 & 0 \\
0 & 0 & 0 \end{array} \right),\
\lambda_2= \left( \begin{array}{ccc}
0 & -i & 0 \\
i & 0 & 0 \\
0 & 0 & 0 \end{array} \right),\
\end{equation} 

 \begin{equation} \nonumber
\lambda_3= \left( \begin{array}{ccc}
1 & 0 & 0 \\
0 & -1 & 0 \\
0 & 0 & 0 \end{array} \right),\  
\lambda_4= \left( \begin{array}{ccc}
0 & 0 & 1 \\
0 & 0 & 0 \\
1 & 0 & 0 \end{array} \right),\ 
\end{equation}

\begin{equation} \nonumber
\lambda_5= \left( \begin{array}{ccc}
0 & 0 & -i \\
0 & 0 & 0 \\
i & 0 & 0 \end{array} \right),\  
\lambda_6= \left( \begin{array}{ccc}
0 & 0 & 0 \\
0 & 0 & 1 \\
0 & 1 & 0 \end{array} \right)\ 	
\end{equation}

\begin{equation} \nonumber
\lambda_7= \left( \begin{array}{ccc}
0 & 0 & 0 \\
0 & 0 & -i \\
0 & i & 0 \end{array} \right),\  
\lambda_8=\frac{1}{\sqrt{3}} \left( \begin{array}{ccc}
1 & 0 & 0 \\
0 & 1 & 0 \\
0 & 0 & -2 \end{array} \right).\
\end{equation}

\begin{table}[h] 
\caption{\label{t4} Structure constant for SU(3)}
	\centering
		\begin{tabular}{|c|c|c|c|c|c|c|c|c|c|}
		\hline
ijk& 123 & 147 & 156 & 246 & 257 & 345 & 367 & 458 & 678\\ \hline
$f_{ijk}$ & 1 & $\frac{1}{2}$ & -$\frac{1}{2}$ & $\frac{1}{2}$ & $\frac{1}{2}$ & $\frac{1}{2}$ & -$\frac{1}{2}$ & $\frac{\sqrt{3}}{2}$ & $\frac{\sqrt{3}}{2}$
\\	\hline  
  		\end{tabular}
	\textsl{}
\end{table}

One does note that Gell-Mann matrices are generalisation of Pauli matrices;

\begin{equation} \nonumber
\sigma_1= \left( \begin{array}{cc}
0 & 1 \\
1 & 0 \end{array} \right),\  
\sigma_2= \left( \begin{array}{cc}
0 & -i \\
i & 0  \end{array} \right).\
\sigma_3= \left( \begin{array}{cc}
1 & 0 \\
0 & -1  \end{array} \right).\
\end{equation}

Mathematically, quark fields transforms as the fundamental representation of color group SU(3). An infinitesimal element of the group is represented by the transformation,
 
\begin{eqnarray}
x^\prime =&&S x \\
	S=&& e^{-\frac{i}{2}\omega_\alpha \lambda_\alpha} \approx 1-\frac{i}{2}\omega_\alpha \lambda_\alpha
\end{eqnarray}

\noindent where $\omega_\alpha (\alpha=1,2..8)$ are arbitrary infinitesimal real numbers.

\subsection{Lattice QCD}

Schematically, QCD Lagrangian has the form,

\begin{equation}
\mathcal{L}=-\frac{1}{4}F^{\mu\nu}_a F^a_{\mu\nu}+ \sum_{flavors}[ i\bar{\psi}\gamma^\mu(\partial_\mu-ig\frac{\lambda_a}{2}A^a_\mu)\psi
-m \psi \bar{\psi}]
\end{equation}

\noindent with,

\begin{equation}
F^a_{\mu\nu}=\partial_\mu A^a_\nu-\partial_\nu A^a_\mu + g f^a_{bc} A_\mu^b A_\mu^c
\end{equation}

\noindent here 
$A^a_\mu$ is the Gluon Gauge field of color a (a=1,2,...8),and $m$ is the 'bare' quark mass, $f_{abc}$ is the structure constant of the Group and $\psi$ the quark spinors, 

\begin{equation}
\psi= \left( \begin{array}{c}
\psi_r  \\
\psi_g  \\
\psi_b  \end{array} \right)\ 
\end{equation}

Though the Lagrangian looks simple, it is not possible to solve it analytically. Only in the high momentum regime, it can be solved perturbatively.
Perturbative approach however fails in the low momentum regime. The reason being the running of the coupling constant.

\begin{figure}[t]
 \center
 \resizebox{0.50\textwidth}{!}
 {\includegraphics{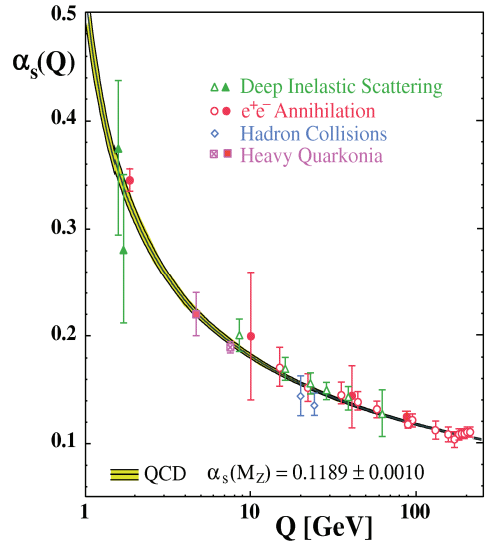}}
\caption{Running strong coupling constant.}
 \label{F9}
\end{figure}

Running coupling constant reflect the change in underlying force law, as the energy/momentum scale, at which physical processes occur, varies. As an example, an electron in short distance scale can appear to be composed of electron, positron and photons. The coupling constant has to be {\em renormalised} to incorporate the change as the scale of physical processes varies. In QED, effective coupling constant at the scale $q$ can be written as, 

\begin{equation}
\alpha_{eff}(q^2)=\frac{e^2}{4\pi}=\frac{\alpha(q_0^2)} 
{1-\frac{\alpha(q_0^2)}{3\pi} \log\frac{q^2}{q_0^2}} 
\end{equation}

QED coupling increase as the momentum scale is increased. In the other words, effective electric charge becomes much larger at small distances.

In QCD coupling constant, at the momentum scale q is,

\begin{equation}
\alpha_s(q^2)=\frac{g^2}{4\pi}=\frac{\alpha_s(\Lambda^2)} {1+\frac{\alpha_s(\Lambda^2)}{4\pi}
(11-\frac{2N_f}{3} ) \ln\frac{q^2}{\Lambda^2}} 
\end{equation}

\noindent where $N_f$ is the number of flavors and $\Lambda\approx$200 MeV is the QCD scale parameter. The coupling constant thus increases as the momentum scale $q$ decreases.
Perturbative expansion in terms of coupling constant will not converge. In Fig.\ref{F9}, I have shown the experimentally measured values of $\alpha_s$. Experimental measurements agree closely with QCD predictions.

One possible way to obtain equation of motion is to simulate QCD on a lattice, i.e. solve the Lagrangian numerically. In lattice simulation, the space-time is discretized to reduce the infinite degrees of freedom of 'Field variables' to a finite and (numerically) tractable number. One immediately notices that due to finite dimension of the lattice, Lorentz invariance is broken. Gauge invariance however, is kept explicitly, by parallel transportation of the gauge fields between adjacent lattice sites. In the continuum limit, lattice spacing $a\rightarrow 0$, Lorentz invariance can be restored.

In the following, I briefly discuss some aspects of lattice QCD. Lattice QCD is intimately related to Feynman's path integral formulation of Quantum mechanics.  Below I briefly sketch the ideas behind the path integral method and parallel transport. For more informative exposure on lattice QCD, see 
\cite{Greiner},\cite{Kogut:1982ds},\cite{Karsch:2001cy},\cite{Lepage:1998dt}.

\subsubsection{Path integral method}

Richard Feynman is one of the most celebrated physicists of twentieth century. Apart from the path integral formulation of Quantum mechanics, he made pioneering contributions in Quantum electrodynamics, superfluidity and particle physics. He invented the diagrammatic approach of QED (the Feynman diagrams). In 1965, Feynman, along with Julian Schwinger and Sin-Itiro Tomonaga, was awarded Nobel prize for their contributions in QED.

\begin{figure}[h]
 \center
 \resizebox{0.35\textwidth}{!}
 {\includegraphics{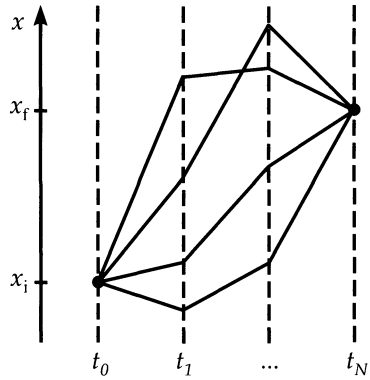}
 }
\caption{Particle trajectories or paths between ($x_i,t_0$) and ($x_f,t_N$) for discrete time steps.}
 \label{F10}
\end{figure}

Consider the propagation of a particle from position $x_i$ at time $t_0$ to  
position $x_f$ at time $t_N$. For a given trajectory, $(x,\dot{x})$, the action is,
 
\begin{equation}
S=\int_{t_0}^{t_N} dt L(x,\dot{x}),  
\end{equation}

\noindent $L(x,\dot{x})$ is the Lagrangian.
Path integral method states that the transition probability from ($x_i,t_0$) to
($x_f,t_N$) can be expressed as the weighted sum of all the possible paths or trajectories, 

\begin{equation}
\la x_f(t_N)|x_i(t_0)\ra \sim \sum_{paths(P)} exp[iS(x(t),\dot{x}(t))]
\end{equation}

Let us discretize the time interval into N steps, $t_N-t_0=N\Delta t$. In Fig.\ref{F10}, the discretized paths are shown . One understands that for small enough time steps, any continuous path can be adequately traced. 
Now, one can sum the trajectories at a particular time step, say $t_n$.

\begin{equation}
 \sum_{P(t=t_n)} e^{iS(x(t),\dot{x}(t))} \sim \int_{-\infty}^\infty dx(t_n) e^{iS(x(t_n),\dot{x}(t_n))}
\end{equation}

The procedure can be repeated for each time steps. In the limit $N\rightarrow \infty$,

\begin{equation}
\la x_f(t_N)|x_i(t_0)\ra \sim \mathcal{N} \int \prod^{N-1}_{n=1} e^{iS(x(t_n),\dot{x}(t_n))}
\end{equation}

$\mathcal{N}$ is some normalisation.

It is easy to extend the formalism to fields. Consider a one dimensional  field $\phi(x,t)$. Again, consider the transition amplitude for the field $\phi_i(x,t_i)$  to $\phi_f(x,t_f)$,

\begin{equation}
\la \phi_i(x,t_i)|\phi_f(x,t_f)\ra \sim \sum_{\phi_P} e^{iS( \{ \phi_P \}, \{\partial_\mu \phi_P\}) }
\end{equation}

As before, we discretize the time intervals in $N$ steps. Additionally, we discretize the space coordinates into $N_x$ steps. Note that space is infinite
dimension. Thus, discretization can only be an approximation of infinite space.

\begin{eqnarray}
\la \phi_i(x,t_i)|\phi_f(x,t_f)\ra  \sim \mathcal{N} 
\lim_{N_x,N\rightarrow \infty} \int \prod_{m=1}^{N_x} \prod_{n=1}^{N-1}  d\phi(x_m,t_n) e^{iS( \{ \phi_(x_m,t_n) \}, \{\partial_\mu \phi_(x_m,t_n)\}) }\nonumber \\
\end{eqnarray}

 It is convenient to make a Wick's rotation, $t=-i\tau$ so that the space is Euclidean. Then,
 
 \begin{equation}
 iS=i\int dtd^3x \mathcal{L}=- \int d\tau d^3x \mathcal{L}_E=-S_E
 \end{equation}

In terms of the Euclidean action ($S_E$), the transition probability can be written as,

\begin{eqnarray}
\la \phi_i(x,t_i)|\phi_f(x,t_f)\ra  &\sim& \mathcal{N} 
\lim_{N_x,N\rightarrow \infty} \int \prod_{m=1}^{N_x} \prod_{n=1}^{N-1}  d\phi(x_m,t_n) e^{-S_E( \{ \phi_(x_m,t_n) \}, \{\partial_\mu \phi_(x_m,t_n)\}) }\nonumber \\
&\sim& \mathcal{N}  \int \mathcal{D}[\phi] e^{-S_E} 
\end{eqnarray}

$\mathcal{D}[\phi]$ is the shorthand notation of the integration measure,

 \begin{equation}
 \mathcal{D}[\phi]=\prod_{m=1}^{N_x} \prod_{n=1}^{N-1}  d\phi(x_m,t_n)
\end{equation}

Now in statistical mechanics, central problem is to compute the partition function, defined as,

\begin{equation}
Z=\sum_\phi \la \phi(x)|e^{-\beta H}|\phi(x)\ra,
\end{equation}

\noindent the summation is over all the possible state $|\phi(x)\ra$. $\beta=1/T$ is the inverse temperature. It can be rewritten as,

\begin{eqnarray}
Z&=&\sum_\phi \la e^{-\beta H}\phi(x)|\phi(x)\ra \nonumber\\
&=&\sum_\phi \la \phi(x,t=-i\beta)|\phi(x,t=0)\ra,
\end{eqnarray} 

The partition function in statistical mechanics then corresponds to the path integral formulation for the transition probability.

\begin{equation}
Z=\mathcal{N} \int \mathcal{D}[\phi] e^{-S_E}
\end{equation} 

This is an important realisation. All the tools of statistical mechanics can be applied to field theory problems. Expectation value of any observable can be obtained as,

\begin{equation}
\la \mathcal{O} \ra = \frac{1}{Z}\int \mathcal{D}[\phi] e^{-S_E} \mathcal{O}
\end{equation}

\subsubsection{Parallel transport}

One of the problems in general relativity is the derivative of a vector (or more generally, a tensor) quantity. In flat space-time, derivative of a vector can be computed easily,

\begin{equation}
{V^\prime}^\mu(x)=\lim_{h\rightarrow 0}\frac{V^\mu(x+h)-V^\mu(x)}{h} 
\end{equation}

\begin{figure}[t]
 \center
 \resizebox{0.35\textwidth}{!}
 {\includegraphics{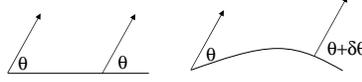}}
\caption{In a parallel transport, tangent angle of a vector is changed in curved space-time, but not in a flat-space-time. } 
 \label{F11}
\end{figure}

However, in a curved space-time, since the metric tensor $g^{\mu\nu}$ depend on space, additional terms arise. This can be understood from Fig.\ref{F11}. In flat space, a vector at $x$, when transported to $x+h$, the tangent angle remains the same. But, in a curved space, the tangent angle is changed. In general relativity, this is accommodated by defining covariant (or semicolon) derivative,

\begin{equation} \label{eq5.42a}
\partial_{;\mu}V^\nu=\partial_\mu V^\nu+ \Gamma^\mu_{\nu\alpha}V^\alpha
\end{equation}

\noindent where $\Gamma^\mu_{\nu\alpha}$ is the Christoffel symbol and is defined as,

\begin{equation}
\Gamma^\mu_{\nu\alpha}=\frac{1}{2}g^{\mu m}
\left (\frac{\partial g_{m\nu}}{\partial x^\alpha}+\frac{\partial g_{m\alpha}}{\partial x^\nu}
-\frac{\partial g_{\nu\alpha}}{\partial x^m} \right )
\end{equation}

 $\Gamma^\mu_{\nu\alpha} V^\alpha$ in Eq.\ref{eq5.42a} accounts for the change in the vector's coordinate representation during the transport ($\Gamma^\mu_{\nu\alpha}=0$ is flat space time).

Covariant derivative,

\begin{equation} \label{eq5.23}
D^\mu \phi(x) = [\partial^\mu + i e A^\mu(x) ]\phi(x)
\end{equation}

\noindent defined in Eq.\ref{eq4.16}, is analogous to parallel transport, $ieA^\mu(x)$ is the change in the field's representation during transport from $x^\mu$ to $x^\mu+dx^\mu$. Then,

\begin{eqnarray}
\phi(x^\mu+dx^\mu)&=&\phi(x^\mu)+dx^\mu D_\mu \phi(x^\mu)\nonumber \\
&=& \phi(x^\mu)+dx^\mu (\partial_\mu+ieA_\mu) \phi(x^\mu) \nonumber \\
&=&dx^\mu \partial_\mu \phi 
+[1+ieA_\mu dx^\mu) ]\phi(x^\mu) \label{eq5.24}
\end{eqnarray}

The first term in Eq.\ref{eq5.24} is essentially a translational term. The 2nd term containing the $A^\mu$ describe the transport of gauge field between two close points $x$ and $x+dx$.
For infinitesimal distance, the 2nd term in Eq.\ref{eq5.24} can be written as,
$e^{ieA_\mu dx^\mu}\phi(x^\mu)$. By repeated application of infinitesimal transport,
the current (gauged) value of phase of the wave function $\phi$, at the 4-dimentional space-time  point $y$ is related to its value at some reference point $x$ by the parallel transport,

\begin{equation}\phi(y)=e^{ie\int_x^y d\zeta^\mu A_\mu(\zeta)} \phi(x), \end{equation}

\noindent  the integration in the exponent goes along some path $C_{xy}$ that connects $x$ and $y$. For the non-abelian gauge group SU(3), a quark can alter its color under parallel transport. Then for SU(3) gauge fields, the exponential or the phase factor is a 3x3 unitary matrix. An extension of the above equation can be written as,  

\begin{equation}\phi(y)=P e^{ig\int_x^y d\zeta^\mu A^\alpha_\mu(\zeta) L_\alpha} \phi(x) \end{equation}

The symbol $P$ means path ordering. To construct the matrix of parallel transport at finite distance, one has to subdivide the path $C_{xy}$ into small parts and form ordered product of parallel transport along these small parts:
 
\begin{equation} U=P e^{ig\int_x^y d\zeta^\mu A^\alpha_\mu(\zeta) L_\alpha}  = \prod_\zeta (1 + i g d\zeta^\mu A^\alpha_\mu(\zeta) L_\alpha ) \end{equation}

$U$ is the path dependent representation of an element of the gauge group G (presently SU(3)).

\subsection{Lattice formulation of QCD}

\begin{figure}[h]
 \center
 \resizebox{0.45\textwidth}{!}
 {\includegraphics{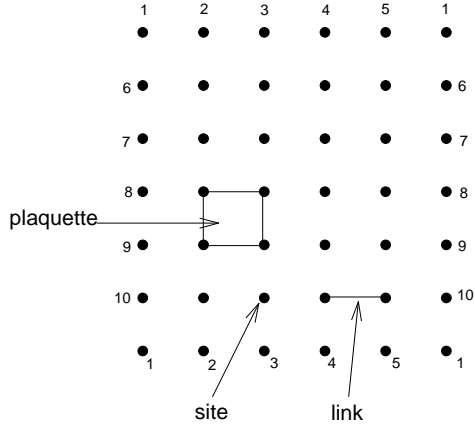}}
\caption{\label{F12}Schematic representation of a lattice, in two dimensions.} 
\end{figure}

Lattice is a regular set of space-time points. A schematic representation of a lattice in two dimensions is given is Fig.\ref{F12}. For our purpose, we define,

(i) site (node): the lattice points, characterised by the coordinate $x$, generally in unit of the lattice spacing.

(ii) Link: shortest distance connecting two sites, characterised by coordinates and direction,

(iii)plaquette: elementary square bounded by 4-links, characterised by coordinates and two directions.

In general, one also imposes periodic boundary conditions for bosons, $x_{n+1}=x_1$ and anti-periodic  boundary condition for fermions, $x_{n+1}=-x_1$

 Lattice QCD simulations are computer intensive.
Total number of degrees of freedom is very large on lattice. The fermions are defined on the nodes (site), $\psi^a_\alpha(i)$, where $a(=1,2,3)$ is the color index   and $\alpha(=1,2,3,4)$ is the Dirac index. They are complex, requiring 24 real variables per node.   One associates the gauge fields with the links, $U^{ab}_{i\rightarrow j}$, where $i$ and $j$   are the neighbouring points and a,b are color indices. $U$ is a unitary $3\times3$ matrix, a total of 9 complex variables times   four possible directions, i.e. 72 real variables per node for the link variables. In total in each node we have (24$N_f$+72) variables. For two flavor QCD, even a small lattice $16^4$ will deal with $7864320$ real variables. Effectively, one has to compute a $7864320$ fold integration.

The relation between the matrices $U$ and the gauge field $A_\mu^\alpha$ is the following,

\begin{equation}
U_\mu(x)=[exp(i\hat{A}_\mu dx^\mu )], \hat{A}_\mu=\sum_\alpha \lambda_\alpha  A^\alpha_\mu
\end{equation}

 $U_\mu(x)$ is the SU(3) matrix attached to the lattice link connecting the sites at  $x$ and $x+dx$, in the direction $\mu$.  Inverse of the matrix
 connects the sites in the   opposite direction, 

\begin{equation}
U_{-\mu}(x+dx)=U^{-1}_\mu(x)=U^\dagger_\mu(x)
\end{equation}

In lattice QCD, one evaluates the partition function,

\begin{equation}
Z=\int d[U] d[\psi] d[\psi^\prime] e^{-S_E(U,\psi,\psi^\prime)}
\end{equation}

\noindent where the action $S_E(U,\psi,\psi^\prime)=\int d^4x \mathcal{L}(\psi,\psi^\prime,A)$ and $d[\psi]=\prod_n \psi_n$ represents all the possible paths.  

\begin{figure}[h]%
 \center
 \resizebox{0.35\textwidth}{!}
 {\includegraphics{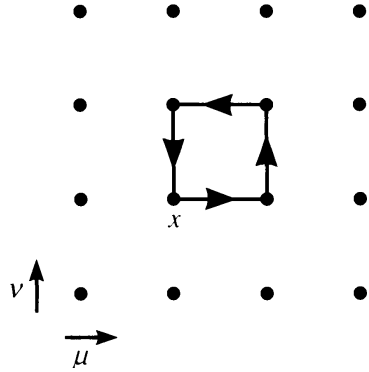}}
\caption{\label{F13}A plaquette on the lattice. The unit vectors, $\mu$, $\nu$ are two generic directions.} 
\end{figure}

Gauge invariance is explicitly maintained In lattice QCD. As mentioned earlier,
quark fields are placed on the nodes and gauge fields are associated with the links.
One then parallel transports the gauge fields from lattice site $n$ to $n+1$, maintaining gauge invariance.  Gauge invariant objects are made from
gauge links between quark and anti-quark or products of gauge fields in a closed loop. In Fig.\ref{F13}, simplest close loop of gauge field is shown. It is called plaquette, product of 4 links connecting 4 adjacent nodes.

\begin{equation}
P_{\mu\nu}(x)=U_\mu(x)U^\dagger_\nu(x) U^\dagger_\mu(x+dx)U_\nu(x+dx)
\end{equation}

Let us consider each term separately (  we have omitted  the $\hat{}$ for ease),

\begin{eqnarray*}
U_\mu(x)&\approx& exp(iaA_\mu(x+ a e_\mu/2)\\
&\approx& exp(ia[A_\mu(x)+a/2\partial_\mu A_\mu(x) ])\\
U^+_\nu(x) &\approx& exp(-ia[A_\nu(x)+a/2\partial_\nu A_\nu(x) ])\\
U^+_\mu(x+ae_\nu) &\approx& exp[-iaA_\mu(x+a(e_\nu+e_\mu/2))] \\
&\approx& exp[-ia(A_\mu(x)+a\partial_\nu A_\mu(x)+a/2 \partial_\mu A_\mu (x))]\\
U_\nu(x+ae_\mu)&\approx&exp( ia[A_\nu(x)+a\partial_\mu A_\nu(x)+a/2\partial_\nu  A_\nu(x)])
\end{eqnarray*}

Product of the links then gives,

\begin{eqnarray}
P_{\mu\nu}(x)&\approx& exp  (ia^2[\partial_\mu A_\nu (x)-\partial_\nu A_\mu (x)] \nonumber\\
&+&a^2[A_\mu(x),A_\nu(x)]   ) \nonumber \\
&\approx& Lim_{a\rightarrow 0}exp (ia^2F_{\mu\nu}(x)) \nonumber\\
&\approx& 1+ia^2 F_{\mu\nu}-a^4/2 F_{\mu\nu}F^{\mu\nu} + ...
\end{eqnarray}

The term $F_{\mu\nu}$ vanishes when summed over the indices $\mu$ and $\nu$ and one obtain,

\begin{equation}
a^4F_{\mu\nu}F^{\mu\nu} =2[1-P_{\mu\nu}(x)]
\end{equation}

Now the pure gauge action in the continuum, in terms of the scaled field $A^\mu \rightarrow \frac{1}{g}A^\mu$,

\begin{equation}
S= \frac{1}{g^2}\int d^4x \left [-\frac{1}{4} F^\alpha_{\mu\nu} F^{\mu\nu}_\alpha \right ]
\end{equation}

Comparing above two equations, pure gauge action on the lattice can be written as, 
\begin{eqnarray}
S_{G}&=&\frac{1}{g^2}\sum_{\mu,\nu,x} Tr[1-P_{\mu\nu}(x)]
\end{eqnarray}

\subsection{Fermions on lattice}

  Adding quarks to lattice action needs additional effort. Quark fields are defined   on the nodes. Quarks are fermions and obey Pauli exclusion principle. Thus they have to be included as anticommuting Grassmann numbers. Grassmann numbers are mathematical construction such that they are anti-commuting. A collection of Grassmann variables $\theta_i$  are independent elements of an algebra which contains the real numbers that anticommute with each other but commute with ordinary numbers $x$

\begin{eqnarray}
\theta_i \theta_j &=& - \theta_j \theta_i\\
\theta_i x &=& x \theta i\\
\theta_i^2&=&0
\end{eqnarray}

One also note that the operation of integration and differentiation are identical in Grassmann algebra,

\begin{eqnarray}
\int 1 d\theta= 0\\
\int \theta d\theta=1
\end{eqnarray}

Grassmann numbers can always be represented as matrices.
In general, a Grassmann algebra on n generators can be represented by $2^n \times 2^n$ square matrices.

  In continuous Euclidean space-time, a fermion field $\psi(x)$ has the action,
\begin{equation}
\int d^4x \bar{\psi}(x)(i\gamma^\mu D^\mu - m) \psi(x)
\end{equation} 

On the lattice is translate into,
 
\begin{equation}
S_F(U,\psi)= \sum_{x,y} \bar{\psi}(x)M(U,x,y)\psi(y)
\end{equation}

\noindent where $M$ is the Dirac matrix, essentially lattice rendering of the Dirac operator, $\slashed{D}+m$. The functional integral for the partition function then become, 

\begin{equation}
Z= \int  [dU][d\psi][d\bar{\psi}]   e^{-S_G(U)-S_F(U,\psi)}
\end{equation}

Computing numerically with Grassmann variables is non-trivial. One generally integrate out the fermion fields, leaving only the gauge fields, weighted by the determinant of the Dirac matrix $M$,

\begin{equation}
Z= \int  [dU]    e^{-S_G(U)} det[M(U)] \label{eq5.42}
\end{equation}
 
Before proceeding  further, I must mention the well known problem of 'Fermion doubling'. If fermion action is naively discretized on a lattice, spurious states appear. For each fermion on the lattice one obtain $2^{d=4}=16$ fermions.  
There are many ways to formulate Fermion action on a lattice, e.g. Wilson fermions, staggered fermions, domain wall fermions etc. We would not elaborate on them.
We just mention that till today, Fermion action on lattice is inadequately treated. 

\subsection{Metropolis Algorithm}

Partition function in Eq.\ref{eq5.42} is a many fold integration. 
One generally uses Monte Carlo sampling to evaluate the partition function. One such algorithm is by Metropolis. There are several other algorithms also. 
Metropolis algorithm is based on the principle of detail balance. 



Metropolis algorithm proceeds as follows:

(i)start from arbitrary configuration (e.g. randomly distributed),

(ii)looks at the value of the field (say $\phi$) at any given point and change it:
$\phi \rightarrow \phi^\prime$,

(iii)calculate the variation in action :$\delta S=S(\phi^\prime)-S(\phi)$. 
if $\delta S$ is negative, it is a lower energy state and desirable. One replace the old value $\phi$ with the new value $\phi^\prime$. If $\delta S$ is positive, one accepts the new value with the probability $exp(-\delta S)$. 

The procedure, after many iterations will produce a equilibrium distribution.
Any physically relevant observable can be computed from the equilibrium partition function,

\begin{equation}
\la f(U) \ra =\frac{\int d[U] e^{-S_{eq}} f(U) }{\int d[U] e^{-S_{eq}} }
\end{equation}

\subsection{Wilson loop}

\begin{figure}[h]
 \center
 \resizebox{0.45\textwidth}{!}
 {\includegraphics{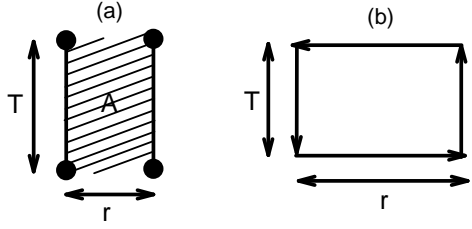}}
\caption{(a) Area spanned by the $q\bar{q}$ pair at a relative distance $r$, as a
function of time, (b) a Wilson loop. } 
 \label{F14}
\end{figure}

Consider a $q\bar{q}$ pair at a distance $r$. A schematic representation of the evolution of the pair is shown in Fig.\ref{F14}a. In quantum mechanics, time evolution of the pair is governed by;

\begin{equation}
\psi(t)=e^{-iE_{q\bar{q}}t} \psi(t=0)
\end{equation}

For confining quark potential ($V(r)\approx k r$), as kinetic energy goes as $1/m$, for  infinitely heavy quarks, $E_{q\bar{q}}\approx k r$. In Euclidean space-time ($t\rightarrow -i\tau$), time evolution of the pair is then governed by,

\begin{equation}
e^{-iE_{q\bar{q}}t}\rightarrow e^{-k r \tau}=e^{-k A}
\end{equation}

\noindent where $A$ is the area spanned by the $q\bar{q}$ system during its evolution.  

The Wilson loop is defined as the trace of the gauge fields along the world line. A typical Wilson loop is shown in Fig.\ref{F14}b. It is just the product of link variables along the contour 

\begin{equation}
w(r,T)=tr U_1 U_2...U_N
\end{equation}

In the continuum, expectation value of Wilson loop, for large $T$ and $r$ is,

\begin{equation}
\la w(r,T) \ra \sim  \la e^{-i\int_c dx^\mu A_\mu} \ra \sim e^{-k.(Area)}
\end{equation}

The area law is a manifestation of confinement.

\subsection{Lattice QCD at finite temperature}

QCD at finite temperature can be simulated on a lattice where one of the 4-dimension, say the time, is much smaller than the others. In the limit where the space dimensions go to infinity, but the time remains finite, the value of the temperature can be related to the time size,

\begin{equation}
Time = \frac{1}{Temperature}
\end{equation}

Finite temperature QCD is then studied on a anisotropic lattice with, 

\begin{equation}
N_t <<N_x=N_y=N_z
\end{equation}
 
\begin{figure}[h]
 \center
 \resizebox{0.35\textwidth}{!}
 {\includegraphics{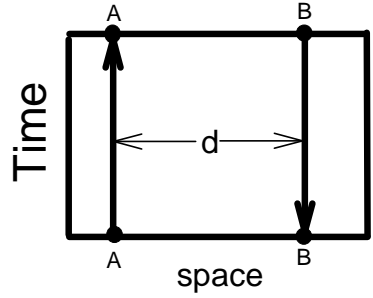}}
\caption{Schematic representation of two Polyakov loop separated by a distance d.} 
 \label{F15}
\end{figure}

 The central role in QCD at finite temperature is played the trace of the product $U_{x,\mu}$ along a line parallel to the time axis (see Fig.\ref{F15}). The trace is called Polyakov loop. 

Consider two Polyakov loop separated by the distance $d$. Gauge invariance is ensured by periodicity of boundary condition which allows us to 'close' the loops.
The points denoted by A  are physically same points due to boundary conditions. The correlation of the two loops as a function of their separation $d$ decreases as,

\begin{equation}
C(d)\sim e^{-E_{q\bar{q}}(d).t}\sim e^{-E_{q\bar{q}}(d)/T}
\end{equation}

\noindent where $E_{q\bar{q}}(d)$ is the potential energy of the quark pair. Now imagine that one separates the two loop more and more such that one of the loop goes out of the lattice volume. Then one measure $E_{q\bar{q}}(d\rightarrow \infty)$, i.e. energy of a free quark. Therefore, the expectation value of 'one' Polyakov loop behaves as,

\begin{equation}
\la L \ra \sim e^{-E_{q\bar{q}}(d=\infty)/T}
\end{equation}

Polyakov loop can be identified as the order parameter of a confinement-deconfinement phase transition. 

\begin{eqnarray}
\text{confinement}: E_{q\bar{q}}(d=\infty) &=& \infty \Rightarrow \la L \ra = 0 \nonumber \\
\text{deconfinement}: E_{q\bar{q}}(d=\infty) &=& finite \Rightarrow \la L \ra \neq 0 \nonumber\\
\end{eqnarray}

Now when ever there is a phase transition, some internal symmetry is broken. What is the symmetry broken in confinement-deconfinement phase transition? QCD has a hidden, discrete symmetry called $Z(3)$ symmetry. 
To understand the symmetry, let us define:

$Z(G)$: Centre of a group G is the set of elements that commute with every elements of G,

\begin{equation}
Z(G)=[ Z \in G | zg=gz, g \in G ]
\end{equation}
 
For SU(3), the center of group Z(3) has elements, $(1,e^{i2\pi/3},e^{i 3\pi/3})$.
One understand Z(3) symmetry as the group of discrete rotation around the unit circle in the complex plane. The Euclidean action is invariant under these group of rotation, but Polyakov loop is not.
The issue of confinement-deconfinement is then related to breaking of Z(3) symmetry. In the confined phase $\la trL \ra$=0 and $Z(3)$ symmetry is preserved. In the deconfined phase, $\la trL \ra \neq$0 and $Z(3)$ symmetry is broken.

\subsection{Some results of lattice simulations for QCD equation of state}

Several groups worldwide are involved in lattice simulations. 
Since these simulations are costly, some groups have merged their resources to form bigger group.
In the following I will discuss some representative results of lattice QCD. They are from Wuppertal-Budapest collaboration  \cite{Borsanyi:2011bn},\cite{Borsanyi:2010cj}. However, similar results are obtained in simulation by other groups e.g. HotQCD \cite{Cheng:2007jq}.

As indicated above, in lattice QCD, one calculate the partition function,
 
\begin{equation}
Z= \int  [dU]    e^{-\beta S_G(U)} \prod_q det[M(U,m_q)] 
\end{equation}

\noindent where $S_G$ is the Gauge action, $\beta$ is related to the gauge coupling, $\beta=1/g^2$ and $M$ is the Dirac matrix, $m_q$ is the quark mass for flavor $q$. Once the partition function is known, all the thermodynamic variables can be calculated using the thermodynamic relations.


 \begin{figure}[t]
\begin{minipage}{15pc}
\includegraphics[width=15pc]{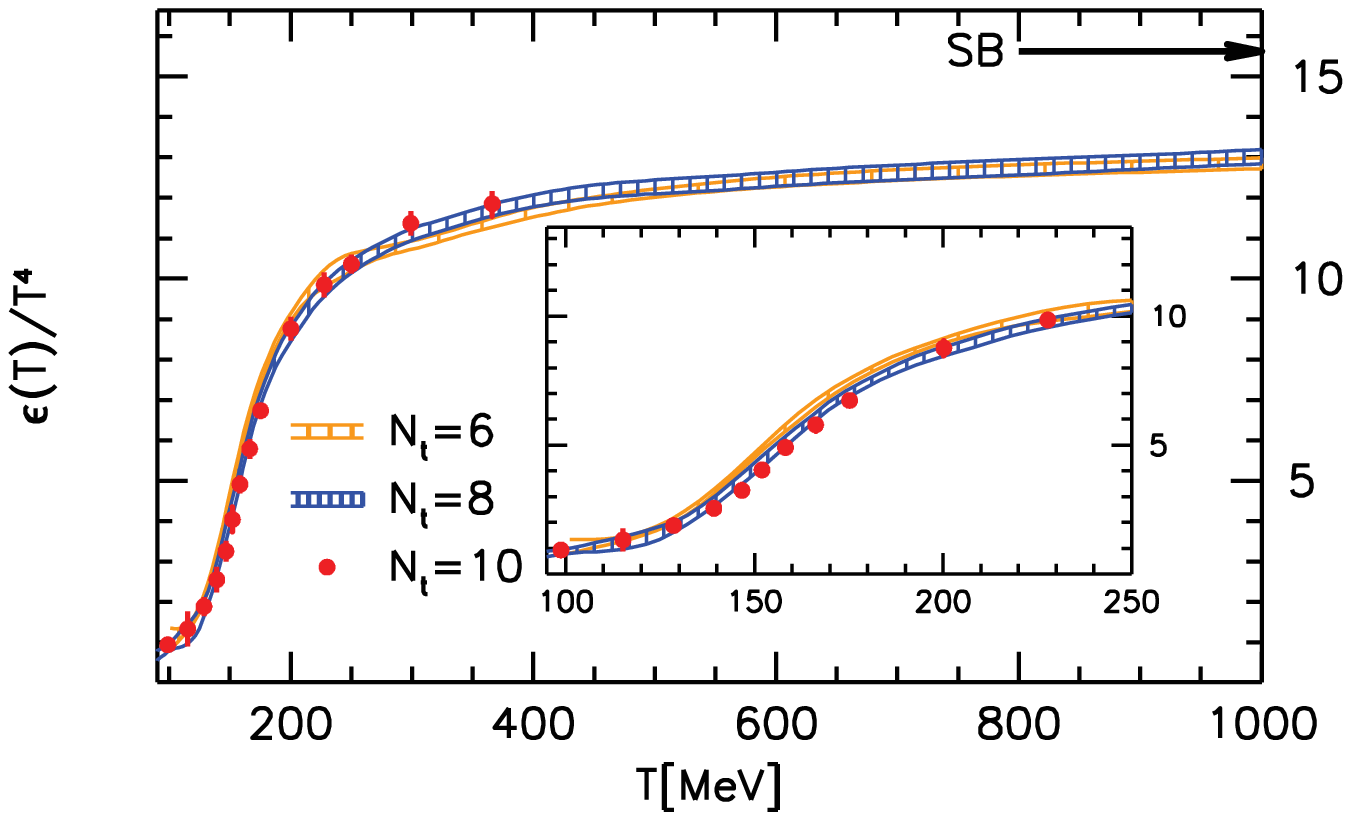}
\end{minipage}\hspace{0.5pc}%
\begin{minipage}{15pc}
\includegraphics[width=15pc]{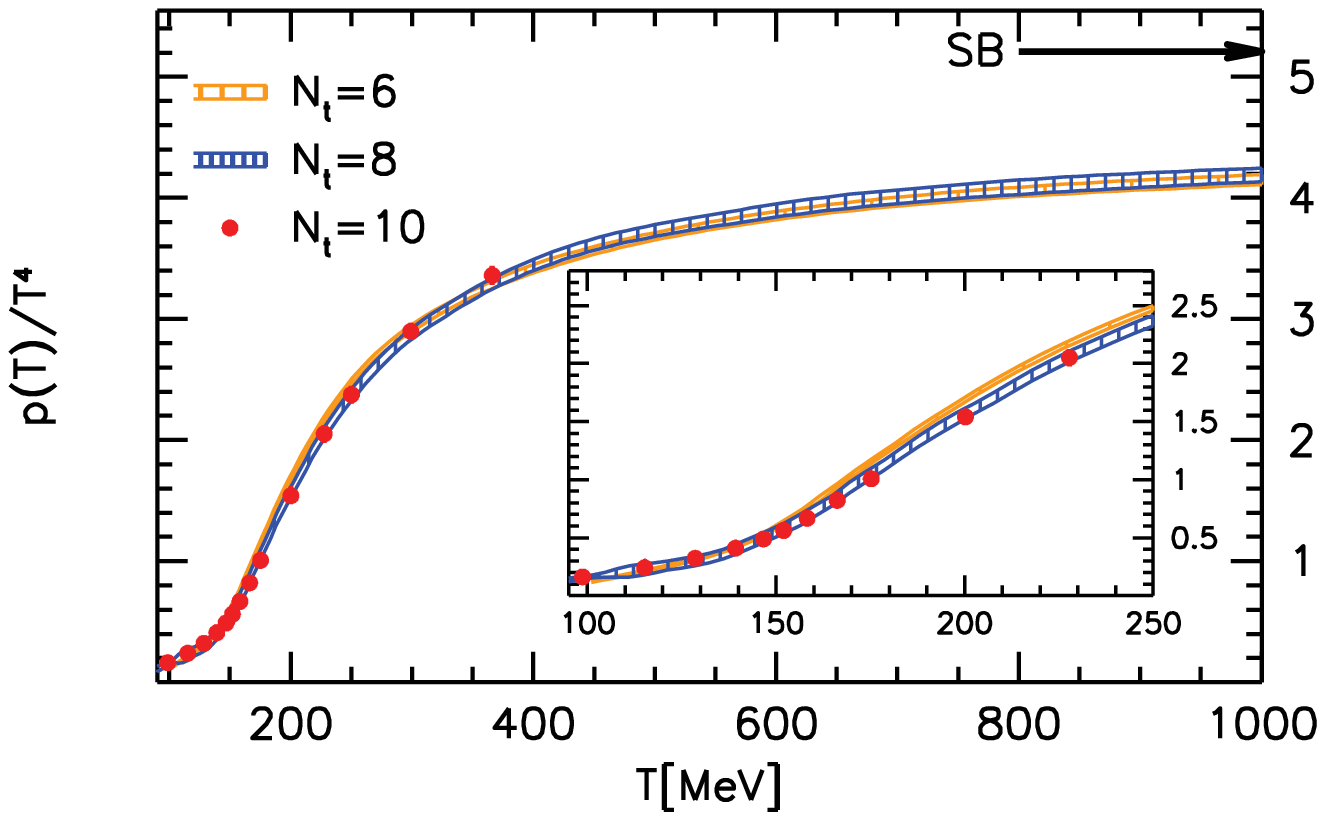}
\end{minipage}\hspace{0.5pc}%
 \caption{\label{F16} (left panel) Lattice simulations for energy density   as a function of temperature. (right panel) Lattice simulations for pressure as a function of temperature. } 
\end{figure}

In Fig.\ref{F16}, Wuppertal-Budapest simulations for energy density ($\varepsilon$)   and pressure ($p$),  as a function of temperature is shown.  
 One notes that $\varepsilon/T^4$  sharply rises over a narrow temperature range 150-200 MeV. At large temperature, it saturates. Very similar behavior is seen in simulated pressure, $p/T^4$ saturates at large $T$. In Fig.\ref{F16}, the Stefan-Boltzmann limit is indicated. Simulated $\varepsilon/T^4$ as well as $p/T^4$, though saturates, remains below the  Stefan-Boltzmann limit. If we believe that at high temperature QCD matter exists as QGP, its constituents are not free, they are interacting. This is the reason QGP is call strongly interacting QGP (sQGP).  

 \begin{figure}[t]
\begin{minipage}{15pc}
\includegraphics[width=15pc]{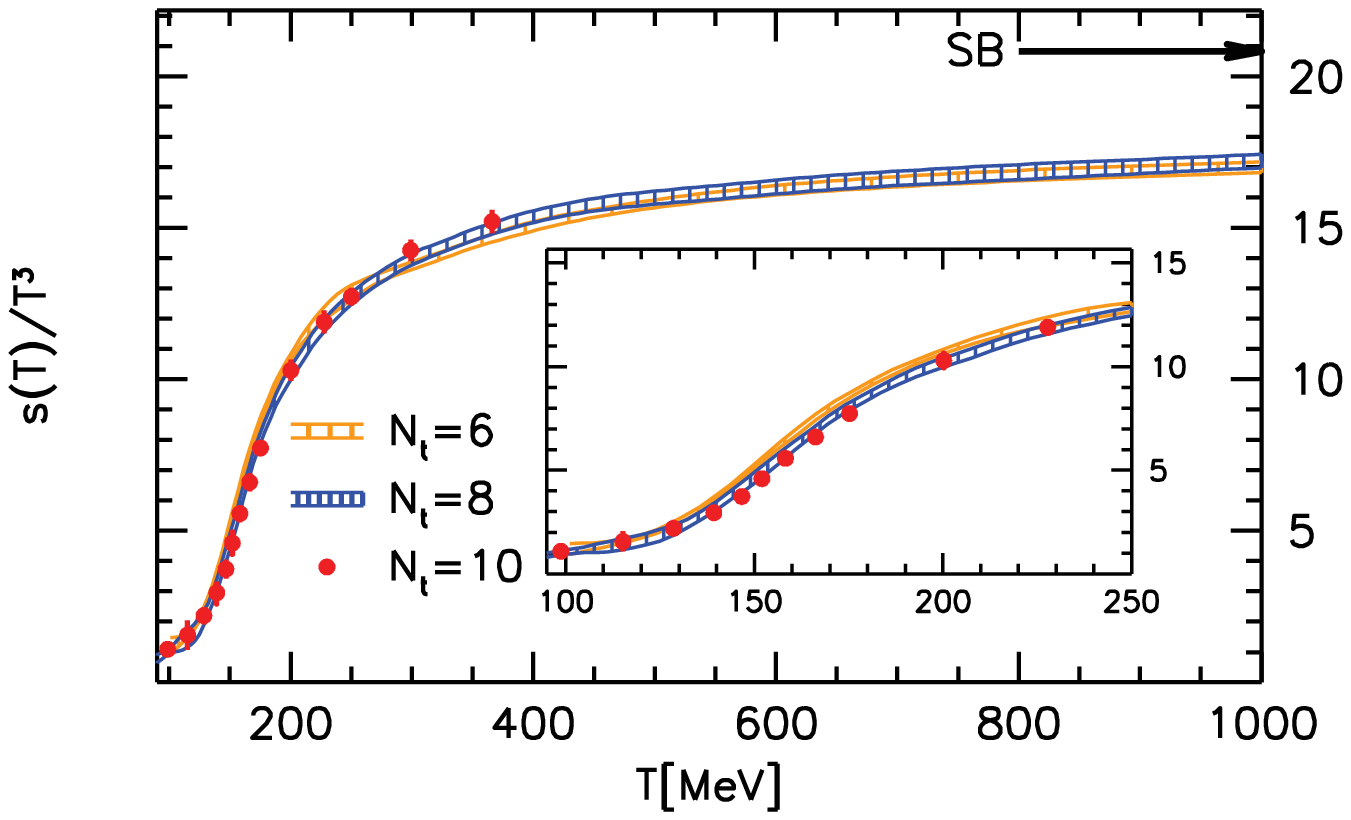}
\end{minipage}\hspace{0.5pc}%
\begin{minipage}{15pc}
\includegraphics[width=15pc]{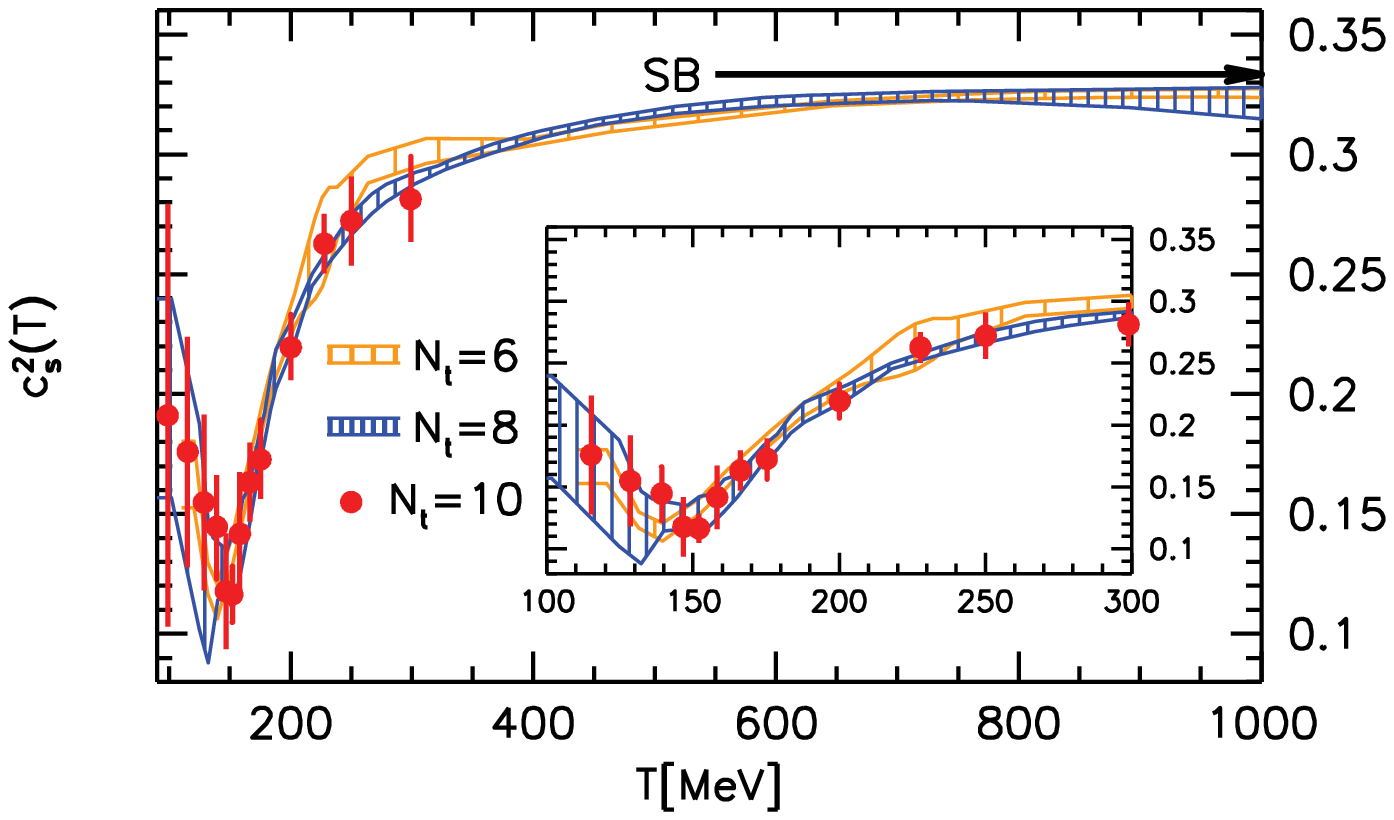}
\end{minipage}\hspace{0.5pc}%
 \caption{\label{F17}(left panel) Lattice simulations for entropy density   as a function of temperature. (right panel) Lattice simulations for speed of sound as a function of temperature. } 
\end{figure}

In the left panel of Fig.\ref{F17}, Wuppertal-Budapest simulation for entropy density is shown. Entropy density over cube of the temperature  also increases rapidly over a narrow temperature range $T\approx$=150-200 MeV. At large temperature, it saturates below the Stefan-Boltzmann limit. $\varepsilon/T^4$, $p/T^4$ or $s/T^3$ are effectively proportional to the degeneracy of the medium.
Temperature dependence of thermodynamic variable, e.g. energy density, pressure and entropy density thus indicate that effective degrees of freedom rapidly changes across the narrow temperature range T=150-200 MeV.  In the right panel of Fig.\ref{F17} variation of square of speed of sound ($c_s$) with temperature is shown. Speed of sound shows a dip around temperature $T\approx$150 MeV.

\begin{figure}[t]
 \center
 \resizebox{0.5\textwidth}{!}
 {\includegraphics{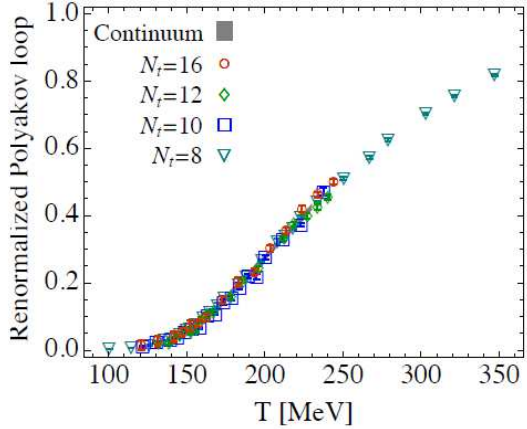}}
\caption{\label{F18}Renormalised Polyakov loop on the lattice. } 
\end{figure}

In Fig.\ref{F18}, renormalised Polyakov loop $L_{ren}$ on the lattice is shown. 
From a small value ($\approx$0) at low temperature, $L_{ren}$ increase   at high temperature. However, the increase is not rapid, rather smooth and over a large interval of temperature. Smooth change of $L_{ren}$ indicates that the confinement-deconfinement phase transition is not a true phase transition, rather a cross-over. The cross-over temperature can be identified with the pseudo-critical temperature for the transition. It can be found by computing the inflection point of $L_{ren}$ ( 
an inflection point, curvature of a curve changes sign). For Wuppertal-Budapest simulation, cross-over temperature is $T_c\approx$ 160 MeV.

\subsection{Chiral phase transition}

We have talked about QCD confinement-deconfinement phase transition. However, QCD has a well known phase transition called 'Chiral phase transition'. Chirality means 'handedness'. Handedness can be understood from the helicity concept. Let us  define the helicity operator,

\begin{equation}
h={\bf J} \cdot \bf{\hat p}=({\bf L}+{\bf s}) \cdot \bf{\hat p}={\bf s} \cdot \bf{\hat p}
\end{equation}
 
\noindent $h$ is the projection of spin on the momentum direction. For spin half fermions, helicity operator will have two eigen values, $+1/2$ and $-1/2$. A particle with helicity +1/2 (-1/2) is called right (left) handed particle.

\begin{figure}[t]
 \center
  \includegraphics[scale=0.4]{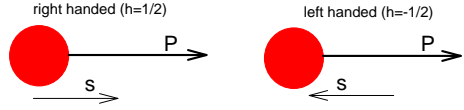}
\caption{Two particles with helicity +1/2 and -1/2 is shown. }
 \label{F19}
\end{figure}

In Fig.\ref{F19}, two particles with helicity +1/2 and -1/2 is shown. One understands that for massive particles helicity is not a good quantum number.
Massive particle will move with finite speed $v < c$ and one can go to frame from where particle will move backward and helicity will be reversed. However,   massless particles moves with speed $c$ and helicity is a good quantum number for massless particles.

Concept of chirality is more abstract. Consider a Dirac field $\psi$ for massless particle. 
The Lagrangian is,

\begin{equation}
\mathcal{L}=i \bar{\psi}\gamma^\mu \partial_\mu \psi 
\end{equation}

For the sake of completeness, we note that $\bar{\psi}=\psi^\dagger \gamma_0$. We also  list the $\gamma$ matrices,

\begin{equation} 
\gamma= \left( \begin{array}{cc}
0 & \sigma \\
-\sigma & 0 \end{array} \right),\  
\gamma^0= \left( \begin{array}{cc}
I & 0 \\
0 & -I  \end{array} \right),\
\gamma_5= \left( \begin{array}{cc}
0 & I \\
I & 0  \end{array} \right).\
\end{equation}

and,

\begin{equation}
\gamma_5=\gamma_5^\dagger=i\gamma^0\gamma^1\gamma^2\gamma^3
\end{equation}

$\gamma$ matrices obey the anticommutation relations,

\begin{equation}
\{\gamma^\mu,\gamma^\nu \}=2g^{\mu\nu}, \{\gamma^\mu,\gamma^5 \}=0, 
\end{equation}

Consider the following transformation,

\begin{equation}
\Lambda_V: \psi \rightarrow e^{-i\frac{\sigma}{2}\Theta} \psi=(1-i\frac{\sigma}{2}\Theta) \psi
\end{equation}

\noindent $\sigma$ is the Pauli matrices and $\Theta$ is the rotation angle. This is the general structure of a  unitary transformation.
The conjugate field transforms under $\Lambda_V$ as,

\begin{equation}
\Lambda_V: \bar{\psi} \rightarrow e^{+i\frac{\sigma}{2}\Theta} \bar{\psi}=(1+i\frac{\sigma}{2}\Theta) \bar{\psi}
\end{equation}

The Lagrangian is invariant under the transformation $\Lambda_V$.

\begin{eqnarray}
i\bar{\psi} \slashed{\partial} \psi &\rightarrow&  i \bar{\psi}  \slashed{\partial}  \psi  -i\Theta \left (\bar{\psi} i\slashed{\partial}\frac{\sigma}{2}\psi
- \bar{\psi} i\frac{\sigma}{2} \slashed{\partial}\psi \right )\\
&=&i\bar{\psi} \slashed{\partial} \psi
\end{eqnarray}

One say that the vector current $V^a_\mu=\bar{\psi}\gamma_\mu \frac{\sigma^a}{2} \psi$ is conserved.

Let us now consider the following transformation,

 \begin{eqnarray}
\Lambda_A: \psi  &\rightarrow&  e^{-i\gamma_5 \frac{\sigma}{2}\Theta} \psi=\left (1-i\gamma_5\frac{\sigma}{2}\Theta \right ) \psi \\
\Rightarrow \bar{\psi}  &\rightarrow&  e^{-i\gamma_5 \frac{\sigma}{2}\Theta}   \bar{\psi}=\left (1-i\gamma_5\frac{\sigma}{2}\Theta \right ) \bar{\psi}
 \end{eqnarray}

\noindent where anti-commutation relation $\gamma_0\gamma_5=-\gamma_5\gamma_0$ is used. The Lagrangian for massless Dirac particle transforms as,

\begin{eqnarray}
i\bar{\psi} \slashed{\partial} \psi &\rightarrow& i \bar{\psi}  \slashed{\partial}  \psi  -i\Theta \left (\bar{\psi} i\partial_\mu \gamma^\mu \frac{\sigma}{2} \psi
+ \bar{\psi}\gamma_5 \frac{\sigma}{2} i\partial \gamma^\mu \psi \right ) \nonumber \\
&=&i\bar{\psi} \slashed{\partial} \psi
\end{eqnarray}

\noindent the 2nd term vanishes due to the anti-commutation relation $\{\gamma_5,\gamma_\mu \}=0$. The Lagrangian for massless Dirac particle is also invariant under the transformation $\Lambda_A$, with conserved 'Axial Current', $A^a_\mu=\bar{\psi}\gamma_\mu\gamma_5 \frac{\sigma}{2}\psi$.

Let us introduce the mass term in the free Dirac Lagrangian,

\begin{equation}
\delta \mathcal{L}=-m\bar{\psi} \psi
\end{equation}

and see how it transforms under $\Lambda_V$ and $\Lambda_A$.

\begin{eqnarray}
\Lambda_V: m\bar{\psi} \psi &\rightarrow& e^{+i\frac{\sigma}{2}\Theta} \bar{\psi} e^{-i\frac{\sigma}{2}\Theta} {\psi}= m\bar{\psi} \psi\\
\Lambda_A: m\bar{\psi} \psi &=& m\bar{\psi} \psi -2im\Theta (\bar{\psi} \frac{\sigma}{2} \gamma_5 \psi)
 \end{eqnarray}
 
Thus   for   massless Fermions, Dirac Lagrangian is invariant under the transformation, $\Lambda_A$ and $\Lambda_V$, i.e. vector and axial vector currents are conserved. This symmetry is called Chiral symmetry and its group structure is $SU(2)_V\times SU(2)_A$. For massive Dirac particles only the vector current is conserved. 

Chiral transition is signaled by the quark condensate $\la \bar{\psi}\psi \ra$. In a chiral symmetric phase,  $\la \bar{\psi}\psi \ra=0$. In the chiral symmetry broken phase $\la \bar{\psi}\psi \ra \neq0$.
In QCD, quarks masses are small but non-zero. Chiral symmetry is broken and  quark condensate $\la \bar{\psi}\psi \ra \neq 0$. However, at sufficiently high temperature, quark mass decreases and condensate $\la \bar{\psi}\psi \ra \rightarrow 0$ and one says that chiral symmetry is restored. 

\begin{figure}[t]
 \center
  \includegraphics[scale=0.7]{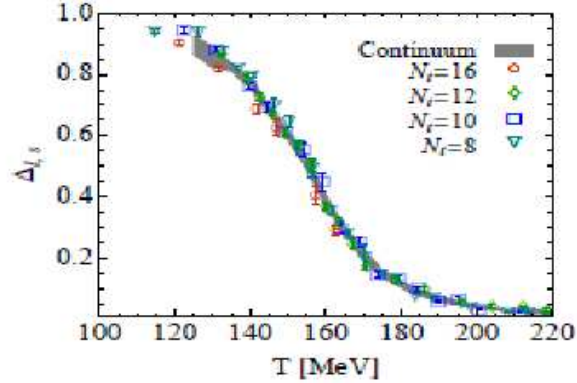} 
\caption{Subtracted Chiral condensate in lattice.} 
 \label{F20}
\end{figure}

\begin{table}[h] 
\caption{ \label{table5} Some key properties of chiral and deconfinement transitions in QCD}
	\centering
		\begin{tabular}{|c|c|c|}
		\hline
   & Chiral Phase  & deconfinement phase  \\  
   &   transition & transition\\  \hline   
quark mass & 0 & $\infty$ \\ \hline 
symmetry & chiral symmetry & Center group symmetry \\ \hline
order parameter & quark condensate & Polyakov loop \\ \hline
 \end{tabular}
	\textsl{}
\end{table}

In Fig.\ref{F20}, lattice simulation for quark condensate is shown. 
Generally to remove various uncertainties associated with lattice simulations, one   calculate a subtracted quark condensate,

\begin{equation}
\Delta_{l,s}=\frac {\la \bar{\psi}\psi \ra_{l,T} - \la \bar{\psi}\psi \ra_{s,T} } {\la \bar{\psi}\psi \ra_{l,0}- \la \bar{\psi}\psi \ra_{s,0} }, l=u,d
\end{equation}

From the inflexion point of $\Delta_{l,s}$ one computes chiral transition occur at $T_c\approx$160 MeV.
In Wuppertal-Budapest simulations,  both confinement-deconfinement phase transition and chiral transition occur approximately at the same temperature   However, the two transitions are unrelated.
Some key properties of chiral transition and deconfinement transition is listed in table.\ref{table5}.   

\subsection{Nature of QCD phase transition}

\begin{figure}[t]
 \center
 \resizebox{0.35\textwidth}{!}
 {\includegraphics{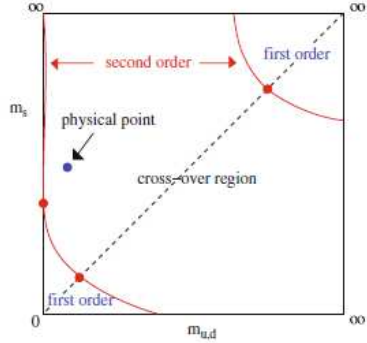}}
\caption{Current understanding about nature of confinement-deconfinement phase transition as a function of quark mass $m_u$,$m_d$ and $m_s$.} 
 \label{F21}
\end{figure}

In Fig.\ref{F21}, the current understanding \cite{Laermann:2003cv} about the nature of confinement-deconfinement phase transition, in a baryon free matter, as a function of quark mass $m_u$,$m_d$ and $m_s$ is shown. The results can be summarised as follow: 

(i) In a pure gauge theory ($m_q \rightarrow \infty$), the transition is 1st order.

(ii) For $m_q\rightarrow 0$, the Lagrangian is chirally symmetric and there is a chiral symmetry restoration phase transition. It is also 1st order.

(iii) For $0 < m_q < \infty$, there is neither confinement-deconfinement phase transition nor a chiral symmetry restoring phase transition. The system undergoes a cross-over transition. The order parameter, e.g. Polyakov loop, or the susceptibility shows a sharp temperature dependence and it is possible to define a pseudo critical cross-over transition temperature.

\subsubsection{QCD phase diagram at finite baryon density}

\begin{figure}[t]
 \center
 \resizebox{0.35\textwidth}{!}
 {\includegraphics{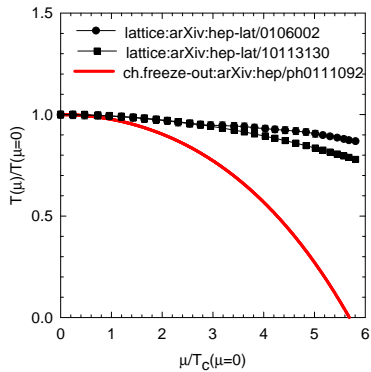}}
\caption{Lattice QCD calculation of QCD phase diagram, in  imaginary chemical potential method \cite{Fodor:2001pe}  and Taylor expansion method \cite{Kaczmarek:2011zz}  are shown. The red line is the chemical freeze-out curve obtained in a statistical model \cite{Cleymans:2006qe}.} 
 \label{F22}
\end{figure}

At finite baryon density, Fermion determinant is complex and standard technique of Monte-Carlo importance sampling fails. Several techniques have been suggested to circumvent the problem, (i) reweighting \cite{Fodor:2001pe},\cite{Fodor:2004nz} , (ii) analytical continuation of imaginary chemical potential 
\cite{deForcrand:2002ci}, \cite{D'Elia:2002gd} and (iii)Taylor expansion \cite{Allton:2003vx},\cite{Gavai:2003mf}. These methods has been used to locate the phase boundary in $T-\mu_B$ plane, $\mu_B=3\mu_q$ is the baryonic chemical potential. The calculations suggest that   the curvature parameter
 in the expansion,
 
 \begin{equation} \label{eq1}
\frac{T_c(\mu_B)}{T_c(\mu_B=0)}=1-\kappa \left (\frac{\mu_B}{T_c(\mu_B=0)} \right )^2
\end{equation}

\noindent is small    \cite{Philipsen:2008gf}. As an example, in Fig.\ref{F22},
QCD phase diagram obtained in the analytical continuation method \cite{Fodor:2001pe} (the filled circles) and in Taylor expansion    \cite{Kaczmarek:2011zz} (the filled squares) are shown. Both the methods gives nearly identical phase diagram for $\mu_B/T_c(\mu_B=0) < 3 GeV$, curvature parameter is small, $\kappa\approx 0.006$. At larger $\mu_B$, they differ marginally. 

From theoretical considerations, QCD phase transition is expected to be 1st order in baryon dense matter. 
Since at $\mu_q\approx 0$ deconfinement transition is a cross-over, one expect a QCD critical end point (CEP) where the 1st order transition line ends up at the cross over. Location of the QCD critical end point is of current interest.  At the critical end point, the first order transition becomes continuous,  resulting in long range correlation and fluctuations at all length scales. Mathematically, it is true thermodynamic  singularity.  

Experimental signature of QCD critical end point is tricky. Since at CEP, fluctuations exists at all length scale, one expects these fluctuations to percolate in the observables. Event-by-event fluctuations of baryon number, charge number can possibly signal a QCD CEP.  

In Fig.\ref{F22}, chemical freeze-out curve \cite{Becattini:2005xt},\cite{Cleymans:2006qe}, obtained in statistical model analysis of particle ratios are shown (the red line). Curvature of the chemical freeze-out curve is factor of 4 larger than the curvature in the QCD phase diagram. Small curvature of the freeze-out curve, compared to the chemical freeze-out is interesting. Experimental signal of critical end point will get diluted as the deconfined medium produced at the critical end point will evolve longer to reach chemical freeze-out. Fluid will have more time to washout any signature of CEP.

\section{Color Glass Condensate}\label{sec6}

In ultra-relativistic heavy ion collisions deconfined medium, called QGP can be produced. Theoretical considerations, however indicate that prior to QGP, a new form of matter,'Color Glass Condensate (CGC) '\cite{Gyulassy:2004zy},\cite{McLerran:2008es} may be formed. I briefly describe here the beautiful concept behind the color glass condensate.
According to theory, the new form of matter (CGC) controls the high energy limit of the strong interaction  \cite{Gyulassy:2004zy},\cite{McLerran:2008es} and  should describe, (i) high energy cross-sections, (ii) distribution of produced particles in high energy collisions, (iii) distribution of small $x$ particles in a hadron and (iv) initial conditions for heavy ion collisions.

As we know hadrons consist of gluons, quarks and anti-quarks. 
Constituents of hadrons e.g. quarks, gluons are generically called parton (the parton name was given by Feynman, while Murray Gellman picked the word 'quark'
from the sentence 'Three quarks for Muster Mark' in James Joyce book, 'Finnegans Wake'). 
At very high energy hadron wave function has contributions from partons, e.g. gluons, quarks and anti-quarks. A convenient variable to measure contribution of constituents to hadron wave function is the fraction of the momentum carried by the constituent (Bjorken $x$ variable),

  \begin{figure}[t]
 \center
 \resizebox{0.50\textwidth}{!}
 {\includegraphics{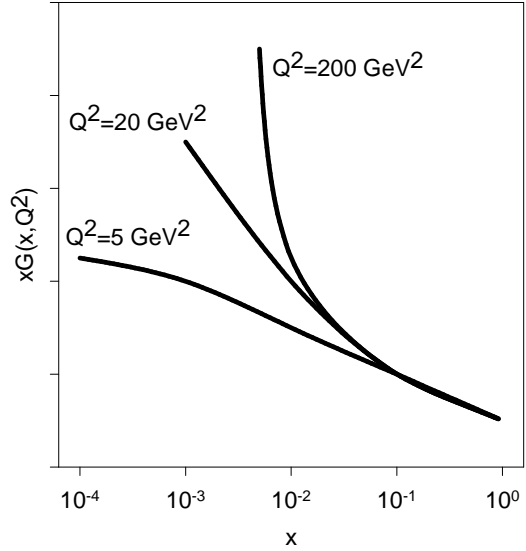}}%
\caption{gluon density measured in HERA in three momentum scale.} \label{F32}
\end{figure}

\begin{equation}\label{eq4.1}
x=E_{constituent}/E_{hadron}
\end{equation}

Probability $G(x)$ to obtain a parton with momentum fraction $x$ and $x+dx$ is generally called  the parton distribution function. Parton distribution function depend weakly on the resolution scale $Q^2$. One can write the density of small
$x$ partons as, 

\begin{equation}\label{eq4.2}
\frac{dN}{dy}=xG(x,Q^2)
\end{equation}

 In Fig.\ref{F32}, gluons distribution function as measured in HERA (Hadron Electron Ring Accelerator) is shown. One observes that gluon density rapidly increases at small $x$. It is also an increasing function of the  resolution scale ($Q^2$). Increase in gluon density at small $x$ is commonly referred to as the small $x$ problem. It means that
if we view the proton head on with increasing energy, gluon density grows.  
QCD is asymptotically free theory, coupling constant 
decreases at short distances. As the density increases, typical separation between the gluons decreases, strong coupling constant gets weaker. Then higher the density,
the  gluons interact more weakly. However, density can not be increased indefinitely, it will then lead to infinite scattering amplitude and violate the unitary bound (unitary bound is a constrain on quantum system, that sum of all possible outcome of evolution of a quantum system is unity).  One then argues that as the gluon density increases repulsive gluon interaction become important and in the balance, gluon density saturates. The saturation density will corresponds to a saturation momentum scale $Q_{sat}$. Qualitatively, one can argue as follows: imagine a proton is being packed with fixed size gluons. Then after a certain saturation density or the closepack density, repulsive interaction will take over and no more gluon can be added to the proton. Naturally, the saturation density depend on the gluon size,
for a smaller size gluon, the saturation density will increase. Then there is a characteristic momentum scale $Q_{sat}$ which corresponds to inverse of the smallest size gluon which are close packed. Note that saturation scale only tells that gluon of size $1/Q_{sat}$ has stopped to grow. It does not mean that number of gluons   stopped to grow. 

It is very reasonable to assume that some effective potential $V$ describe the
system of gluons. If phase space density of gluons is denoted by $\rho$,

\begin{equation}
\rho= \frac{1}{\pi R^2} \frac{dN}{dy d^2p_T},
\end{equation}

\noindent at low density, the system will wants to increase the density and $V\sim -\rho$. On the other hand , repulsive interaction balance the inclination to condensate, $V_{repulsion} \sim \alpha_s \rho^2$. These contributions balance each other when $\rho \sim 1/\alpha_s$. Density scaling as inverse of interaction strength is characteristic of condensate phenomena such as super conductivity. 

Phase space density $\rho=\frac{1}{\pi R^2}\frac{dN}{dy d^2p_T} \sim 1/\alpha_s$ can be integrated to obtain saturation momentum scale ($Q_s$),

\begin{equation}
\frac{1}{\pi R^2}\frac{dN}{dy} \sim \frac{1}{\alpha_s} Q_s^2
\end{equation}


The origin of the name 'Color Glass condensate' is now clear. The word color refers to Gluons which are colored. The system is at very high density, hence the word condensate.  
The matter is of glassy nature. Glasses are disordered systems, which 
behave like liquid on long time scale and like solid on short time scale.
The word 'glass' arise because the gluons evolve on time scale long compared to their natural time scale $1/Q_{sat}$. The small $x$ gluons are produced from gluons at larger values of $x$. Their (the fast gluons) time scale is Lorentz diluted and can be approximated as a static fields. This scale is transferred to the small $x$ gluons. The small $x$ gluons then can be approximated as static classical fields.

 \begin{figure}[t]
 \center
 \resizebox{0.30\textwidth}{!}
 {\includegraphics{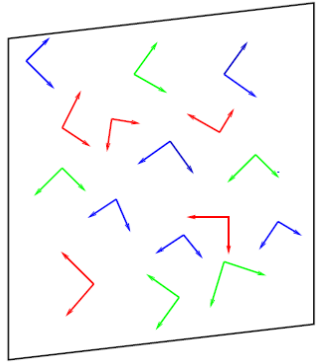}}
\caption{CGC as high density gluon fields on a two dimensional sheet travelling near speed of light.}
 \label{F24}
\end{figure}

CGC acts as a infrared cut off when computing total multiplicity. For momentum scale $p_T > Q_{sat}$, produced particles are incoherent and ordinary perturbation applies. For momentum scale $p_T \leq Q_{sat}$, the produced particles are in a coherent state, which is color neutral on the length scale $1/Q_{sat}$.  

One may wonder about the quarks degrees of freedom. At high energy, gluon density grows faster than quark density and distribution is overwhelmingly gluonic.
Fields associated with CGC can be treated as a classical fields.  Since they arise from fast moving partons, they are plane polarised, with mutually orthogonal color magnetic and electric fields perpendicular to the direction of motion of the hadron. They are also random in two dimensions (see Fig.\ref{F24}). 

\begin{figure}[t]
\begin{minipage}{15pc}
\includegraphics[width=15pc]{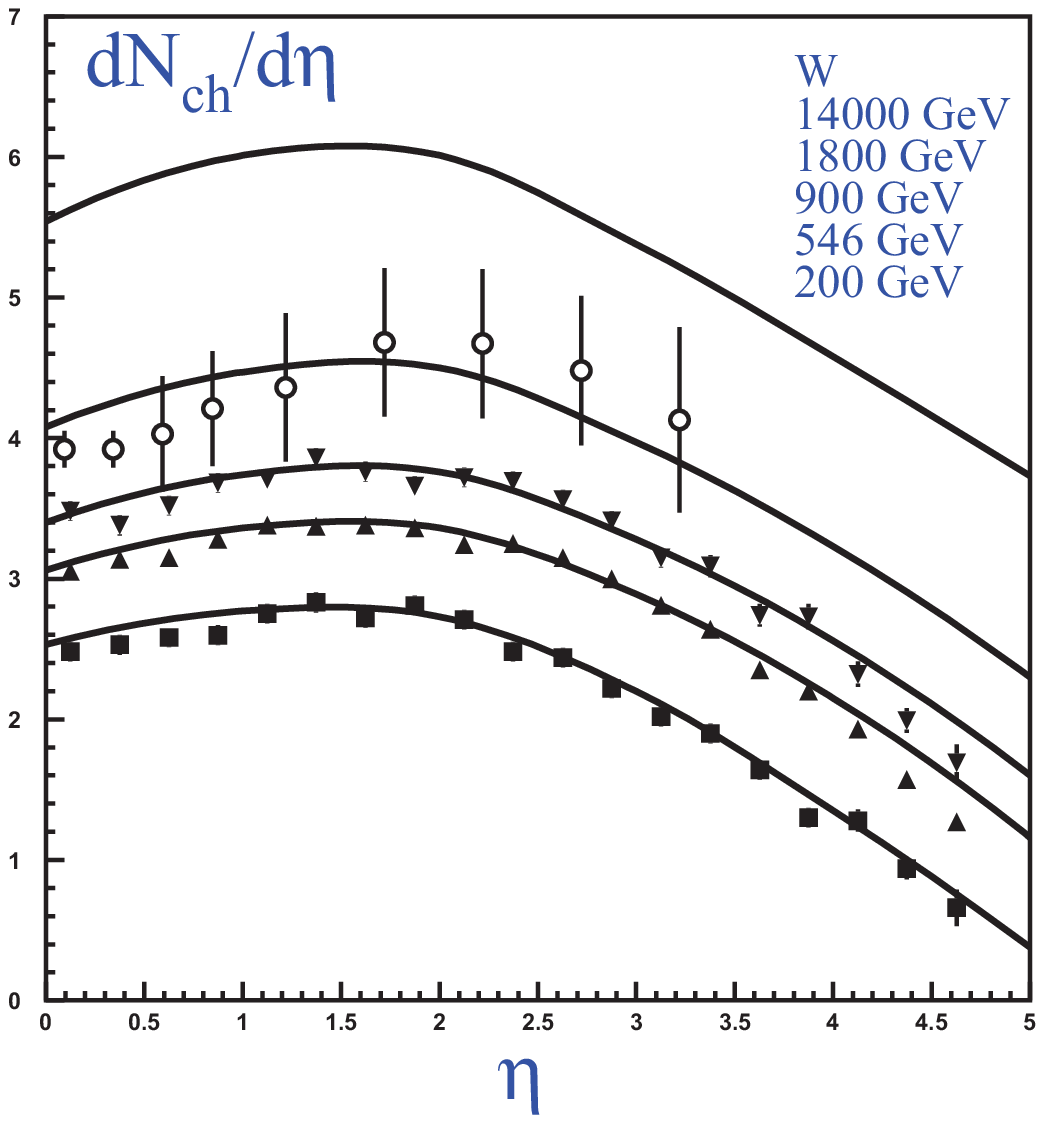}
\end{minipage}\hspace{0.5pc}%
\begin{minipage}{15pc}
\includegraphics[width=15pc]{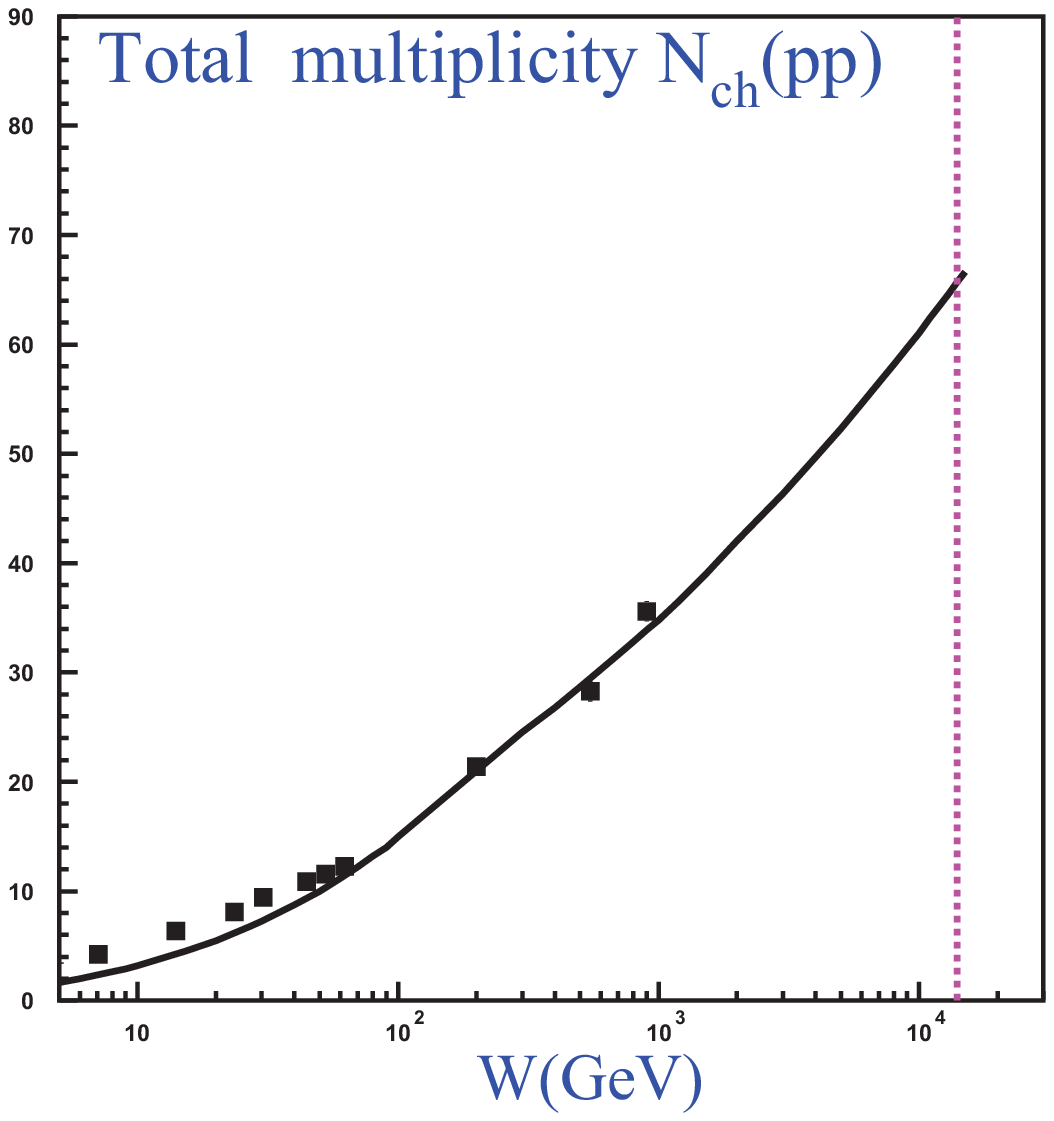}
\end{minipage}\hspace{0.5pc}%
 \caption{\label{F25a} (left panel) CGC model predictions for rapidity dependence $dN/d\eta$ of charged hadron multiplicities in proton - proton (antiproton)
collisions as a function of the pseudorapidity at different energies. The data are taken from
Ref.\cite{cgc_Eidelman},  (right panel) Energy dependence of total multiplicity in proton - proton (antiproton) collisions. The vertical
dotted line marks the LHC energies for proton-proton collisions (W = 14000GeV ). collisions.
The experimental data are taken from Ref.\cite{cgc_back}.} 
\end{figure}

There are many successful application of CGC model in explaining various experimental results. For completeness purpose, I will show two results obtained in CGC model \cite{Kharzeev:2004if}. In Fig.\ref{F25a}, in two panels, rapidity density of charged particles
in pp collisions and energy dependence of charge multiplicity are shown. The solid
lines in the figure are obtained in a CGC based model. It no small wonder, that
CGC based model can explains the data. Such an description to the data, from a first
principle model was not available earlier.

\section{Relativistic kinetic Theory}\label{sec7}

QGP is a macroscopic system. Properties of many-body system depend on: (i) interaction of the constituent 
particles and (ii) external constraints. One characterises the system    in terms of macroscopic 
state variables, e.g. particle density, temperature etc. and of the
characteristic microscopic parameters of the system. One then tries to understand certain equilibrium/non-equilibrium properties of the macroscopic system.
In kinetic theory this programme is realised by means of a statistical
description, in terms of 'one-particle distribution function' and its transport
equation.  From the transport equation, on the basis of conservation laws, hydrodynamic
theory of perfect fluid can be constructed. Supplementing the
conservation laws with entropy law, hydrodynamics for dissipative fluid 
is constructed.

In the following, we briefly discuss relativistic Boltzmann or the kinetic equation. We then show that basic equations for hydrodynamics are obtained by coarse graining Boltzmann transport equations. Most of the discussions are from
\cite{groot}.

\subsection{Some basic definitions in kinetic theory}

(1) Distribution function, $f(x,p)$: in kinetic theory, a macroscopic system is generally studied in terms of the distribution function, 
 $f(x,p)$.  $f(x,p)d^3xd^3p$  is defined as the average number of particles  in small volume  $d^3x$, at time $t$,
with momenta between ${\bf p}$, ${\bf p+dp}$. 
 
It is implicitly understood that particle content in the volume element  
$d^3x$ is large enough to apply concepts of statistical physics, yet, $d^3x$ small in macroscopic scale.

(2) Particle four-flow $N^\mu$:   is defined as the 1st moment of the distribution function.

\begin{eqnarray}
N^\mu(x) = \int \frac{d^3p}{p^0} p^\mu f(x,p)
\end{eqnarray}

4-components of particle 4-flow can be identified as follows:

\begin{eqnarray}
\mbox{Particle density:}&& N^0(x)=\int d^3p  f(x,p) \\
\text{particle flow:}&& N^i(x)=\int d^3p \left (\frac{p^i}{p^0}\right )  f(x,p)\nonumber \\
&=&\int d^3p u^i f(x,p), i=1,2,3
 \end{eqnarray}

\noindent where we have introduced the velocity $\bf{u}=\bf{p}/p^0$.

(3) Energy-momentum tensor $T^{\mu\nu}$:   is the 2nd moment of
the distribution function. 
 
\begin{eqnarray}
T^{\mu\nu}(x) = \int \frac{d^3p}{p^0} p^\mu p^\nu f(x,p)
\end{eqnarray}

The components can be identified as follows:

\begin{eqnarray*}
\text{energy density:}&&T^{00}(x)=\int d^3p p^0 f(x,p)\\
\text{energy flow:}&&T^{0i}(x)=\int d^3p p^0 u^i f(x,p), i=1,2,3\\
\text{momentum density:}&&T^{i0}(x)=\int d^3p p^i f(x,p), i=1,2,3\\ 
\text{momentum flow or}&&\\
\text{pressure tensor:}&&T^{ij}(x)=\int d^3p p^i u^j f(x,p), i=1,2,3\\
\end{eqnarray*}

(4) Entropy four-flow  $S^\mu$:

\begin{equation}
S^\mu(x) = - \int \frac{d^3p}{p^0} p^\mu  f(x,p) [\log f(x,p) -1]
\end{equation}
 
 $f(x,p)$ is a dimensionful quantity (dimension=$fm^{-3}GeV^{-3}$). To make it dimensionless, one generally multiply with $h^3$ and subtract unity. Note that absolute value of entropy is not measurable, only change in entropy is measurable. Then the observables remain unaffected.

(5) Hydrodynamic four-velocity, $u^\mu$: in each space-time point a time-like vector is defined,

\begin{equation}
u^\mu(x)u_\mu(x) =1
\end{equation}

In the local rest frame, $u^\mu =(1, 0, 0, 0)$. 

with help of  $u^\mu$ one defines a tensor quantity,

\begin{equation}
\Delta^{\mu\nu}(x) = g^{\mu\nu} - u^\mu u^\nu
\end{equation}

It is called projector,  annihilates that part of the 4-vector parallel to 
$u^\mu$,

\begin{equation} 
 \Delta^{\mu\nu} u_\nu = 0 
 \end{equation}

 Choices of hydrodynamic four-velocity:

(a) Eckart's definition: Hydrodynamic four velocity is related to the
particle four flow  $N^{\mu}$,

\begin{equation} 
 u^\mu =\frac{N^\mu}{\sqrt{N^\nu N_\nu}}
\end{equation}

(b) Landau and Lifshitz definition: $u^\mu$ is related to the flow of
energy,

\begin{equation} 
 u^\mu = \frac{T^{\mu\nu} u_\nu }{u_\rho T^{\rho\sigma} u_\sigma }
\end{equation}

In the study of high energy heavy ion collisions, central rapidity region is essentially particle free. It is difficult to define hydrodynamics four velocity according to Eckart's definition.  Landau-Lifshitz choice of hydrodynamic velocity is preferred as it is related to energy flow.

 \subsection{Physical quantities of a simple system}
  With the help of hydrodynamic four velocity $u(x)$ one can define relevant macroscopic quantities, e.g. particle density, energy density, heat flow, the pressure tensor and entropy density, in 
a covariant manner.

(i) particle density is the density of particles in the rest frame $u(1,0,0,0)$,

\begin{equation} \label{eq8.9}
n=N^\mu u_\mu
\end{equation}

(ii) energy density of the particles in the rest frame,

\begin{equation}\label{eq8.10}
e=u_\mu T^{\mu\nu} u_\nu
\end{equation}

(iii) heat flow: the difference of energy flow and flow of enthalpy $h$ carried by the particles, 

\begin{equation}\label{eq8.11}
q^\mu=(u_\nu T^{\mu\sigma}-h N^\sigma ) \Delta^\mu_\sigma
\end{equation}

where enthalpy  per particle is defined as,
\begin{equation}
h=\frac{e+p}{n} 
\end{equation}

\noindent with $p$ the local hydrostatic pressure, to be defined shortly. Heat flow has the property that it is transverse to the hydrodynamic 4-velocity,

\begin{equation}
q^\mu u_\mu=0 
\end{equation}

(iv) Pressure tensor:

\begin{equation}
P^{\mu\nu}=\Delta^\mu_\sigma T^{\sigma\tau}\Delta^\nu_\tau
\end{equation}

It is symmetric when $T^{\mu\nu}$ is symmetric. In the local rest frame, it is purely spatial,

\begin{equation}
P^{00}_{LR}=0; P^{0i}_{LR}=P^{i0}_{LR}=0; P^{ij}_{LR}=T^{ij}, i,j=1,2,3 
\end{equation}

Pressure tensor has a 'reversible' and an 'irreversible' part,

\begin{equation}\label{eq8.16}
P^{\mu\nu}=\Delta^\mu_\sigma T^{\sigma\tau}\Delta^\nu_\tau=-p\Delta^{\mu\nu} + \Pi^{\mu\nu}
\end{equation}

\noindent $p$ is the hydrostatic pressure. The quantity $\Pi^{\mu\nu}$ is called the viscous pressure tensor. 
Writing Eqs.\ref{eq8.10},\ref{eq8.11} and \ref{eq8.16} in a slightly different manner, we find,
 
\begin{eqnarray}
e&=&u_\mu T^{\mu\nu} u_\nu \\
q^\mu+h\Delta^{\mu\nu}N_\nu&=&u_\nu T^{\nu\sigma} \Delta^\mu_\sigma \\
-p\Delta^{\mu\nu}+\Pi^{\mu\nu}&=& \Delta^\mu_\sigma T^{\sigma\tau}\Delta^\nu_\tau
\end{eqnarray}

The terms can be collected to obtain the expression for energy-momentum tensor,

\begin{equation}
T^{\mu\nu}=T^{\mu\nu}_{rev} + T^{\mu\nu}_{irr},
\end{equation}

\noindent with the 'reversible' and 'irreversible' parts,

\begin{eqnarray}
 T^{\mu\nu}_{rev}&=&e u^\mu u^\nu -p \Delta^{\mu\nu}\\
  T^{\mu\nu}_{irr}&=&[(q^\mu+h\Delta^{\mu\sigma}N_\sigma) u^\nu
 +(q^\nu+h\Delta^{\nu\sigma}N_\sigma) u^\mu]  +\Pi^{\mu\nu}
\end{eqnarray}

As mentioned earlier, two choices of hydrodynamic velocity is popular. In the Eckart frame, $u^\mu=N^\mu/\sqrt{N^\nu N_\nu}$ and $\Delta^{\mu\nu}N_\nu=0$.
Heat flow is,

\begin{equation}
q^\mu=u_\nu T^{\nu\sigma}\Delta^\mu_\sigma
\end{equation}

\noindent and the irreversible part of the energy-momentum tensor reduces to,
 
\begin{eqnarray}
T^{\mu\nu}_{irr}&=&[q^\mu u^\nu +q^\nu u^\mu]  +\Pi^{\mu\nu}
\end{eqnarray}

In the  Landau-Lifshitz frame, $u^\mu=T^{\mu\nu}u^\nu/\sqrt{u_\rho T^{\rho\sigma}T_{\sigma\tau}u^\tau}$. Using the property, $\Delta^{\mu\nu}u_\nu=0$,  
heat flow is,

\begin{equation} \label{eq8.27}
q^\mu=-h \Delta^{\mu\nu}N_u
\end{equation}

The irreversible part of energy-momentum tensor then  has the simplified form,

\begin{eqnarray} \label{eq8.28}
T^{\mu\nu}_{irr}&= \Pi^{\mu\nu}
\end{eqnarray}


In Landau-Lifshitz frame, the energy momentum tensor does not contain heat flow $q^\mu$. However, as manifest in Eq.\ref{eq8.27},  heat flow is not zero. It is manifested  in particle flow. Using the projector,
$\Delta^{\mu\nu}$, the particle four flow can be split into two part, in the direction of four velocity and in the direction perpendicular to it,

\begin{equation} \label{eq8.31}
N^\mu =nu^\mu + V^\mu=nu^\mu + \Delta^{\mu\nu}N_\nu
\end{equation}

In the Landau-Lifshitz frame, 

\begin{equation} \label{eq8.32}
q^\mu=-h\Delta^{\mu\nu} N_\nu \Rightarrow V^\mu=\frac{q^\mu}{h}
\end{equation}

It is interesting to note that in the Eckart's frame, $u^\mu \propto N^\mu$,
particle flow is in the direction $u^\mu$. While energy-momentum tensor explicitly contain heat flow,  $V^\mu$  is identically zero.

The irreversible part of the energy-momentum tensor $T^{\mu\nu}_{irr}$ leads to dissipation. In a realistic system,  $T^{\mu\nu}_{irr}$ is never identically zero. However, if it is neglected, the fluid is called ideal or inviscid fluid.

\subsection{Relativistic Kinetic (TRANSPORT) Equation:}

In kinetic theory, macroscopic system is described in terms of one body distribution function $f(x,p,t)$. Kinetic or transport equation give the  space-time development of the distribution function. It was originally derived by 
Boltzmann and called Boltzmann equation.
Ludwig Eduard Boltzmann (1844-1906) was an Austrian Physicist. He made major contributions in statistical thermodynamics. He was an early proponent of atomic theory, when atomic theory has not gained ground.  During his life time, his works were not appreciated. Famous physicists like Ernest Mach, Wilhelm Ostwald opposed his atomic view. Rejection of his views made Boltzmann depressed and in 1906, he committed suicide. In the following, we derive the relativistic version of the Boltzmann equation. For more complete exposure to Boltzmann equations see 
 \cite{groot}.
 
\subsubsection{Transport Equation without collisions:}

We have defined particle 4-flow,
\begin{eqnarray}
N^\mu = \int \frac{d^3p}{p^0} p^\mu f(x,p)
\end{eqnarray}

The time component ($\mu=0$) is the particle density  and the space  components 
($\mu=1,2,3$) are the particle flow, both measured with respect to the observer frame of reference. If $d\sigma_\mu$ is a oriented three surface element of a plane space-like surface (i.e. a surface whose tangent vector is time) and $\Delta^3 \sigma$ is a small segment situated as x, then we can contract $N^\mu$
with $d\sigma_\mu$ to obtain the scalar quantity.

\begin{eqnarray}
\Delta N(x)&=&\int_{\Delta^3\sigma} d\sigma^3_\mu N^\mu(x) 
=\int_{\Delta^3\sigma} \int d^3\sigma_\mu \frac{d^3p}{p^0}
p^\mu f(x,p)
\end{eqnarray}

In Lorentz frame, $d^3\sigma_\mu=(d^3x,0,0,0)$ is purely time like and,

\begin{equation}
\Delta N(x)=\int_{\Delta^3x} \int d^3x d^3p f(x,p)
\end{equation}

$\Delta N(x)$ is just the number of particles in a volume element ${\Delta^3x}$ 
In Minkowski space, particle is represented by a world line. $\Delta N$
thus represent the av. no. of world lines crossing the 3-segment $\Delta^3\sigma$.

\begin{equation}
\Delta N(x,p)=\int_{\Delta^3\sigma} \int_{\Delta^3p} d^3\sigma_\mu \frac{d^3p}{p^0}
p^\mu f(x,p)
\end{equation}

$\Delta N(x,p)$: the av. no. of particle world lines crossing a segment
$\Delta^3\sigma$ with momenta in the range $\Delta^3p$ around $p$.\\
Some time later, same particles will cross a surface element $\Delta^3\sigma^\prime$. Then we have the identity,

 \begin{eqnarray}
 \int_{\Delta^3\sigma} \int_{\Delta^3p}d^3\sigma_\mu \frac{d^3p}{p^0}
p^\mu f(x,p)
=\int_{\Delta^3\sigma^\prime} \int_{\Delta^3p}d^3\sigma_\mu \frac{d^3p}{p^0}
p^\mu f(x,p)   
 \end{eqnarray}
 
Consider the 4-volume $\Delta^4x$ enclosed by surface $\Delta^3\sigma$ and
$\Delta^3\sigma^\prime$ and the surface of the tube of the world lines.  No particle
world lines cross the tube surface. Thus net flow of particle through the surface
$\Delta^3x$ of 4-volume $\Delta^4x$ vanishes.

\begin{eqnarray}
  \int_{\Delta^3\sigma} \int_{\Delta^3p} d^3\sigma_\mu \frac{d^3p}{p^0}
p^\mu f(x,p) =0
\end{eqnarray} 

Apply  Gauss theorem,  

\begin{eqnarray}
   \int_{\Delta^4x} \int_{\Delta^3p} d^4x \frac{d^3p}{p^0}
p^\mu \partial_\mu f(x,p) =0
\end{eqnarray}

Since $\Delta^4x$ and $\Delta^3\sigma$ arbitrary

\begin{eqnarray}
p^\mu \partial_\mu f(x,p) =0
\end{eqnarray}
 
 This is the Boltzmann transport equation for a collisionless system.

\subsubsection{Transport equation with collisions:}

Number of particles in ranges $\Delta^4x$ and $\Delta^3p$
changes due to collisions. The amount of change can be written as,
 
\begin{eqnarray}
\Delta^4x \frac{\Delta^3p}{p^0} C(x,p)
\end{eqnarray} 

where $C(x,p)$ is an invariant function, whose form is to be found.
We make the following assumptions,

(a)Only two-particle interactions (dilute system)

(b)Molecular Chaos   (absence of particle correlations).

(c) $f(x,p)$ vary slowly in space-time.

Consider a collision of two particles 
 
$$(p_1^\mu,p_2^\mu) \rightarrow  ({p_1^\prime}^\mu,{p_2^\prime}^\mu)$$

 According to molecular chaos hypothesis, av. no. of such collisions
in $\Delta^4x$ is proportional to:

$$d^3p_1 f(x,p_1)  d^3p_2 f(x,p_2) d^4x$$

The proportionality factor, 

$$\frac{W(p_1p_2|p^\prime_1p^\prime_2)}{p^0_1p^0_2
{p^\prime}^0_1{p^\prime}^0_2}$$

is the transition rate. Note we have neglected difference in space-time coordinate
in $f(x,p_1)$ and $f(x,p_2)$ (assumption (c) slow variation of $f(x,p)$).

The average number of particles lost through collisions is then,

\begin{eqnarray}
\frac{1}{2}d^4x \frac{d^3p_1}{p^0_1}\int\frac{d^3p_2}{p^0_2}
 \frac{d^3p^\prime_1}{{p^\prime}^0_1}\frac{d^3p^\prime_2}{{p^\prime}^0_2}
f(x,p_1)f(x,p_2) \times W(p_1p_2|p^\prime_1p^\prime_2)
\end{eqnarray}

In a similar manner, gain term due to  restituting collisions can  be calculated.

Boltzmann Equation with collision is

\begin{eqnarray} \label{eq8.39}
p^\mu \partial_\mu f 
&=&\frac{1}{2} \int\frac{d^3p_2}{p^0_2}
 \frac{d^3p^\prime_1}{{p^\prime}^0_1}\frac{d^3p^\prime_2}{{p^\prime}^0_2}
[f^\prime_1f^\prime_2 W(p^\prime_1p^\prime_2|p_1p_2)  -
 f_1f_2 W(p_1,p_2|p^\prime_1p^\prime_2)] \nonumber \\
 &=&C[f]  \label{eq8.21}
\end{eqnarray}
 
 The transition rate in Eq.\ref{eq8.39} can be related with cross-section of the process, $1+2\rightarrow 2+3$ and  the collision term $C[f]$ can be represented in several other forms. There is established procedure for solving Boltzmann transport equation, e.g. Chapman-Enskog method or Grad's 14-moment method. 
 However, in the present course, those methods  will not be discussed.

\subsubsection {H-theorem:}

Boltzmann transport equation is manifestly time irreversible. However, the  microscopic interactions are reversible. How this qualitative change came? 
One of the major assumption in transport equation is the molecular chaos hypothesis:  colliding particles are un-correlated. 
This hypothesis
makes a distinction of past and future. Relativistic transport equation thus
describe a irreversible process. This property is manifests most clearly in
Boltzmann H-theorem. In simple terms, H-theorem (or 2nd law of thermodynamics) states that entropy production
at any space-time point is never negative, $\partial_\mu S^\mu \geq 0$.

H-theorem also defines the equilibrium state of a macroscopic system.
 In the equilibrium state,  
 
 \begin{equation}
 \partial_\mu S^\mu = 0
 \end{equation}
 
This condition with transport equation determine the equilibrium distribution function, it is called Juttner distribution.

\subsubsection{Equilibrium distribution function}

If the macroscopic system is in local thermal equilibrium, at each space-time point $x$, we can specify, in addition to hydrodynamic velocity $v(x)$, a temperature $T(x)$ and for each particle species a chemical potential $\mu_i(x)$, which control the
particle density at $x$. Equilibrium distribution function can be obtained from kinetic theory under the condition that $\partial_\mu s^\mu=0$, i.e. it is the
distribution which extremises the entropy-four flow. 

Lorentz-covariant equilibrium distribution function can be written as,

\begin{eqnarray}
f_{i,eq}(x,p)&=&\frac{g_i}{e^{ [p^\mu u_\mu(x)-\mu_i(x)]/T(x)} \pm 1} \nonumber\\
&=&g_i \Sigma_{i=1}^\infty (\mp)^{n+1} e^{-n[p.u(x)-\mu(x)]/T(x)} \label{eq8.43}
\end{eqnarray}

$g_i$ is the degeneracy of the particle, the factor $p.u$ in the exponent is the energy of the particle in the local rest frame ($p.u \rightarrow p^0=E$ when
$u^\mu \rightarrow (1,0)$). The plus and minus sign in the denominator accounts for proper quantum statistics of the particle species, $(+)$ for fermions and $(-)$ for bosons.

\subsection{Conservation equations:} 

An important property of the collision term in Boltzmann equation is

\begin{eqnarray} \label{eq8.49}
\int \frac{d^3p}{p^0} \psi(x,p) C(x,p) =0 \end{eqnarray}
 
with $\psi(x,p)=a(x)+b_\mu(x)p^\mu$. $\psi(x,p)$ is generally called summational invariant.

Eq.\ref{eq8.49} can be used to derive conservation equations. For example, consider the summational invariant $\psi(x,p)=a(x)$. One obtain,

\begin{eqnarray*}
0&=&\int \frac{d^3p}{p^0} a(x) C(x,p) = \int \frac{d^3p}{p^0} a(x)p^\mu \partial_\mu f(x,p)\\
\end{eqnarray*}

Since, we have defined particle 4-current as,
$N^\mu = \int \frac{d^3p}{p^0} p^\mu f(x,p)$,
above equation can be written as the macroscopic conservation law of total particle number,

\begin{equation}
\partial_\mu N^\mu =0
\end{equation}

In a system where number of particles of each component is conserved separately, one can write,

\begin{equation}
\partial_\mu N_k^\mu =0, K=1,2,...N.
\end{equation}

For summational invariant $\psi(x)=b_\mu(x) p^\mu$, one obtain

\begin{eqnarray*}
0&=&\int \frac{d^3p}{p^0} b_\mu(x) p^\mu C(x,p) = \int \frac{d^3p}{p^0} b_\mu(x)p^\mu p^\nu \partial_\nu f(x,p)\\
\end{eqnarray*}

Energy-momentum tensor is defined as, $T^{\mu\nu}=\int \frac{d^3p}{p^0}p^\mu p^\nu f(x)$ and above equation then gives the energy-momentum conservation law,

\begin{equation}
\partial_\mu T^{\mu\nu}=0
\end{equation}

For a system with a singly conserved charge, the five equations,

\begin{eqnarray}
\partial_\mu N^\mu &=&0, \label{eq8.35} \\
 \partial_\mu T^{\mu\nu}&=&0 \label{eq8.36} 
\end{eqnarray}

\noindent govern the motion of the fluid. They are called hydrodynamic equations.
They must be supplemented by the H-theorem or the 2nd law of thermodynamics,

\begin{equation}
\partial_\mu S^\mu \ge 0
\end{equation}  

Explicit decomposition of energy-momentum tensor and particle four flow are given earlier. For completeness, I repeat them here,

In Landau-Lifshitz frame, 

\begin{eqnarray}
N^\mu &=&nu^\mu + \frac{q^\mu}{h} \label{eq8.58} \\
T^{\mu\nu}&=&eu^\mu u^\nu -p\Delta^{\mu\nu}+ \Pi^{\mu\nu} \label{eq8.59}
\end{eqnarray}

In Eckart's frame, 

\begin{eqnarray}
N^\mu &=&nu^\mu  \label{eq8.60}\\
T^{\mu\nu}&=&eu^\mu u^\nu -p\Delta^{\mu\nu}+[q^\mu u^\nu +q^\nu u^\mu]  +\Pi^{\mu\nu} \label{eq8.61}
\end{eqnarray}

It is convenient to split the pressure tensor $\Pi^{\mu\nu}$ into a traceless part $\pi^{\mu\nu}$ and the reminder,

\begin{equation}\Pi^{\mu\nu}=\pi^{\mu\nu}- \Pi \Delta^{\mu\nu}\end{equation} 

The traceless part $\pi^{\mu\nu}$ is the shear stress tensor. The reminder $\Pi$ is the bulk viscous pressure.  

When the energy-momentum tensor of a closed macroscopic system contains the dissipative terms, $q^\mu$, $\Pi$, $\pi^{\mu\nu}$, the system is said to be in non-equilibrium. The one body distribution function deviates from the equilibrium distribution function given in Eq.\ref{eq8.43}.
In the course of time, system approaches equilibrium. This approach is dominated by two concepts, thermodynamic forces (the derivatives of macroscopic variables characterising the system) and dissipative flows, heat flow ($q^\mu$), viscous flow $\Pi^{\mu\nu}$. Flows tend to reduce the non-uniformities in the system. Phenomenologically, to a good approximation, flows are linearly related to the thermodynamic forces. The proportionality constants are called transport coefficients. In ideal or inviscid fluid approximation, transport coefficients are exactly zero. Entropy is maximised $\partial_\mu S^\mu=0$. 
Ideal fluid approximations have been widely used to model relativistic heavy ion collisions. For ideal fluid, the particle 4-flow and energy-momentum tensor can be decomposed as,

\begin{eqnarray}
N^\mu&=& nu^\mu\\
T^{\mu\nu}&=&(e+p)u^\mu u^\nu - p g^{\mu\nu} \label{eq8.38}
\end{eqnarray}

\noindent where, n is the particle density, $e$ is the energy density, $p$ hydrostatic pressure and $u$ the four velocity, with the constraint, $u^2=1$..
One notes that the 5 conservation equations contain 6 unknowns. They are the density $n$, energy density $e$, pressure $p$ and 3-components of hydrodynamic velocity $u$ (  $u$ is constrained by definition, $u^2=1$).  System is closed only with an equation of state, $p=p(e,n)$. Equation of state in an important input of hydrodynamic calculations. Note that hydrodynamic equations are macroscopic in nature. The equations do not depend on the constituents of the system. Only the  equation of state connects the macroscopic state to the microscopic constituents. A hydrodynamic model has the advantage that one can include the phenomena of phase transition through the equation of state.

In general, the fluid is not an ideal. Energy-momentum tensor contains the  dissipative flows $q^\mu$, $\Pi$ and $\pi^{\mu\nu}$.
In kinetic theory, it is possible to relate the    dissipative flows  with gradients of state variables i.e. thermodynamic forces. Phenomenologically, if the departure from ideal fluid is small, entropy current can be expanded in terms of small deviations, $(\delta N=N^\mu-N^\mu_0$, $\delta T^{\mu\nu}=T^{\mu\nu}-T^{\mu\nu}_0)$. The subscript $0$ denotes   the equilibrium value. If the expansion contains only the terms first order in $\delta N$ and $\delta T^{\mu\nu}$ one obtain the 'first order' theory of dissipative hydrodynamics. The
procedure yield the 'constitutive' relations for heat flow, bulk viscous pressure and shear stress tensor,

\begin{eqnarray}
\Pi&=&-\zeta \nabla^\mu u_\mu \label{eq8.63} \\
q^\mu&=&\kappa (\nabla^\mu T - \frac{T}{e+p}\nabla^\mu p)\label{eq8.64}\\
\pi^{\mu\nu}&=&2\eta <\nabla^\mu u^\nu > \label{eq8.65}
\end{eqnarray}

\noindent where the angular bracket define a symmetric, traceless tensor,

\begin{equation}
<\nabla^\mu u^\nu>=\left [\frac{1}{2}(\Delta^\mu_\sigma \Delta^\nu_\tau +
\Delta^\mu_\tau \Delta^\nu_\sigma ) - \frac{1}{3}\Delta^{\mu\nu} \Delta_{\sigma\tau} \right ] \nabla^\sigma u^\tau
\end{equation}

In Eqs.\ref{eq8.63}, \ref{eq8.64},\ref{eq8.65}, $\zeta$ is the bulk viscosity coefficient, $\kappa$ the heat conductivity and $\eta$ is the shear viscosity coefficient. The 'constitutive' relations can be inserted in the energy-momentum tensor and hydrodynamic equations Eqs.\ref{eq8.58},\ref{eq8.59} or
Eqs.\ref{eq8.60},\ref{eq8.61} can be solved in principle.
However, first order theory of dissipative hydrodynamics suffers from the problem of causality. They are acausal. Also, there may be the problem of instability. The causality problem is removed if the expansion of entropy current contains terms in second order in deviations $\delta N$ and $\delta T^{\mu\nu}$. The procedure gives rise to the relaxation equations for the dissipative flows. In addition to the conservation equations, relaxation equations for the dissipative flows need to be solved. 

\begin{eqnarray}
\Pi&=&-\zeta \nabla^\mu u_\mu -\tau_\Pi D\Pi \\
q^\mu&=&\kappa (\nabla^\mu T - \frac{T}{e+p}\nabla^\mu p) -\tau_q Dq^\mu \\
\pi^{\mu\nu}&=&2\eta <\nabla^\mu u^\nu >  - \tau_\pi D\pi^{\mu\nu} 
\end{eqnarray}

$D=u^\mu \partial_\mu$ is the convective time derivative.  $\tau_\Pi$, $\tau_q$ and $\tau_\pi$ are relaxation time for the bulk viscous pressure, conductivity and shear stress tensor. Transport coefficients and as well as the relaxation times, in principle, can be calculated in kinetic theory. However, for strongly interacting system like QGP, calculations are very complex. I may note here that I have written the simplest possible relaxation equation. Relaxation equations can contain additional terms.  

There are 10 independent dissipative flows. 
The shear stress tensor $\pi^{\mu\nu}$ is symmetric, traceless and transverse to hydrodynamic velocity ($u_\mu \pi^{\mu\nu}=0$). $\pi^{\mu\nu}$ has 6 independent components. Heat flow $q^\mu$
is also transverse to  hydrodynamic velocity ($u_\mu w^\mu=0$) and has 3 independent components. And there is the bulk viscous pressure. 
In second-order hydrodynamics, in addition to the 5 conservation equations, 10 
relaxation equations for dissipative flows need to be solved.  
 
Above discussions about dissipative hydrodynamics are rather sketchy. For more information one can see \cite{IS79}\cite{Muronga:2001zk}\cite{Muronga:2006zw} \cite{Heinz:2005bw}\cite{Chaudhuri:2006jd}\cite{Romatschke:2009im}\cite{Denicol:2010xn}.

\section{Hydrodynamic model for heavy ion collisions}\label{sec9}

\subsection{Different stages of HI collisions:}\label{sec8}

We want to study properties of QGP. QGP existed in early universe. It is also possible that it exists at the core of neutron stars. However, QGP at the early universe or at the core of neutron stars are not accessible for study. Theoretical considerations led us to believe that QGP can be produced in laboratory by colliding heavy nuclei at very high energy. A nucleus-nucleus collision is a well established tool to study properties of nuclear matter. Recent experiments at Relativistic Heavy Ion Collider (RHIC) at Brookhaven National Laboratory and
Large Hadron Collider (LHC) at CERN, strongly suggests that QGP is formed in high energy nuclear collisions.

Let us qualitatively discuss the collision process with increasing collision energy. 
In very low energy collisions, nucleus as a whole interacts. Indication are obtained from, say giant dipole excitations, where the compound nucleus undergoes  dipole oscillation (proton and neutron fluid oscillate out of phase). One can also excite the nucleus and populate various excited states. As the energy is increased, nucleons in the nucleus start to interact,
one can see production of new particles, e.g. $\pi$. At still higher energy, quarks inside the nucleons will interact. Here also production of different particle species will be observed. However, in contrast to low/medium energy nuclear reactions, where one can describe pA/AA collisions entirely in terms of NN collisions, in relativistic energy, such a description will fail. 

\begin{figure}[t]
 \center
 \resizebox{0.5\textwidth}{!}
 {\includegraphics{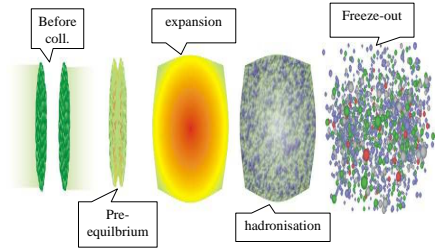}}
   \vspace{1.0cm}
\caption{different stages of nuclear collisions.} \label{F25}
\end{figure}

 A nucleus-nucleus collision at relativistic energy passes through different stages. Initial collisions are expected to be in the partonic level. Schematic picture of different staged of the collisions are shown in Fig.\ref{F25}. One can broadly classify different stages;

(i)Pre-equilibrium stage: initial partonic collisions produce a fireball in a highly excited state.
In all possibility, the fireball is not in equilibrium. Constituents of the system
collide frequently to establish a 'local' equilibrium' state. The time takes
to establish local equilibrium is called thermalisation time. It is an important parameter. Explicit comparison of hydrodynamic models with experiments indicate that in $\sqrt{s}_{NN}$=200 GeV Au+Au collisions, local equilibrium is established sufficiently fast,
in the time scale 0.5-1.0 fm. However, it is not understood how such fast equilibration can be achieved. In classical electromagnetic plasma,  self excited
transverse modes exist in plasmas with momentum anisotropy \cite{Weibel:1959zz}. They are called Weibel instability. The instability can grow very fast. The   instability reduces the momentum anisotropy. It is believed that
in QGP,  non-abelian version of the Weibel instability establishes rapid   thermalisation \cite{Randrup:2003cw}.

(ii) expansion stage: In the equilibrium or the thermalised state, the system has thermal pressure which acts against the surrounding vacuum. The system then undergoes collective (hydrodynamic) expansion. As the system expands, its density (energy density) decreases and the system cool. The expansion and cooling 
is governed by the energy-momentum conservation equations, which involve an equation of state $p=p(e,n_B)$. Now if there is a QCD phase transition in the model, then below the critical energy density $\epsilon_{cr}\approx  1GeV/fm^3$,
or critical temperature $T_{cr}\approx 200 MeV$, the partons (quarks and gluons) will convert to hadrons. In the hadronisation stage, over a small temperature interval, entropy density will decrease very fast. Since total entropy can not decrease, it implies that the fire ball will expand rapidly, while temperature remains approximately constant.    
If the transition is 1st order, there will be mixed phase, where QGP and hadronic resonance gas can co-exist. In the mixed phase, speed of sound  
$c_s^2=\frac{dp}{d\epsilon} \approx 0$. In 2nd or cross-over transition, there will not be any mixed phase, even then, near the transition, equation of state is soft and speed of sound is small (see Fig.\ref{F17}). Then collective flow will not grow much during the transition period. Ultimately all the partonic matter will be converted into hadronic matter.  

(iii) Freeze-out: Hadronic matter will also be in thermal equilibrium. 
Constituent hadrons will collide to maintain local equilibrium. The system
will expand and cool. A stage will come when inelastic collisions, in which 
hadron changes identity, become too small to keep up with expansion. The stage is called chemical freeze-out. Hadron abundances will remain fixed after the chemical freeze-out. However, due to elastic collisions, local equilibrium can still be maintained and system will cool and expands with fixed hadron abundances. 
Eventually a stage will comes when average distance between the constituents will be larger than strong interaction range. Collisions between the constituents will be so infrequent that   'local' thermal equilibrium can not be maintained. The hydrodynamic description will break down. The hadrons decouple or   freeze-out. It is called kinetic freeze-out.  Hadrons from the freeze-out surface will be detected in the detector.

Hydrodynamics provides a simple, intuitive description of relativistic
heavy-ion collisions. Hydrodynamic models requires the assumption of local (thermal) equilibrium, i.e. one assume that at each space time point $x$ of the fluid,  a small region can be considered where equilibrium is achieved, a temperature $T(x)$ can be defined. As discussed above, from the equilibrium stage to the kinetic freeze-out, relativistic heavy ion collisions can be modeled by hydrodynamics.


A variety of experimental data  from Relativistic Heavy Ion Collider (RHIC) experiments,   are  successfully explained in ideal hydrodynamical models  \cite{Kolb:2001qz}\cite{Kolb:2003dz}. Some problems however remained. For example, ideal hydrodynamics description of experimental data   becomes poorer  as the collisions become more and more peripheral. Also, contrary to the experimental data, in ideal hydrodynamics,
elliptic flow (it is an important observable in high energy heavy ion collisions and will be discussed later) continue to increase with transverse momentum.
Deficiencies of ideal hydrodynamic models are to some extent corrected in dissipative hydrodynamics \cite{Teaney:2003kp}\cite{Chaudhuri:2008sj}.%


\subsection{Hydrodynamic equations}

In this section, I describe the general procedure followed in hydrodynamic modeling of heavy ion collisions.
For simplicity, we assume an ideal fluid with a single conserved charge (e.g. baryon density). The  five conservation equations,

\begin{eqnarray}
\partial_\mu N^\mu &=&0, \\
 \partial_\mu T^{\mu\nu}&=&0
\end{eqnarray}

\noindent govern the motion of the fluid.  Given an initial configuration of the fluid and an equation of state $p=p(e,n_B)$, the equations can be solved numerically to obtain the   space-time 
evolution of the fluid. 

In heavy ion collisions, appropriate coordinates for solving hydrodynamic equations are ($\tau,x,y,\eta$) rather than ($t,x,y,z$).

\begin{eqnarray}
x^\mu=(t,x,y,z) &\rightarrow& x^m=(\tau,x,y,\eta) \\
t=\tau \cosh \eta;&& \tau=\sqrt{t^2-z^2} \\
z=\tau \sinh \eta :&& \eta=\frac{1}{2}\ln \left (\frac{t+z}{t-z}\right )
\end{eqnarray}

In $(\tau,x,y,\eta)$ coordinate system the metric is,

\begin{eqnarray}
ds^2=g_{\mu\nu}dx^\mu dx^\nu&=&dt^2-dx^2-dy^2-dz^2\\
 &=& d\tau^2 -dx^2 -dy^2 -\tau^2 d\eta^2
\end{eqnarray}

and ,

\begin{equation}
g^{\mu\nu}=diag(1,-1,-1,-1/\tau^2).
\end{equation}

One note that the space-time is not flat anymore, it is curved. Accordingly, one need 'affine connections' or the Christoffel symbols,

\begin{equation}
\Gamma^i_{jk}=\frac{1}{2}g^{im}
\left (\frac{\partial g_{mj}}{\partial x^k}+\frac{\partial g_{mk}}{\partial x^l}
-\frac{\partial g_{jk}}{\partial x^m} \right )
\end{equation}

In ($\tau$,x,y,$\eta$) coordinate only non-zero Christoffel symbols are,

\begin{equation} 
\Gamma^\tau_{\eta\eta}=\tau, \Gamma^\eta_{\tau\eta}=1/\tau
\end{equation}
 
Covariant (semicolon) derivative of a contravariant tensor is given by,
 
\begin{eqnarray*}
A^i_{;p} =&& \frac{\partial A^i}{\partial x^p} + \Gamma^i_{pm}A^m\\
A^{ik}_{;p}=&&\frac{\partial A^{ik}}{\partial x^p} +\Gamma^i_{pm}A^{mk}
+\Gamma^k_{pm}A^{mi}
\end{eqnarray*}

Five 
conservation Equations in ($\tau$,x,y,$\eta$) coordinate system can be easily derived. For 'ideal fluid', they are, 

\begin{eqnarray}
(i) N^\mu_{;\mu}=0&=& N^\tau_{;\tau}+ N^x_{;x}+ N^y_{;y}+ N^\eta_{;\eta} \nonumber \\
  &=&
(\partial_\tau N^\tau+ \Gamma^\tau_{\tau m} N^m)   
+ \partial_x N^x+\partial_y N^y+
(\partial_\eta N^\eta +\Gamma_{\eta m}^\eta N^m \nonumber\\
  &=&
\partial_\tau N^\tau+    
+ \partial_x N^x+\partial_y N^y+
\partial_\eta N^\eta +\frac{1}{\tau} N^\tau \label{eq9.11}\\
(ii) T^{\mu \tau}_{;\mu}=0&=& T^{\tau \tau}_{;\tau}+ T^{\tau x}_{;x}+ T^{\tau y}_{;y}+ T^{\tau \eta}_{;\eta}\nonumber\\
  &=&
(\partial_\tau T^{\tau \tau}+2 \Gamma_{\tau m}^\tau T^{m\tau})   
+ \partial_x T^{\tau x}+\partial_y T^{\tau y}\nonumber \\
&+&(\partial_\eta T^{\tau\eta} +\Gamma_{\eta m}^\tau T^{m\eta}+\Gamma_{\eta m}^\eta T^{m\tau})\nonumber\\
  &=&
\partial_\tau T^{\tau \tau}  
+ \partial_x T^{\tau x}+\partial_y T^{\tau y}+
\partial_\eta T^{\tau \eta}+\tau T^{\eta \eta} +\frac{1}{\tau} T^{\tau\tau}\label{eq9.12}\\
(iii) T^{\mu x}_{;\mu}=0&=& T^{\tau x}_{;\tau}+ T^{x x}_{;x}+ T^{x y}_{;y}+ T^{\eta x}_{;\eta}\nonumber\\
  &=&
\partial_\tau T^{\tau y}  
+ \partial_x T^{x y}+\partial_y T^{yy}+
\partial_\eta T^{y\eta}+\frac{1}{\tau} T^{\tau x}\label{eq9.13}\\ 
(iv) T^{\mu y}_{;\mu}=0&=& T^{\tau y}_{;\tau}+ T^{x y}_{;x}+ T^{y y}_{;y}+ T^{\eta y}_{;\eta}\nonumber\\
  &=&
\partial_\tau T^{\tau y}  
+ \partial_x T^{x y}+\partial_y T^{yy}+
\partial_\eta T^{y\eta}+\frac{1}{\tau} T^{\tau y}\label{eq9.14}\\ 
(v) T^{\mu\eta}_{;\mu}=0&=& T^{\tau\eta}_{;\tau}+ T^{x\eta}_{;x}+ T^{y\eta}_{;y}+ T^{\eta\eta}_{;\eta}\nonumber\\
 &=&
\partial_\tau T^{\tau\eta}   
+ \partial_x T^{x\eta}+\partial_y T^{y\eta}+
\partial_\eta T^{\eta\eta} + \frac{3}{\tau} T^{\eta \tau} \label{eq9.15}
\end{eqnarray}

In ideal hydrodynamics, or in 1st order hydrodynamics with dissipation, the five partial equations noted above are to be solved simultaneously to obtain space-time evolution of the fluid. Flux corrected SHASTA algorithm \cite{Boris} can be used to solve those equations. In 2nd order dissipative hydrodynamics however, 10 additional relaxation equations have to be solved simultaneously.


For illustrative purpose, let us specialize in one-dimensional, Bjorken scaling flow \cite{Bjorken:1982qr}. In one dimension expansion, hydrodynamic four velocity is $u^\mu=(1,0,0,0)$ and relevant energy-momentum components are,

\begin{equation}
T^{\tau\tau}=e; T^{\tau x}=p; T^{\tau y}=p; T^{\eta\eta}=p/\tau^2
\end{equation}

Inserting the values in Eq.\ref{eq9.12} we obtain,

\begin{eqnarray}
0&=&\partial_\tau T^{\tau \tau}  
+ \partial_x T^{\tau x}+\partial_y T^{\tau y}+
\partial_\eta T^{\tau \eta}+\tau T^{\eta \eta} +\frac{1}{\tau} T^{\tau\tau} \nonumber \\
&=&\partial_\tau T^{\tau \tau}  
 +\tau T^{\eta \eta} +\frac{1}{\tau} T^{\tau\tau}=\frac{\partial e}{\partial \tau} +\frac{e+p}{\tau} \label{eq8.17}
\end{eqnarray}

Using entropy density $s=\frac{e+p}{T}$, the Eq.\ref{eq8.17} can be recast in to

\begin{equation}
\frac{d\tau s}{d\tau}=0,
\end{equation}

\noindent which can be solved as, $s\tau=constant$. One dimensional flow is isentropic. In ideal gas, $s \propto T^3$ and we get the well known $T^3$ law for Bjorken scaling expansion,

\begin{equation}  
T_i^3 \tau_i =T_f^3 \tau_f
\end{equation}

Given the fluid temperature at initial time $\tau_i$, its value at a subsequent time is easily obtained. Similarly, one can also solve for the number conservation equation. For fluid velocity $u^\mu=(1,0,0,0)$, $N^\mu=(n,0,0,0)$. From Eq.\ref{eq9.11} we obtain,
 
\begin{eqnarray}
0&=&\partial_\tau N^\tau+    
+ \partial_x N^x+\partial_y N^y+
\partial_\eta N^\eta +\frac{1}{\tau} N^\tau \nonumber\\
&=&\partial_\tau n +\frac{1}{\tau} n 
\end{eqnarray}

The Eq. can be solved to give,

\begin{equation}
n_f=n_i\frac{\tau_i}{\tau_f}\\
\end{equation}

Noting that density is inversely proportional to volume, one find the scaling law for volume expansion,

\begin{equation}
V_f=V_i\frac{\tau_f}{\tau_i}\\
\end{equation}

\subsection{Cooper-Frye prescription for particle distribution}

Hydrodynamic equations  give the space-time evolution of the fluid till a freeze-out condition. One obtain information about the fluid energy density (or temperature) and velocity at the freeze-out. The information needs to be converted into particle distribution. This is done through Cooper-Frye prescription \cite{Cooper:1974mv}.

Consider a three dimensional hypersurface $\Sigma(x)$ in 4-dimensional space-time and count the number of particles crossing the hypersurface.  Let $d\sigma_\mu$ is an infinitesimal element perpendicular to $\Sigma(x)$ and directing outward. If $j^\mu$ is the current of particles, then the scalar product $d\sigma_\mu j^\mu$, gives the number of particles crossing the infinitesimal surface $d\sigma$. The total number crossing the hyper surface $\Sigma$ is,

 \begin{equation} N=\int_\Sigma d\sigma_\mu j^\mu = \int_\Sigma d\sigma_\mu \left (\frac{1}{(2\pi)^3} \int \frac{d^2p}{E} p^\mu f(x,p) \right )
 \end{equation}
 
\noindent with $f(x,p)$ the one body distribution function.

 In differential form, 
 
\begin{equation} \label{eq9.25}
E\frac{dN}{d^3p}=\frac{dN}{dyd^2p_T}=\frac{1}{(2\pi)^3} \int_\Sigma d\sigma_\mu   p^\mu f(x,p)  
\end{equation}

This is the Cooper-Frye prescription \cite{Cooper:1974mv}  for the invariant distribution of particles. Let us evaluate the term $d\sigma_\mu p^\mu$.


In $(\tau,x,y,\eta)$ coordinates, the freeze-out surface can be parameterised as,

\begin{equation}
\sigma^\mu=(\tau_f(x,y)\cosh \eta,x,y,\tau_f(x,y) \sinh \eta),
\end{equation}

\noindent and the normal vector on the hyper surface is,

\begin{equation}
d\sigma^\mu=(\cosh \eta,-\frac{\partial\tau_f}{\partial x},\frac{\partial\tau_f}{\partial y},-\sinh \eta) \tau_f dx dy d\eta,
\end{equation}

4-momentum $p^\mu$ can be parameterised as,

\begin{equation}
p^\mu=(m_T \cosh y, p_x, p_y, m_T \sinh y)
\end{equation}

The volume element

\begin{equation}
p^\mu d\sigma_\mu=(m_T \cosh (y-\eta)-p_x \frac{\partial\tau_f}{\partial x} - p_y \frac{\partial\tau_f}{\partial y}) \tau_f dx dy d\eta \label{eq9.29}
\end{equation}


$f$ in \ref{eq9.25} is the distribution function. In ideal hydrodynamics, the $f$ corresponds to Lorentz-covariant equilibrium distribution function, 

\begin{eqnarray}
f_{i,eq}(x,p)&=&\frac{g_i}{e^{ [p^\mu u_\mu(x)-\mu_i(x)]/T(x)} \pm 1} \nonumber\\
&=&g_i \sum_{i=1}^\infty (\mp)^{n+1} e^{-n[p^\mu u_\mu(x)-\mu(x)]/T(x)}
\end{eqnarray}

$g_i$ is the degeneracy of the particle, the factor $p.u$ in the exponent is the energy of the particle in the local rest frame ($p^\mu u_\mu \rightarrow p^0=E$ when
$u^\mu \rightarrow (1,0)$). The plus and minus sign in the denominator accounts for proper quantum statistics of the particle species, $(+)$ for fermions and $(-)$ for bosons.

Let us evaluate the term $p^\mu u_\mu$. The fluid 4-velocity can be parameterised as,

\begin{eqnarray}
u^\mu=\gamma_T(\cosh \eta, v_x, v_y, \sinh \eta),\\ 
\gamma_T=\frac{1}{\sqrt{1-v_T^2}}= \frac{1}{\sqrt{1-v_x^2-v_y^2}}
\end{eqnarray}

The scalar product $p.u$ in the equilibrium distribution function is then easily calculated as,

\begin{equation}\label{eq9.35}
p^\mu u_\mu=\gamma_T (m_T \cosh(y-\eta) - v_x p_x-v_y p_y)
\end{equation}

Eqs.\ref{eq9.29},\ref{eq9.35} completely specifies the Cooper-Frye invariant distribution Eq.\ref{eq9.25}.

In dissipative hydrodynamics, distribution function $f$ deviates from the equilibrium value,
 
\begin{equation}
f(x,p)=f_{eq}[1+\phi(x,p)]
\end{equation}

Since distribution function is a scalar, the deviation function must be written as sum of scalars constructed from $p^\mu$, $u^\mu$ and dissipative flows. It can be parameterised as,

\begin{equation}
\phi(x,p)=A(x,p)\Pi + B(x,p) p_\mu q^\mu + C(x,p) p_\mu p_\nu \pi^{\mu\nu} 
\end{equation}

The coefficients A, B and C can be determined from the condition of the fit,
that the number density and energy density is determined by the equilibrium distribution functions,

\begin{eqnarray}
\int \frac{d^3p}{p^0}p^\mu u_\mu f_{eq} \phi &=&0 \\
\int \frac{d^3p}{p^0} (p^\mu u_\mu)^2 f_{eq} \phi &=&0 
\end{eqnarray}

I will not discuss in detail, but the deviation function for shear viscosity can be written as,

\begin{eqnarray}
\phi_{shear}=C(x,p)p_\mu p_\nu \pi^{\mu\nu}
=\frac{1}{2((e+p)T^2} p_\mu p_\nu \pi^{\mu\nu}
\end{eqnarray}

In \cite{Monnai:2009ad}, deviation function for bulk viscosity is obtained from Grads 14 moment method,

\begin{eqnarray}
\phi_{bulk}=A(x,p)\Pi=D_0 p^\mu u_\mu + B_0 p^\mu p^\nu \Delta_{\mu\nu} + \tilde{B}_0 p^\mu p^\nu u_\mu u_\nu
\end{eqnarray}

expressions for $D_0$, $B_0$ and $\tilde{B}_0$ can be found in  \cite{Monnai:2009ad}.


\subsection{Initial conditions for hydrodynamic analysis}

One understands that hydrodynamics is an initial value problem. For example, in Bjorken one dimensional hydrodynamics, given the density/temperature at some initial time $\tau_i$ (the time beyond which hydrodynamics is applicable, itself a parameter of the model), density/temperature evolution of the fluid can be obtained. A kinetic freeze-out condition also required to define the freeze-out surface such that Cooper-Frye prescription can yield particle's invariant distribution. One simple procedure to implement kinetic freeze-out is to assume a fixed freeze-out temperature $T_F$.
Depending on the model, experimental data are fitted with $T_F$=100-140 MeV.

In a more general system, one has to initialise the (baryon) number density $n(x,y,\eta)$, energy density $e(x,y,\eta)$ and velocity ${\bf u}(x,y,\eta)=\gamma {\bf v}(x,y,\eta)$  distributions at the initial time $\tau_i$. Indeed, one of the aims of hydrodynamic analysis of heavy ion collisions in ultra-relativistic collisions is to obtain the initial conditions of the produced fluid, by comparing hydrodynamic simulations with experimental data.
As discussed earlier, experimental results are given in terms of collision centrality. One then tries to parameterise the initial condition in terms of impact parameter, such that once the parameters are fixed at some particular collision centrality, it can predict for other collision centralities.     

In ($\tau$,x,y,$\eta$) coordinate, for the initial energy density, a common practice is to assume a  factorised form,

\begin{equation} e(x,y,\eta)=\varepsilon(x,y)H(\eta)\end{equation}

\noindent $\varepsilon(x,y)$ being the initial energy density in the transverse plane and $H(\eta)$ in the direction of (spatial) rapidity. One can use a Gaussian distribution for $H(\eta)$. Transverse energy distribution $\varepsilon(x,y)$ can be conveniently parameterised in a Glauber model or in color glass condensate (CGC) model.  The number density distribution or the velocity distribution at the initial time can be similarly parameterised. In general, one assume zero initial fluid velocity at the initial time, though it is possible that fluid have non-zero velocity, especially near the surface of the fluid. The reasoning is simple. Fluid constituents can have random velocity. In the interior of the fluid, the random velocities will balance to produce net zero velocity. However, near the surface random velocities will not be balanced.

  
\subsubsection{Glauber model initial condition}

In section.\ref{sec3}, I have discussed the Glauber model. Expressions for the number of participant nucleons and number of binary collisions, in impact parameter $\bf{b}$ collisions were obtained.

\begin{eqnarray} 
N_{coll}({\bf b})&=&AB \sigma_{NN} \int d^2s T_A({\bf b}) T_B({\bf b-s})\\
N_{part}({\bf b})&=&A\int d^2sT_A({\bf s})(1-[1-\sigma_{NN}T_{B}({\bf b-s})]^{B}) \nonumber \\
&+&
B\int d^2s T_B({\bf b-s})(1-[1-\sigma_{NN}T_{A}({\bf s})]^{A}) 
\end{eqnarray}

From the above equations,  , transverse profile of binary collision number and participant numbers, in impact parameter ${\bf b}$ collisions can be easily obtained as,

\begin{eqnarray}
N_{coll}(x,y)&=&AB\sigma_{NN} T_A(x+{\bf b}/2,y) T_B(x-{\bf b}/2,y)\\
N_{part}(x,y)&=&A T_A(x+{\bf b}/2,y)(1-[1-\sigma_{NN}T_{B}(x-{\bf b}/2,y)]^{B}) \nonumber \\
&+&
B  T_B(x-{\bf b}/2,y)(1-[1-\sigma_{NN}T_{A}(x+{\bf b}/2,y)]^{A})  \label{eq9.43}
\end{eqnarray}
 
Comparison of hydrodynamic simulations with experimental data indicate that a combined profile,

\begin{equation}
e(x,y) \propto[(1-f)N_{part}(x,y) + f N_{coll}(x,y)] =e_0 [(1-f)N_{part}(x,y) + f N_{coll}(x,y)]
\end{equation}

\noindent with $f\approx 0.1-0.2$ best explains the data. Once the proportionality factor $e_0$ is fixed in a given collision centrality, the impact parameter dependence of the model allow one to predict for the energy density distribution at other collision centralities.

\subsubsection{CGC initial condition}

In section.\ref{sec6}, I have discussed Color Glass Condensate (CGC).
CGC is a quantum mechanical state of matter at high energy. The earliest state of matter produced in high energy nucleus-nucleus collisions,   may not be much different from this quantum mechanical state. At very early time, CGC evolves in to a distribution of gluons. Later these gluons thermalise and form  QGP. CGC models have been used extensively to model the transverse energy distribution of the initial QGP fluid 
in hydrodynamic models \cite{Hirano:2004en}\cite{Luzum:2008cw}\cite{Roy:2012jb} 
In the following, I briefly describe the procedure to obtain the initial condition in high energy nuclear collisions in the  KLN (Kharzeev-Levin-Nardi) \cite{Kharzeev:2002ei} approach to CGC. 

The number of gluons produced 
in  the $k_T$ factorisation formula is given by,

\begin{eqnarray}
\frac{dN_g}{d^2r_T dY}&=&\frac{4\pi^2 N_c}{N_c^2-1}\int \frac{d^2p_T}{p_T^2} \int 
d^2k_T \alpha_s(k_T)  \nonumber  \\
&\times&  \phi_A(x_1,p_T^2,r_T)\phi_B (x_2,(p_T-k_T)^2,r_T ) \label{eq13.4}
\end{eqnarray}

\noindent where $p_T$ and $Y$ are the transverse momentum and rapidity of the produced gluon. $x_{1,2}=\frac{p_T}{\sqrt{s}} e^{\pm Y}$ is the momentum fraction of colliding gluons ladders at c.m. energy $\sqrt{s}$. $\alpha_s(k_T)$  is the strong coupling constant at the momentum scale $k_T$.  The unintegrated gluon distribution functions $\phi_A$ in Eq.\ref{eq13.4} are related to the gluon density in a nucleus at the transverse position $r_T(=x,y)$,

\begin{equation}
xG(x,Q^2)=\int^{Q^2} d^2k_T d^2r_T \phi_A(x,k^2_T;r_T)
\end{equation}

In principle, unintegrated gluon distribution function should be a solution of non-linear quantum evolution equation e.g. JIMWLK equation \cite{JalilianMarian:1997jx}\cite{JalilianMarian:1997gr}\cite{Iancu:2001ad}\cite{Iancu:2000hn}. In Kharzeev-Levin-Nardi (KLN) approach \cite{Kharzeev:2002ei}  approach (which captures the essential features of the gluon saturation), the unintegrated gluon distribution functions   are taken as \cite{Kharzeev:2002ei}\cite{Drescher:2006pi},

\begin{equation}
\phi_A (x,k_T^2,r_T) \sim \frac{1}{\alpha_s(Q_s^2)} \frac{Q_s^2}{max(Q_s^2,k_T)}  
\end{equation}

\noindent where $Q_s$ is saturation momentum at the given momentum fraction $x$ and at the transverse position $r_T$.  



In the KLN approach, the saturation scale in AB collision is parameterised as \cite{Kharzeev:2002ei}\cite{Drescher:2006pi},

\begin{equation}
Q_{s, A(B)}(x,r_T)= 2 GeV^2 \left ( \frac{N^{A(B)}_{part}(r_T)}{1.53} \right )
 \left (\frac{0.01}{x} \right )^\lambda
\end{equation}

The form $Q_s(x)\sim x^{-\lambda}$, with $\lambda\approx0.2-0.3$ is motivated from DIS experiments.
$N^{A(B)}_{part}$ in the above equation is the transverse density of participant nucleons, which can be calculated in a Glauber model (e.g. see Eq.\ref{eq9.43}).

\begin{equation}
N^A_{part}(r_T)=
A T_A(x+{\bf b}/2,y)(1-[1-\sigma_{NN}T_{B}(x-{\bf b}/2,y)]^{B})
\end{equation}

In the CGC model, the transverse energy density should follow Eq.\ref{eq13.4}. However, Eq.\ref{eq13.4}  is valid in the time scale $\tau_s\sim \frac{1}{Q_s}$, when the medium may not be in thermal equilibrium. One assumes that  the medium undergoes one dimensional Bjorken (longitudinal, isentropic) expansion during the period $\tau_s$ to $\tau_i$. The density at the time $\tau_i$, when hydrodynamic become applicable is easily obtained as $n(\tau_i)= \frac{\tau_s}{\tau_i} n(\tau_s)$. The transverse energy density profile at the initial time $\tau_i$ is then, 

\begin{equation}
e(x,y,b)= e_0 \left [\frac{dN_g}{dxdy dY} \right ]^{4/3}
\end{equation} 
 
\noindent with $e_0$ a normalising factor, which is to be fixed from experimental data.

\begin{figure}[t]
 \center
  \includegraphics[scale=0.6]{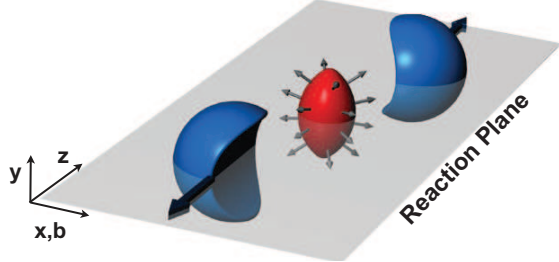}
\caption{\label{F28}Schematic picture of a non-zero impact parameter collision.}
\end{figure}

\subsection{Collective flow}

In relativistic heavy ion collisions, one of the important observables is the azimuthal distribution of produced particles.  In Fig.\ref{F28},  geometry of a collision at non-zero impact parameter collision is shown. The overlap region of the two nuclei is the participant region, where most of the collisions occur. The target and projectile remnants on the periphery acts as spectator.    It is obvious from Fig.\ref{F28}, that in non-zero impact parameter collision, the participant or the reaction zone in coordinate   space   do not posses azimuthal symmetry. Multiple collisions among the constituent particles translate this spatial anisotropy into momentum anisotropy of the produced particles.
The observed momentum anisotropy is called collective flow and has a natural explanation in a hydrodynamic model \cite{Kolb:2003dz}. 
In the following we briefly discuss collective flow phenomena. More detailed expositions can be found in \cite{Poskanzer:1998yz}\cite{Voloshin:2008dg}\cite{Ollitrault:1992bk}.

 
 Momentum anisotropy is best studied by decomposing the invariant distribution in a   Fourier series. For example, the momentum integrated invariant distribution of a particle can be expanded as,

\begin{equation}\label{eq10.1}\frac{dN}{d\phi}=\frac{N}{2\pi}[1+ 2\sum_n v_n cos[n(\phi-\psi) ], n=1,2,3...\end{equation} 
 

\noindent   $\phi$ is the azimuthal angle of the detected particle and 
$\psi$ is the  plane of the symmetry of initial collision zone. For smooth initial matter distribution, plane of symmetry of the collision zone coincides with the reaction plane $\Psi_{RP}$ (the plane containing the impact parameter and the beam axis). The sine terms are not present in the expansion due to symmetry with respect to the reaction plane. 

Flow coefficients $v_n$ are easily obtained, 

\begin{equation}
v_n=\la cos(n\phi-n \psi) \ra= \frac{\int d\phi \frac{dN}{d\phi} cos(n\phi-n\psi)}
{\int d\phi \frac{dN}{d\phi}}, n=1,2,3...
\end{equation}

$v_1$ is called (integrated) directed flow, $v_2$ (integrated) elliptic flow, $v_3$  (integrated) triangular flow, $v_4$ (integrated) hexadecapole flow etc.

Similar to Eq.\ref{eq10.1}, one can Fourier expand the invariant distribution 

  \begin{equation}\label{eq10.3}E\frac{d^3N}{d^3p}=\frac{1}{2\pi}
 \frac{d^2N}{p_T dp_T dy} [1+ 2\sum_n v_n cos[n(\phi-\psi) ], n=1,2,3...\end{equation}

\begin{figure}[t]
 \center
  \includegraphics[scale=1.0]{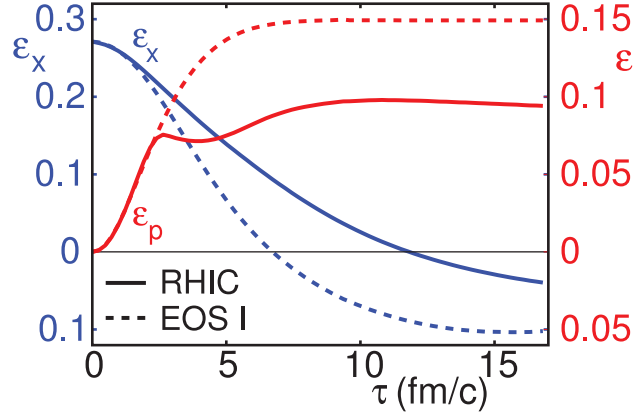}
\caption{\label{F30} temporal evolution spatial eccentricity ($\varepsilon_x$)
and momentum anisotropy ($\varepsilon_p$) with time \cite{Kolb:2003dz}.}
\end{figure}

\noindent and differential flow coefficients are obtained as,

\begin{equation}
v_n(p_T,y)=\la cos(n\phi-n \psi) \ra= \frac{\int d\phi \frac{d^3N}{p_T dp_T d\phi dy} cos(n\phi-n\psi)}
{\int d\phi \frac{d^3N}{p_T dp_T d\phi dy}}, n=1,2,3...
\end{equation}

Second flow coefficient has been studied extensively in RHIC and LHC energy collisions. Finite-non-zero value of $v_2$ is thought to be direct signature of production of thermalised medium.   

Elliptic flow in heavy ion collisions is best understood in a hydrodynamic model \cite{Kolb:2003dz}. Elliptic flow measure the momentum anisotropy.  In non-zero impact parameter collisions, the reaction zone is spatially asymmetric.   Spatial asymmetry of the initial reaction zone can be quantified in terms of eccentricity, defined as,

\begin{equation}
\varepsilon=\frac{\la y^2-x^2 \ra}{\la y^2+x^2 \ra}
\end{equation}

\noindent where $\la .. \ra$ indicate energy/entropy density weighted averaging.
In non-zero impact parameter collision, initial eccentricity is non-zero, positive. If a thermalised medium is produced in the reaction zone, due to thermodynamic pressure, the medium will expand against the outside vacuum. One can immediately see that pressure gradient will be more along the minor axis than along the major axis. Due to differential pressure gradient, as the system evolves with time, eccentricity will reduce. Momentum distribution of particles are isotropic initially. If momentum anisotropy is measured as,

\begin{equation}
\varepsilon_p=\frac{\int dxdy [T^{xx}-T^{yy}]}{\int dxdy [T^{xx}+T^{yy}]}
\end{equation}

\noindent initially $\varepsilon_p$ will be zero. However, as the fluid evolves, 
rescattering of particles will introduce asymmetry and $\varepsilon_p$ will grow. It is expected to saturate beyond certain time, when reaction zone 
attains azimuthal symmetry. In that sense, elliptic flow is self quenching phenomena, driving force of the flow (the reaction zone asymmetry) continuously reduces as the flow grow.  In Fig.\ref{F30}, ideal hydrodynamic model simulations for temporal evolution of spatial eccentricity and momentum anisotropy are shown. They follow our expectations. 

 The second harmonic coefficient or the elliptic flow ($v_2$) has been studied extensively in $\sqrt{s}_{NN}$=200 GeV Au+Au collisions at RHIC \cite{PHENIXwhitepaper,STARwhitepaper}. Recently, ALICE collaboration measured elliptic flow in $\sqrt{s}_{NN}$=2.76 TeV Pb+Pb collisions at LHC \cite{Aamodt:2010pa,:2011vk}. Large elliptic flow  has provided compelling evidence that at RHIC and LHC, nearly perfect fluid is produced. Deviation from the ideal fluid behavior is controlled by shear viscosity to entropy ratio ($\eta/s$). Effect of shear viscosity is to dampen the flow coefficients. Elliptic flow   has sensitive dependence on  $\eta/s$. In smooth hydrodynamics,
sensitivity of elliptic flow has been utilised to obtain phenomenological estimates of $\eta/s$ \cite{Luzum:2008cw,Song:2008hj,Chaudhuri:2009uk,Chaudhuri:2009hj,Roy:2011xt,Schenke:2011tv,Bozek:2011wa,Song:2011qa}.
It appears that   QGP viscosity over entropy ratio is close to $\eta/s\approx 1/4\pi$. 

\subsection{Some results of Hydrodynamic simulation of heavy ion collisions}

\begin{figure}[t]
\begin{minipage}{10pc}
\includegraphics[width=10pc]{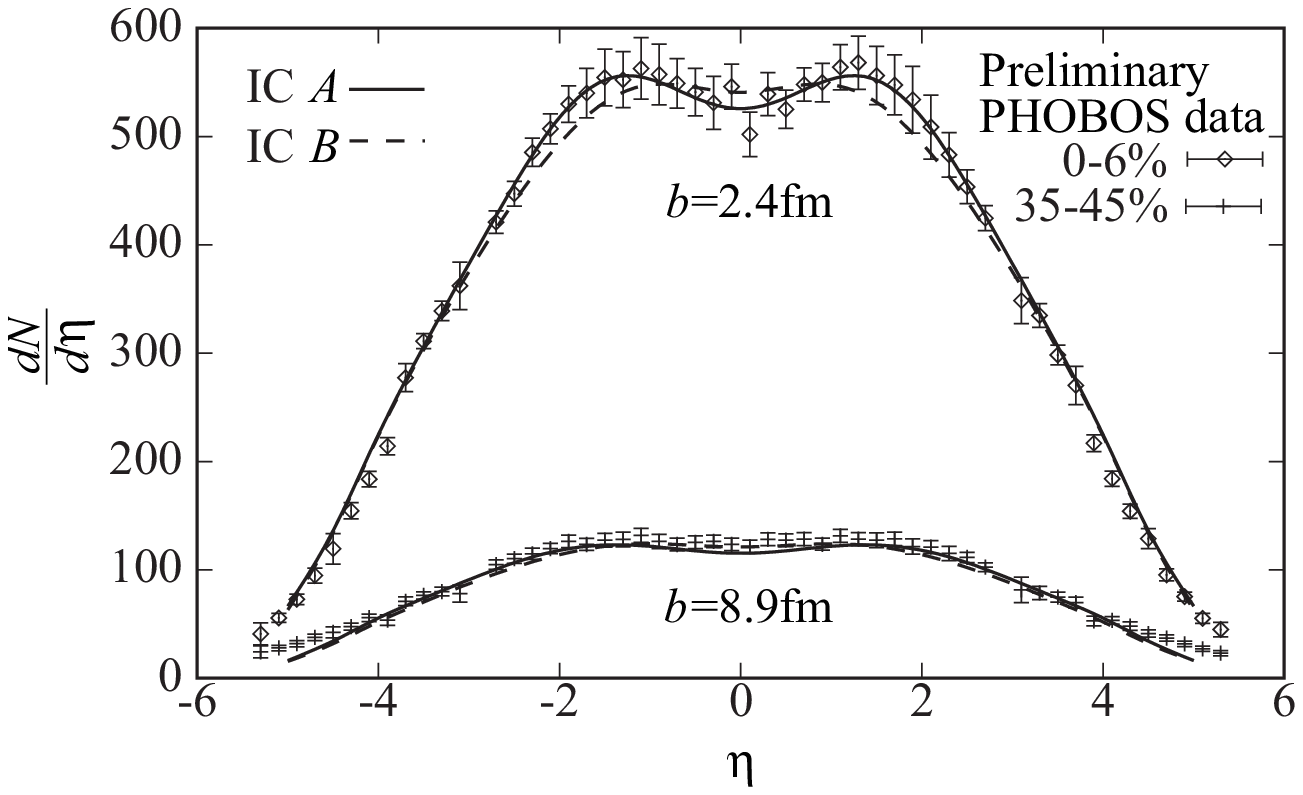}
\end{minipage}\hspace{0.5pc}%
\begin{minipage}{10pc}
\includegraphics[width=10pc]{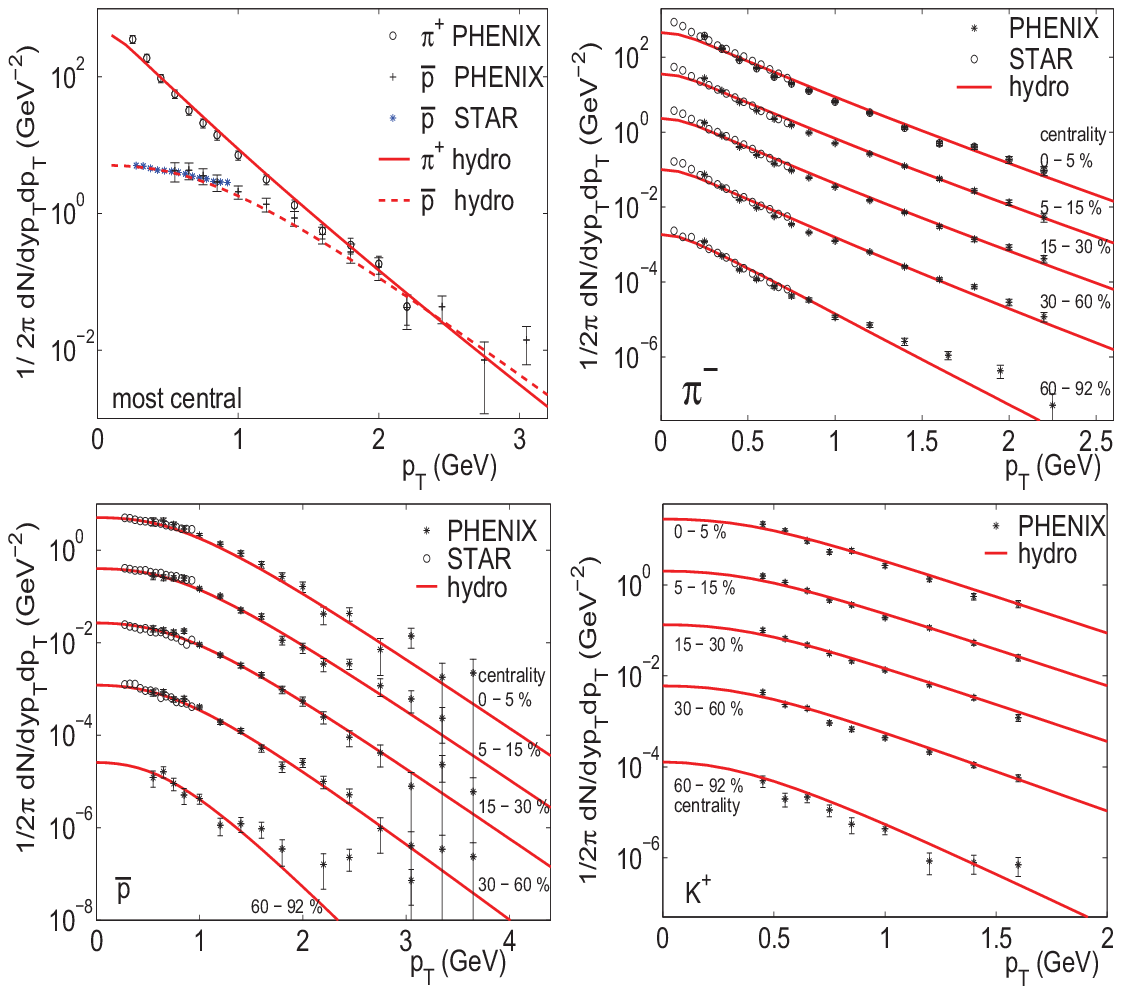}
\end{minipage}\hspace{0.5pc}%
\begin{minipage}{10pc}
\includegraphics[width=10pc]{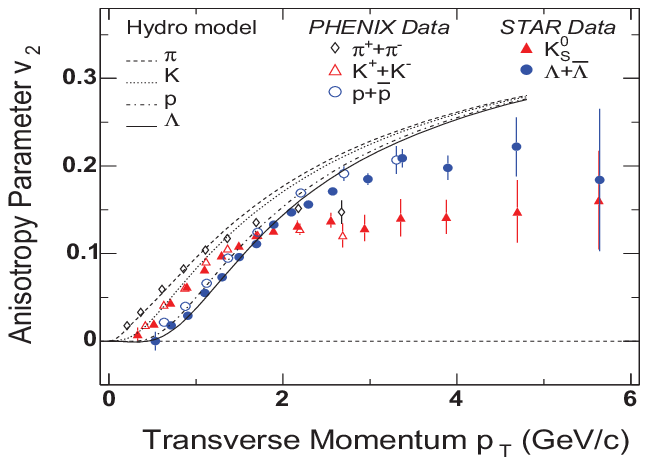}
\end{minipage}\hspace{0.5pc}%
 \caption{\label{F28a}(left panel)Charged particles pseudo  rapidity density in $\sqrt{s}_{NN}$=130 GeV Au+Au collisions, in a central 0-6\% and peripheral 35-45\% collision centralities are compared with hydrodynamic simulation
with two initial condition A and B (for details of the initial conditions see \cite{Hirano}). (middle panel)
transverse momentum spectra of identified particles in $\sqrt{s}_{NN}$=130 GeV Au+Au collisions in different centrality ranges of collisions (see \cite{Kolb:2003dz} for details) and (right panel) differential elliptic flow of identified particles is in $\sqrt{s}_{NN}$=200 GeV Au+Au collisions (see \cite{Heinz:2004ar} for details).} 
\end{figure}

Various authors have simulated heavy ion collisions at relativistic energy. In the following I will show some representative results. In the left panel of Fig.\ref{F28a}, PHOBOS measurements \cite{Back:2001bq} of charged  particles rapidity density in $\sqrt{s}_{NN}$=200 GeV Au+Au collisions, in two different collision centralities are shown. Experimental data are nicely reproduced in hydrodynamic simulations. For details see \cite{Hirano}.  

\begin{table}[h]
\caption{\label{table6a} Central energy density ($\varepsilon_i$) and 
temperature ($T_i$) at the initial time $\tau_i$=0.6 fm/c, for different 
values of viscosity to entropy ratio ($\eta/s$).   The bracketed values are estimated central energy density and temperature in $\sqrt{s}_{NN}$=200 GeV Au+Au collisions \cite{Chaudhuri:2009uk}. Initial time of the simulations is $\tau_i$=0.6 fm. 
} 
  \begin{tabular}{|c|c|c|c|c|}\hline \hline
$\eta/s$         & 0    & 0.08 & 0.12 & 0.16 \\  \hline
$\varepsilon_i$ & $89.2\pm 5.0$ & $78.0\pm 4.0$ & $70.5\pm 3.5$ &  $61.7\pm 3.0$ \\    
 ($\frac{GeV}{fm^3}$)  & ($35.5\pm 5.0$) & ($29.1\pm3.6$) & ($25.6\pm 4.0$) &  ($20.8\pm 2.7$)  \\ \hline
$T_i$   & $486\pm 6$ & $475\pm 5$ & $462\pm 6$ & $447\pm 5$     \\
(MeV)  & ($377\pm 14$) & ($359\pm 12$) & ($348\pm 14$) & ($331\pm 11$) \\  \hline
 \end{tabular}
\end{table}  
In the middle panel of  Fig.\ref{F28a}, fits obtained to the identified particles spectra
in STAR and PHENIX experiments in a hydrodynamic model simulation are shown. Note the quality of fit. Data in the collision centrality 0-5\% to 60-90\% are well explained. Finally, in the right panel of  Fig.\ref{F28a},  elliptic flow of identified particles, as a function of $p_T$ are shown. One observe that
experimental flow of identified particles are mass dependent, more flow for lighter particles than for heavier particles. It is called mass splitting of flow. At high $p_T$ however, effect of mass splitting is reduced. In Fig.\ref{F28a}, the solid lines are hydrodynamic model simulations. Experimental mass splitting of flow is correctly reproduced in hydrodynamic simulations. The agreement with experiment is also
good. The simulations results shown in Fig.\ref{F28a} are for ideal fluid only. As mentioned earlier, there are several simulations with viscous fluid. I will not show the results. I just mention that compared to ideal fluid, a viscous fluid will require less initial energy density or temperature. This is because, entropy is generated during viscous evolution. As an example, in table.\ref{table6a}, I have noted the central energy density and temperature of the fluid obtained from fits to experimental data in $\sqrt{s}_{NN}$=200 GeV Au+Au and $\sqrt{s}_{NN}$=2.76 TeV Pb+Pb collisions. Note that viscous fluid require less energy density or temperature. See \cite{Chaudhuri:2009uk},\cite{Roy:2011xt} for details of the simulations.

\subsection{Event by event hydrodynamics}

Very large number of particles in the final state, in RHIC and LHC energy collisions,
enables experimentalist to analyse the experimental data event-by-event. 
Importance of event-by-event analysis is best explained by the classic example by A. D. Jackson. If in a rainy day, you hold out a sheet of paper outside the window and forget about it for a long time, you find it uniformly soaked. You conclude that the spatial distribution of rain was uniform. However, if you continue to look into the paper, you find that spatial distribution is not uniform in a short time scale. Similarly, high statistics data in a single event may be very different from the data averaged over many events and reveal interesting physics.  More details about event-by-event can be found in \cite{Heiselberg:2000fk}. 

\begin{figure}[t]
 \center
  \includegraphics[scale=0.8]{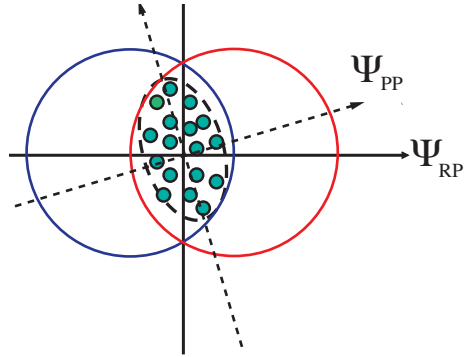}
\caption{Schematic picture of Monte Carlo Glauber simulation for participating nucleons in the transverse plane. The overlap region of participating nucleons are tilted with respect to the reaction plane.} 
 \label{F22a}
\end{figure}

One of the aims of hydrodynamic analysis is to find the initial conditions of the matter produced in heavy ion collisions. Now the initial conditions can fluctuate, event by event. In Fig.\ref{F22a}, a schematic picture for Glauber Monte-Carlo simulation 
for the participating nucleons in the transverse plane is shown. The overlap region of participating nucleons is tilted with respect to the reaction plane. It is an important realisation that azimuth of the particles should be measured with respect to the participating plane angle rather than the reaction plane angle.  

Now,  the positions of the participating nucleons will fluctuate event-be-event, so does the participating plane angle.   
The participating nucleons which determine the symmetry plane ($\psi_{PP}$), will fluctuate around the reaction plane \cite{Manly:2005zy}. As a result odd harmonics, which were exactly zero for smoothed initial distribution, can be developed. Recently, ALICE collaboration has observed odd harmonic flows    in $\sqrt{s}_{NN}$=2.76 TeV Pb+Pb collisions \cite{:2011vk}. In most central collisions, the elliptic flow ($v_2$) and triangular flow ($v_3$) are of similar magnitude. In peripheral collisions however, elliptic flow dominates. Triangular flow is being investigated in event-by-event hydrodynamics \cite{Mishra:2008dm},\cite{Mishra:2007tw},\cite{Takahashi:2009na},\cite{Alver:2010gr},\cite{Alver:2010dn},\cite{Teaney:2010vd},   \cite{Chaudhuri:2011qm},\cite{Chaudhuri:2011pa}. It has been conjectured that third harmonic $v_3$, which is response of the initial triangularity of the medium, is responsible for the observed structures in two particle correlation in Au+Au collisions.  The ridge structure in pp collisions also has a natural explanation if odd harmonic flow develops. It is also expected that triangular flow will be more sensitive to dissipative effects and can constrain the phenomenological estimate of shear viscosity to entropy ratio. However, detailed simulations indicate  that in event-by-event hydrodynamics, sensitivity of $v_2$ and $v_3$ toward viscosity is reduced due to fluctuations \cite{Chaudhuri:2011qm},\cite{Chaudhuri:2011pa}.

\section{Signals of Quark-Gluon-Plasma} 

We expect that QGP is produced in ultra-relativistic heavy ion collisions. However, how do we know it is produced indeed, or, which observable will signal formation of QGP. 
Signals of QGP are very important subject in the study of QGP. Unlike in other phase transitions, in confinement-deconfinement phase transition (or cross over), the constituents of the high temperature phase (QGP) do not exist freely. They are confined within the hadrons. 
The problem is closely related to quark confinement; quarks are unobservable. QGP, even if produced in a collision, is a transient state,  it expands, cools, hadronises, cools further till interactions between the hadrons become too weak to continue the evolution.  Any information about the QGP phase, if produced in high energy nuclear collisions,   has to be obtained from the observed hadrons only. 
Hadronisation is a non-perturbative process. Till now, it is not properly understood. 
Whether or not the hadronisation process erases any memory of the constituent quarks is uncertain. If the hadronisation process erases the memory, from the observed hadrons one can not comment on the initial QGP phase. Present search for QGP at RHIC/LHC is on the premise that the hadronisation process does not erase the memory and from the observed hadrons, one can comment on the possible existence of QGP. In the following, I briefly  discuss some important QGP signals.

\begin{table}[h] 
\caption{ \label{table6}$J/\psi$ mass, radii and formation time from
solution of Schr\"{o}dinger Equation.}
	\centering
		\begin{tabular}{|c|c|c|c|}
		\hline
 &  $J/\psi$ & $\psi^\prime$ & $\chi_c$  \\  \hline
 M(GeV)  & 3.07 & 3.698  & 3.5   \\ \hline
R(rm)  &  0.453& 0.875  & 0.696   \\ \hline
$\tau_F$(fm) & 0.89 & 1.5  & 2.0   \\ \hline		
 \end{tabular}
\end{table}

\subsection{$J/\psi$ suppression}

Charmoniums (bottoniums) are bound state of $c\bar{c}$ ($b\bar{b}$) quarks. Charm and bottom quarks are heavy ($m_{charm}\approx$ 1.15-1.35 GeV, $m_{bottom}\approx$ 4.0-4.0 GeV) and non-relativistic   Schr\"{o}dinger equation can be solved to obtain the bound state properties, with inter-quark potential (Cornell potential)

\begin{equation}\label{s1.1}
V(r)=\sigma r - \frac{\alpha_{eff}}{r},
\end{equation}

\noindent where $r$ is the inter quark separation, $\sigma\approx$ 0.192 $GeV^2$ is the string constant and $\alpha_{eff}\approx$0.471. Quarkonium mass, radius and formation time can be obtained from solution of non-relativistic Schrodinger equation and are given in the table.\ref{table6}.  Now   at   high temperature, 
interaction potential will be screened,

\begin{equation}\label{s1.2}
V(r,T)=\frac{\sigma}{\mu(T)}\left (1-e^{-\mu(T)r} \right ) - \frac{\alpha_{eff}}{r} e^{-\mu(T) r}
\end{equation}

\noindent $\mu(T)$ is the inverse of the screening radius (Debye radius) and is called the screening mass. For $\mu\rightarrow 0$, Eq.\ref{s1.1} is recovered. For $\mu\neq 0$, the screened potential satisfies,

\begin{equation}
Lim_{r\rightarrow 0} [rV(r,T)]  \sim -\alpha
\end{equation}

\begin{figure}[t]
 \center
 \resizebox{0.35\textwidth}{!}
 {\includegraphics{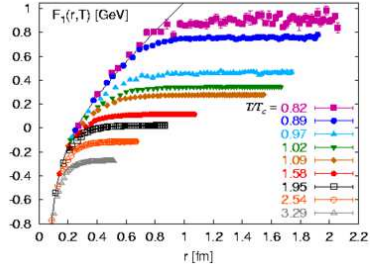}}
\caption{ \label{F31} quark potential as a function of temperature.} 
 
\end{figure}

\noindent the $1/r$ behavior in the short distance limit. For large $r$,

\begin{equation}
Lim_{r\rightarrow \infty} \frac{1}{r}ln\left [\frac{\sigma}{\mu(T)}-V(r,T)\right ]  \sim -\mu(T)
\end{equation}  

\noindent indicating that the range of the binding force decreases exponentially with screening mass.  In Fig.\ref{F31},
inter-quark potential as a function of temperature is shown. The screening mass $\mu$ is as increasing function of temperature $T$. From lattice simulations, $\frac{\mu}{T_c} \approx 4\frac{T}{T_c}$.

In 1986, Matsui and Satz \cite{Matsui:1986dk} suggested that if QGP is formed in nuclear collisions, $J/\psi$, the bound state of $c\bar{c}$ will be suppressed, w.r.t. pp collisions. The idea is simple. In presence of QGP, $J/\psi$ production will be inhibited due to screening of potential. A $c\bar{c}$ pair, which could transform into a $J/\psi$ is now unable to do so. Over  the  years,  several  groups have
measured the $J/\psi$ yield in heavy ion collisions (for a review
of the data prior to RHIC energy collisions, and the interpretations see Refs.  \cite{vo99,ge99}).
In  brief,  experimental  data do show suppression. However, suppression is observed in pA collisions also, where, one does not expect QGP formation. It is understood that in an inelastic collision with nucleons, $J/\psi$'s can be dissociated and lead to suppression. Suppression in pA collisions is termed cold nuclear matter (CNM) effect. It is important to disentangle CNM effect from the experimental data to obtain the suppression due to deconfinement. 

\begin{figure}[t]
 \center
 \resizebox{0.5\textwidth}{!}
 {\includegraphics{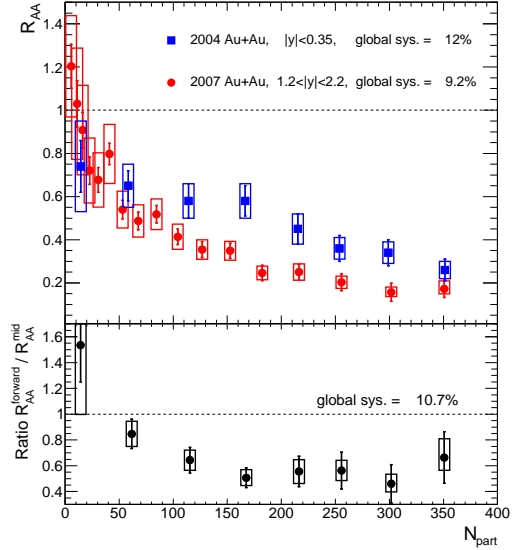}}
\caption{ \label{F30a}Nuclear modification factor $R_{AA}$ for $J/\psi$ in $\sqrt{s}_{NN}$=200 GeV Au+Au collisions.  } 
\end{figure}

PHENIX  collaboration  has made systematic measurements of $J/\psi$ production in nuclear collisions. 
They have measured $J/\psi$ yield in p+p collisions at RHIC to obtain the basic 'reference' invariant yield \cite{Adler:2003qs,Adler:2005ph}. Measurements of $J/\psi$ production  in d+Au collisions
\cite{Adler:2005ph} give reference for cold nuclear matter effects. 
Cold and hot nuclear matter effects are studied  in
Au+Au and Cu+Cu collisions, where yields are measured as a function of collision centrality \cite{Adare:2006ns,Adare:2008sh}. In Fig.\ref{F30a}, PHENIX measurements for nuclear modification factor,

\begin{equation}
R_{AA}=\frac{1}{N_{coll}} \frac {\sigma^{AA\rightarrow J/\psi X}} {\sigma^{pp\rightarrow J/\psi X} }
\end{equation}

\noindent in $\sqrt{s}_{NN}$=200 GeV Au+Au collisions are shown. Data shows suppression. Only CNM effect can not explain the data. If $J/\psi$'s are suppressed in QGP, data are explained.
 At RHIC energy, it has been
argued that rather than suppression, charmonium's will be enhanced
\cite{ Thews:2000rj,Braun-Munzinger:2000px}. 
Due to large initial energy, large number of $c\bar{c}$ pairs will be
produced in initial hard scatterings. Recombination of $c\bar{c}$
can occur enhancing the charmonium production. Apparently, PHENIX data on
$J/\psi$ production in Au+Au   are not consistent
with models which predict $J/\psi$ enhancement.
 
 \begin{figure}[t]
 \center
 \resizebox{0.5\textwidth}{!}
 {\includegraphics{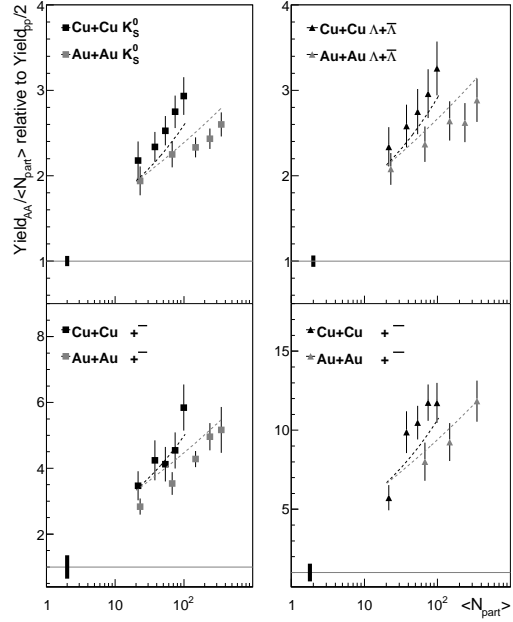}}
\caption{STAR measurements for enhancement factor in $\sqrt{s}_{NN}$=200 GeV Cu+Cu and Au+Au collisions, for various multistrange particles.}
  \label{F33}
\end{figure}

\subsection{Strangeness enhancement}

For long, strangeness enhancement is considered as a signature of QGP formation \cite {Koch:1986ud}. In QGP, strangeness will be more abundant that non-strange quarks. For example, in equilibrating plasma, strange quark density (with degeneracy $2(spin)\times 3(color)$, can be obtained as (see Eq.\ref{eq4.13}),

\begin{equation}
n_s=n_{\bar{s}}=3\times 2 \int  \frac{d^3p}{(2\pi)^3} e^{-\sqrt{p^2+m_s^2}/T}=\frac{3Tm_s^2}{\pi^2}K_2(m_s/T)
\end{equation}

density of non-strange anti-quarks ${\bar q}$ (${\bar q}$ stands for ${\bar u}$ or ${\bar d}$,  in the limit of small mass, is

\begin{equation}
n_{\bar{q}}=3\times 2 \int  \frac{d^3p}{(2\pi)^3} e^{-|p|/T} e^{-\mu_q/T}=e^{-\mu_q/T} \frac{6}{\pi^2}T^3 
\end{equation}

\noindent where $\mu_q=\frac{1}{3}\mu_B$, is the quark chemical potential. The ratio of strange quarks to non-strange quarks,

\begin{equation}
\frac{n_s}{n_q}=\frac{1}{2} \left (\frac{m_s}{T} \right )^2 K_2 \left (\frac{m_s}{T} \right ) e^{\mu_B/3T}
\end{equation}

For $\mu_B>0$, the ratio is greater than one. Strange quarks will produce in more abundance than non-strange quarks. During hadronisation, numerous strange quarks can be bound with available non-strange quarks and strange hadron production will enhance.

One can define a strangeness enhancement factor ($E$) as,

\begin{equation}
E=\frac{\frac{1}{N_{part}} \frac{dN^{AA}}{dy}}{\frac{1}{2} \frac{dN^{pp}}{dy}}
\end{equation}

In Fig.\ref{F33}, experimental data \cite{strange1},\cite{strange2} , in $\sqrt{s}_{NN}$=200 GeV Au+Au collisions for strangeness enhancement are shown. Data do show enhancement.
$\phi$ meson (which is a hidden strange meson) production is also enhanced \cite{Abelev:2007rw}.  
However, it is uncertain whether or not the enhancement is due to increased production in QGP or due to canonical suppression  \cite{Hamieh:2000tk}\cite{Tounsi:2001ck}\cite{Tounsi:2002nd} of strangeness in pp collisions.
The canonical suppression arises from the need to conserve strangeness within a small, local volume, which limit the strangeness production in pp collisions relative to AA collisions. In the language of statistical mechanics, while canonical ensemble is applicable in pp collisions, in AA collisions, grand canonical ensemble is applicable. Assuming the systems correlation volume is proportional to $N_{part}$, canonical framework predict that yield per $N_{part}$ increase with $N_{part}$ as phase space restriction due to strangeness conservation is lifted.

\subsection{Electromagnetic probes}

Photons and dileptons are considered to be important probes for QGP diagnostic. They are called electromagnetic probe as they interact only electromagnetically.  Unlike the hadrons, which are emitted only from the freeze-out surface, photons and dileptons have large mean free path and are emitted from the entire volume.
Total production is then obtained by convoluting their production rate over the 4-volume.

\subsubsection{Photons}

One of the problems with photon as a probe is the large background. For diagnostic purpose, one is interested only in 'direct photons', i.e. photons produced as a result of some collision process. However, QGP is a transient state and it ultimately transforms into hadrons. A large number of hadrons decay into photon. Decay photons constitute the back ground and needed to be eliminated to get the direct photon signal. Experimentally several methods have been devised to eliminate the decay photons, e.g. invariant mass analysis, mixed event analysis etc. However, due to overwhelmingly large number of background photons, the elimination could not be exact and experimental data on 'direct' photon production could not be obtained very accurately. There is definitely scope for further improvement.

In a nucleus-nucleus collision, there are various sources of direct photons. They are briefly discussed below.
 
(i) prompt photons: initial hard collisions produces prompt photons. Prompt photon production in a nucleon-nucleon reaction $a+b\rightarrow \gamma+X$ can be calculated in leading order pQCD,

\begin{eqnarray}
E\frac{d\sigma}{d^3p^\gamma}&=&K\sum_{ij=q,\bar{q},g} \int dx_i dx_j f_{i/a}(x_i,Q^2)f_{j/b}(x_j,Q^2) \nonumber \\
&& \times \delta(\hat{s}+\hat{t}+\hat{u}) \frac{\hat{s}}{\pi}\frac{d\hat{\sigma}}{d\hat{t}}(ij\rightarrow \gamma+X)
\end{eqnarray}

\noindent where $f_{i/a,b}(x,Q^2)$ is the parton distribution function, $\frac{d\hat{\sigma}}{d\hat{t}}$ is the elementary partonic cross section.  
Above equation scaled by the binary collision number gives the prompt photon production in nucleus-nucleus collision.

(ii)Fragmentation photons: initial hard scattered partons may fragment into photon
($q\rightarrow q+\gamma$).

\begin{eqnarray} \label{eq9.10a}
E\frac{d\sigma}{d^3p^\gamma}&=&K\sum_{ijk=q,\bar{q},g} \int dx_i dx_j \frac{dz}{z^2} f_{i/a}(x_i,Q^2)f_{j/b}(x_j,Q^2) D_{\gamma/k}(z,Q_F^2) \nonumber \\
&& \times \delta(\hat{s}+\hat{t}+\hat{u}) \frac{\hat{s}}{\pi}\frac{d\hat{\sigma}}{d\hat{t}}(ij\rightarrow k+l)
\end{eqnarray}

In Eq.\ref{eq9.10a}, $D_{\gamma/k}$ is the fragmentation function. As before, above equation should be scaled by the binary collision number to obtain the fragmentation photons in nucleus-nucleus collision.

(iii)Pre-equilibrium photons: in nucleus-nucleus collisions, an extended QCD medium
is produced. Before achieving local thermal equilibrium the medium is in pre-equilibrium stage. Photons will be emitted in the pre-equilibrium stage also. However, it is difficult to distinguish pre-equilibrium photons from thermal photons. Simulations with Parton Cascade Model \cite{srivastava_photon1} indicate that pre-equilibrium photon production equals the thermal photon at $p_T$=2 GeV. Low $p_T$ photons are predominantly thermal. 

(iv) Thermal photons: photons, emitted from the (locally) equilibrated QGP and hadronic matter are called thermal photons. For QGP diagnostic purpose, these photons are most important. In the following, I will discuss briefly about thermal photons. For more detailed information, see \cite{srivastava_photon2}\cite{Alam:1996fd}\cite{Alam:1999sc}.

In the QGP phase, the important reactions for direct photons are; (i) annihilation:
$q+\bar{q}\rightarrow g+\gamma$, (ii) Compton process: $q+g\rightarrow q+\gamma$ and (iii)  bremsstrahlung: $q+q\rightarrow qq\gamma$. 

In kinetic theory, photon production rate (per unit time per unit volume) from $1+2\rightarrow 3+\gamma$ process can be written as,

\begin{eqnarray}
\mathcal{R}_i&=&\mathcal{N}\int \frac{d^3p_1}{2E_1(2\pi)^3}\frac{d^3p_2}{2E_2(2\pi)^3} f_1(E_1)f_2(E_2) (2\pi)^4
\delta^4(p_1^\mu+p_2^\mu-p_3^\mu-p^\mu) \nonumber \\
&&\times |\mathcal{M}_i|^2 \frac{d^3p_3}{2E_3(2\pi)^3}\frac{d^3p}{2E(2\pi)^3}[1\pm f_3(E_3)]
\end{eqnarray}

\noindent where $\mathcal{M}_i$ is the amplitude for one of the basic process, $f(E)$'s are the Fermi-Dirac or Bose distribution, as appropriate. The $\pm$ in the last term corresponds to Bose enhancement or Pauli blocking. Using the Mandelstam's variable, $s,t,u$, differential rate can be written,

\begin{eqnarray}
E\frac{d\mathcal{R}_i}{d^3p}&=&\frac{\mathcal{N}}{(2\pi)^7}\frac{1}{16E}\int ds dt |\mathcal{M}_i(s,t)|^2 \int dE_1dE_2 f_1(E_1)f_2(E_2) \nonumber \\ 
&&\times [1\pm f_3(E_1+E_2-E)] \theta(E_1+E_2-E) \sqrt{aE_1^2+bE_1+c} \nonumber \\ 
\end{eqnarray}

\noindent where,

\begin{eqnarray}
a&=&-(s+t)^2 \nonumber \\
b&=&2(s+t)(Es-E_2t) \nonumber \\
c&=&st(s+t)-(Es+E_2t)^2
\end{eqnarray}

For massless particles, amplitude $\mathcal{M}$  is related to the differential cross section as,

\begin{equation}
\frac{d\sigma}{dt}=\frac{|\mathcal{M}|^2}{16\pi s^2}
\end{equation}

For Compton and annihilation processes, the differential cross sections are,

\begin{eqnarray}
\frac{d\sigma^{annhilation}}{dt}&=&\frac{8\pi\alpha\alpha_s}{9s^2} \frac{u^2+t^2}{ut}\\
\frac{d\sigma^{Compton}}{dt}&=&\frac{-\pi\alpha\alpha_s}{3s^2} \frac{u^2+s^2}{us}
\end{eqnarray}

$\mathcal{N}=20$ for annihilation process when summing over u and d quarks and $\mathcal{N}=\frac{320}{3}$ for the Compton process. 

Photon production rate from Compton and annihilation processes were first computed in
\cite{Kapusta_photon1}\cite{Kapusta_photon2}\cite{Baier_photon}.  Importance of Bremsstrahlung process was first considered in \cite{aurenche_photon1}\cite{aurenche_photon2}\cite{aurenche_photon3}, however,  Landau-Pomeranchuk-Migdal (LPM) effect (when photon emission is suppressed due to multiple collisions) was neglected. Arnold, Moore and Yaffe \cite{arnold_photon} made a complete calculation in leading order. 
They also provided a simple parameterised form for easy use in hydrodynamics. Below, I list the results.

Leading order photon emission rate from QGP:

\begin{equation}
E^\gamma\frac{dR}{d^3k}=A(k)[\ln(T/m_\infty)+C_{tot}(k/T)]
\end{equation}

\noindent with,
\begin{equation}
C_{tot}=\frac{1}{2}\ln(2k/T)+C_{2\rightarrow 2}(k/T)+C_{brem}(k/T)+C_{annih}(k/T)
\end{equation}

The leading log coefficient $A(k)$ is given by,

\begin{equation}
A(k)=2\alpha  [d_F \sum q_i^2]\frac{m_\infty}{k} n_f(k)
\end{equation}

where $n_f (k) = [exp(k/T)+1]^{-1}$ is the Fermi distribution function, and $d_F$ is the dimension of
the quark representation ($d_F$ = $N_c$ = 3 for QCD). $q_i$=2/3 for up quark and -1/3 for down type quarks.  $m_\infty=g_s^2T^2/3$ is thermal quark mass in  the leading-order, $g_s$ being the strong coupling constant,
($\alpha_s=g_s^2/4\pi$). For two flavor QCD,

\begin{equation}
A(k)=\frac{40\pi T^2}{9} \alpha \alpha_s \frac{n_f(k)}{k}
\end{equation}

$C_{2\rightarrow 2}(k/T)$, $C_{brem}(k/T)$ and $C_{annih}(k/T)$ all involve
multidimensional integrations, which can only be solved numerically. 
Numerical results for QCD plasmas are reproduced quite accurately by the
approximate, phenomenological fits \cite{arnold_photon},

\begin{eqnarray}
C_{2\rightarrow 2}(x)&=& 0.041 x^{-1} - 0.3615 + 1.01 e^{-1.35 x},\\
C_{brem}(x) + C_{annih}(x)&\approx&\sqrt{1+\frac{N_f}{6}}
\frac{0.548 log(12.28 + 1/x)}{x^{3/2}} \nonumber\\
&+&\frac{0.133 x}{\sqrt{1 + x/16.27}}
\end{eqnarray}

\begin{figure}[t]
 \center
 \resizebox{0.5\textwidth}{!}
 {\includegraphics{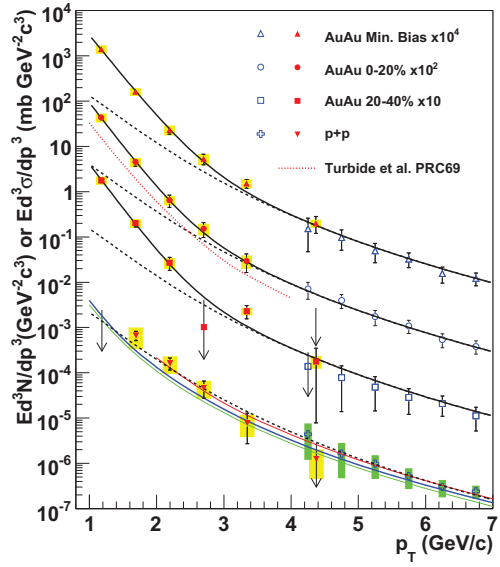}}
\caption{Invariant cross section (p + p) and invariant
yield (Au + Au) of direct photons as a function of $p_T$.
See the text for details.}
  \label{F33a}
\end{figure}

In the hadronic phase, photons are emitted in several reactions. Dominant channels are,
(i)$\pi+\pi\rightarrow \rho+\gamma$, (ii) $\pi+\rho\rightarrow \pi+\gamma$. 
 J. Kapusta, P. Lichard, and D. Seibert \cite{Kapusta_photon1}\cite{Kapusta_photon2} evaluated photon production rate from these channels.   Role of $A_1$ resonance  in photon production was investigated in \cite{Xion_photon}. 
Later, in a more comprehensive analysis \cite{Turbide_photon}, hadronic photon production rate in a meson gas  
consisting of light pseudo-scalar, vector and axial vector mesons
($\pi, K, \rho, K^*, A_1$) was obtained. A parameterised form was also provided. Below, I reproduce the 
parameterised reaction rates.

Photon emission rate from Hadronic phase:

\begin{eqnarray*}
&& E\frac{dR_{\pi^+\rho\rightarrow \pi^+\gamma}}{d^3k}=F^4(E) T^{2.8}\times \nonumber \\
&&exp\left [\frac{-(1.461T^{2.3094}+0.727)}{(2TE)^{0.86}}
+(0.566T^{1.4094}-0.9957)\frac{E}{T} \right ] \\
&& E\frac{dR_{\pi^+\pi\rightarrow \rho^+\gamma}}{d^3k}=F^4(E) \frac{1}{T^5}\times \nonumber \\
&&exp\left [ -(9.314T^{-0.584}-5.328)(2TE)^{0.088}+(0.3189 T^{0.721}-0.8998)\frac{E}{T} \right ] \\
&& E\frac{dR_{\rho\rightarrow \pi\pi\gamma}}{d^3k}=F^4(E) \frac{1}{T^2}\times  
 exp\left [-\frac{(-35.459T^{1.26}+18.827)}{(2TE)^{(-1.44T^{0.142}+0.9996)}}-1.21\frac{E}{T} \right ] \\
&& E\frac{dR_{\pi K^*\rightarrow K\gamma}}{d^3k}=F^4(E) T^{3.75}\times  
exp\left [-\frac{0.35}{(2TE)^{1.05} }+(2.3894T^{0.03435}-3.222)\frac{E}{T} \right ] \\
&& E\frac{dR_{\pi K\rightarrow K^*\gamma}}{d^3k}=F^4(E) \frac{1}{T^3}\times    
 exp\left [-(5.4018T^{-0.6864}-1.51)(2TE)^{0.07}-0.91 \frac{E}{T} \right ] \\
 && E\frac{dR_{\rho K\rightarrow K \gamma}}{d^3k}=F^4(E) T^{3.5}\times \nonumber \\
&&exp\left [-\frac{(0.9386T^{1.551}+0.634) } {(2TE)^{1.01}}
+(0.568T^{0.5397}-1.164)\frac{E}{T} \right ] \\
 && E\frac{dR_{K^*K\rightarrow \pi\gamma}}{d^3k}=F^4(E) T^{3.7}\times  
exp\left [ \frac{(-6.096T^{1.889}+1.0299) } {(2TE)^{(-1613T^{12.162}+0.975) } }
-0.96\frac{E}{T}\right ] 
\end{eqnarray*}

In the above equations, the $E$ and $T$ are in GeV and the rates are in unit of $GeV^{-2}fm^{-2}$.
The dipole form factor $F(E)$ is,

\begin{equation}
F(E)=\left ( \frac{2\Lambda^2}{2\Lambda^2-E}  \right )^2, \Lambda=1 GeV
\end{equation}
 
 \begin{figure}[t]
 \center
 \resizebox{0.5\textwidth}{!}
 {\includegraphics{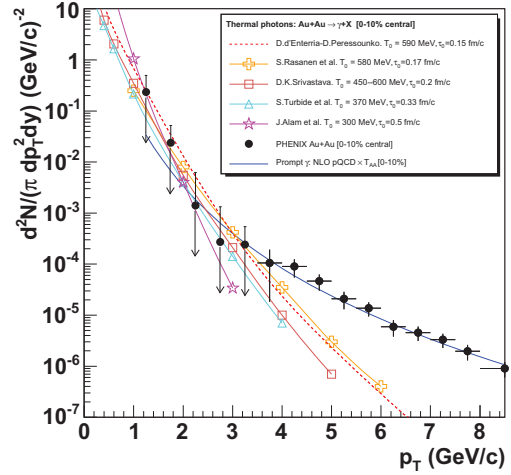}}
\caption{Thermal photon production in 0-10\% Au+Au collisions at
$\sqrt{s}_{NN}$= 200 GeV in different hydrodynamical models   \cite{photon_sr}\cite{photon_ja}\cite{photon_hu}\cite{photon_tu}\cite{d'Enterria:2005vz} 
are compared with experimental data.}
  \label{F34}
\end{figure}

Photon production rate from the QGP phase and the hadronic phase has to be convoluted over the 4-volume
 to obtain total photon production during the evolution of the fireball,
 
\begin{equation}
E\frac{dN^\gamma}{d^3k}=\int d^4x \left ( E\frac{dR}{d^3k} \right )=\int \tau d\tau dx dy d\eta  \left ( E\frac{dR}{d^3k} \right )
\end{equation}

There are several simulations for direct photon production in relativistic heavy ion collisions. For demonstration purpose, I will show a few results.
In Fig.\ref{F33a} transverse momentum dependence of the invariant cross section 
in $\sqrt{s}_{NN}$=200 GeV p+p collisions and  invariant yield in $\sqrt{s}_{NN}$=200 GeV  Au+Au collisions are shown. p+p data are from \cite{Adler:2006yt}. The Au+Au data are from \cite{Adare:2008ab}\cite{Adler:2005ig}. The
three curves on the p+p data represent NLO pQCD calculations. For $p_T >$ 2 GeV, the pQCD calculation is consistent with the p+p data
within the theoretical uncertainties. The dashed curves in Au+Au data are obtained by scaling the photon yield in p+p collisions by the nuclear overlap function $T_{AA}$. At low $p_T<$2.5 GeV, experimental Au + Au data are underpredicted. At low $p_T$, photon production increases faster  
than the binary NN collision scaled p + p cross section. The solid lines in Fig.\ref{F33} are fits with an exponential plus binary collision scaled p+p data. 
 
In Fig.\ref{F34} experimental data on direct photon production in 0-10\% Au+Au collisions are compared with different hydrodynamical model calculation. All the hydrodynamical simulations are comparable to the experimental data and with
each others within a factor of 2. The results confirm the dominance of thermal
radiation in the direct photon spectrum in low and intermediate $p_T$ range.

\begin{figure}[t]
\center
 \resizebox{0.5\textwidth}{!}
 {\includegraphics{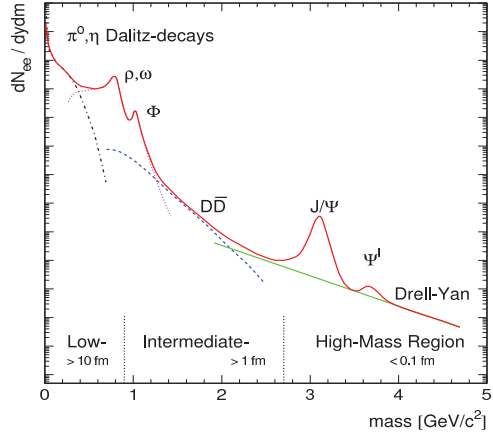}}
\caption{A schematic presentation of invariant mass dependence of dilepton production
in high energy nuclear collisions.} 
  \label{F35}
\end{figure}

\subsubsection{Dileptons}
 
Similar to the photons, dileptons are also emitted at every stage
of heavy ion collisions. In the QGP phase, a $q\bar{q}$ can interact 
to form a virtual photon, which subsequently decays in to a 
lepton pair or   dileptons, $q+\bar{q} \rightarrow \gamma^* \rightarrow l^+ + l^-$ ($l$=$e$ or $\mu$).
In the hadronic phase, dileptons are produced from interactions of charged hadrons
with their anti-particles e.g. $\pi^+ + \pi^- \rightarrow \rho \rightarrow l^+ + l^-$,
  from hadronic decays, e.g. $\pi^0\rightarrow l^++l^-+\gamma$, $\rho \rightarrow l^+ +l^-$, $\omega \rightarrow l^+ +l^-$,
$\phi \rightarrow l^++l^-$ etc.
Dileptons   are also produced in the Drell-Yan process (in the Drell-Yan process, a valence
quark from the projectile hadron interacts with a sea anti-quark
from target hadron to form a virtual photon, which then decays
into a lepton pair). Depending on the dilepton invariant mass ($M$), dilepton emission can be classified into three distinct
regimes. In Fig.\ref{F35}, invariant mass (M) dependence of dilepton production
in high energy nuclear collisions is shown schematically. One can distinguish three distinct regions,
(i) $M\leq M_\phi$(= 1.024 MeV) where dominating source of dilepton production is vector meson decays,
(ii) $M_\phi < M \leq  M_{J/\psi}$	(= 3.1 GeV) where dominant source is the thermal production
from QGP   and
(iii) $M \geq M_{J/\psi}$ dominated by primordial emission, decay of charmoniums etc. 

In the following, I briefly discuss dilepton emission rate in QGP and in hadronic resonance gas. As for photons,
emission rates are to be convoluted over the space-time volume to obtain production yield in nucleus-nucleus
collisions. For more details see \cite{srivastava_photon2}.

Dileptons production rate from a thermal system at temperature $T$, in a reaction, $a^++a^-\rightarrow l^++l^-$,
where $a$ is either a quark or pion, was obtained in \cite{dilepton_ka1}\cite{dilepton_ka2}. From QED, one calculate the cross section for $e^++e^-\rightarrow \mu^+ +\mu^-$,

\begin{equation}
\bar{\sigma}(M)=\frac{4\pi}{3}\frac{\alpha^2}{M^2}\left [ 1+\frac{2m_l^2}{M^2}\right ] 
\left [ 1-\frac{4m_l^2}{M^2}\right ]^{1/2}
\end{equation}

\noindent where $M$ is the invariant mass of $\mu^+\mu^-$ pair, $\alpha$ is the fine structure constant ($\alpha=1/137$) and $m_l$ is the mass of $\mu$. For $q\bar{q}$ annihilation, the above is multiplied by the color factor $N_c$=3, and factor reflecting the fractional charges of quarks. The modified cross section is,

\begin{eqnarray}
\sigma(M)&=&F_q \bar{\sigma}(M)\\ 
F_q&=&N_c(2s+1)^2 \sum_f e_f^2,
\end{eqnarray}

\noindent where $s$ is the spin of quarks, $e_f$ is the fractional charge and the sum is over the quarks flavors.
In the hadronic phase, in the vector meson dominance mode, dilepton production goes via the reaction,
$\pi^+ +\pi^- \rightarrow \rho \rightarrow l^+ +l^-$. The QED cross section is then multiplied by the Breit-Wigner form factor,

\begin{equation}
F_\pi(M)=\frac{m_\rho^4}{(m_\rho^2-M^2)^2+m_\rho^2 \Gamma_\rho^2}, M_\rho\sim 770 MeV, \Gamma_\rho\sim 150 MeV
\end{equation}

Dilepton cross section from $\pi^+\pi^-$ annihilation then become,

\begin{equation}
\sigma_\pi(M)=F_\pi(M) \bar{\sigma}(M)  \left [ 1-\frac{4m_\pi^2}{M^2}\right ]^{1/2}
\end{equation}

Kinetic theory gives the reaction rate (number of reaction per unit time per unit volume)


\begin{eqnarray}
R(a^+a^-\rightarrow l^+l^-)&=&\int \frac{d^3p_1}{(2\pi)^3}\frac{d^3p_2}{(2\pi)^3} f({\bf p}_1) f({\bf p}_2)
\times \sigma(a^+a^-\rightarrow l^+l^-) v_{rel} \label{eq10.27} \nonumber\\
\end{eqnarray}

\noindent where, $f({\bf p})$ is the   occupation probability at momentum $\bf{p}$ and energy $E =\sqrt{M^2+m_a^2}$. Relative velocity $v_{rel}$ can be computed as,

\begin{eqnarray}
v_{rel}&=&\frac{[(p_1.p_2)^2-m_a^4]^{1/2}}{E_1E_2} 
\end{eqnarray}

 Approximating  $f({\bf p})=exp(-E/T)$, and integrating over five of the six variables,
\begin{equation}
R(a^+a^-\rightarrow l^+l^-)=\frac{T^6}{(2\pi)^4} \int_{z_0}^\infty \sigma(z) z^2 (z^2-4z_a^2) K_1(z) dz
\end{equation}

with $z=M/T$, $z_a=m_a/T$ and $K_1$ is the modified Bessel function of the first kind.

Apart from the total number of lepton pairs emitted per unit space-time volume, $R=\frac{dN}{d^4x}$, one is interested in several differential rates. They can be obtained from Eq.\ref{eq10.27} by appropriate change of variables. The rate for producing lepton pairs with invariant mass $M$ is,

\begin{equation}
\frac{dN}{d^4x dM^2}= \frac{\sigma(M)}{2(2\pi)^4} M^3 T K_1(M/T) \left [ 1- \frac{4m_a^2}{M^2}\right ]
 \end{equation}

Production rate of leptons pairs with invariant mass M, momentum p and energy E ($E=\sqrt{p^2+M^2}$)
  can be written as,

\begin{equation}
E\frac{dN}{d^4x dM^2 d^3p}= \frac{\sigma(M)}{4(2\pi)^5} M^2 exp(-E/T) \left [ 1- \frac{4m_a^2}{M^2}\right ]
 \end{equation}

 \begin{figure}[t]
 \center
 \resizebox{1.0\textwidth}{!}
 {\includegraphics{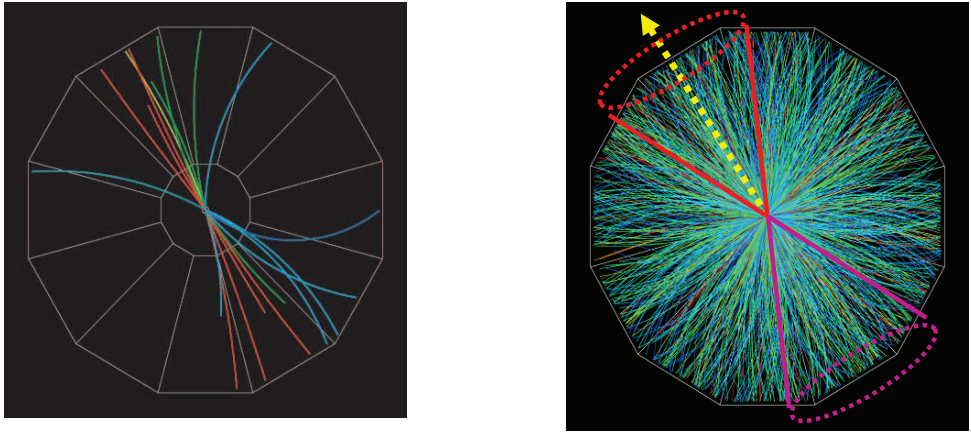}}
\caption{(left panel) An experimental reconstruction of an event in pp collision. Two jet structure is clearly seen, (right panel) same for an event in Au+Au collisions. Notice that due to large number of particles in the final state jet structure is obscured.} 
  \label{F35a}
\end{figure}

Presently, I will not discuss any hydrodynamic simulation for dilepton production. Dileptons have been measured in SPS energy ($\sqrt{s}_{NN}$=17.4 GeV). Hydrodynamical simulations underpredict low invariant mass dilepton yield. In the low invariant mass region, dilepton emission is largely mediated
by $\rho$ meson. Now,   properties of a hadron can change in a medium.
Due to medium effect, $\rho$ meson mass may drop, or its width increase.
These two effects are considered as a likely cause of underestimation of
low mass dilepton yield in hydrodynamic simulations.  Currently, experimental data do not distingush between these two effects. For more detailed account, please see \cite{srivastava_photon2}. I may mention here that dileptons are possibly better suited to probe QGP than direct photons. In contrast to photons which are characterised by the transvesre momentum, dileptons are characterised by two parameters, the transverse momentum and invariant mass. For differential diagnosis,  an increased degree of freedom may be useful.

\subsection{Jet quenching}

Jets are clusters of several hadrons, all moving in approximately the same direction. In hadron-hadron or in $e^+e^-$ collisions one generally observe two jet structure with back to back momenta. In the left panel of Fig.\ref{F35a}, an experimental reconstruction of final state particle trajectories is shown. The two jet structure is clearly evident in p+p collisions. In nucleus-nucleus collisions the jet structure is not obvious due to large number of particles (see the right panel of Fig.\ref{F35a}). However, jets are there and one can use some 'jet algorithm' to find them. 

Jet structure in hadron-hadron collisions can be understood qualitatively in perturbative QCD. 
The jet properties depend in general on two scales, the energy of the jet and its
virtuality, specified by the largest possible transverse momentum of one of its subjets.
The basic two body reaction $1+2\rightarrow 3+4$ is in the partonic (quark or gluon) level. The scattered partons are highly virtual (squared 4-momentum transfer is large) and reduce their virtuality by radiating gluons or by splitting into quark-antiquark pairs. Such a parton branching is governed by Dokshitzer-Gribov-Lipatov-Altarelli-Parisi (DGLAP) equation. Finally the partons fragment into hadrons. The characteristic collimated hadrons from fragmentation of an outgoing parton are called jet. Naturally, the most common structure seen is the two jet event. Three jet events are also seen and results from reactions such as $q\bar{q}\rightarrow q\bar{q}g$. 

From theoretical consideration Xin-Nian Wang and Miklos Gyulassy predicted the Jet quenching phenomena \cite{Wang:1991xy}. They argued that a partonic jet, if travel through a medium,  will lose its energy by gluon emission. The energy degraded parton will ultimately fragment into less number of particles than it would have in absence of a medium. It can be demonstrated as follows:  
In leading order perturbative QCD, production cross-section for the hadron $C$ in $A+B\rightarrow C+X$ reaction can be written as,

\begin{eqnarray}
E\frac{d\sigma}{d^3p_C}&=&K\sum_{ab\rightarrow cd} \int dx_a \int dx_b f_{a/A}(x_a,Q^2)f_{b/B}(x_b,Q^2) \nonumber \\
&& \times \delta(s+t+u) \frac{1}{\pi z_c}\frac{d\sigma}{dt}(ab\rightarrow cd) D_{C/c}(z_c,\mu^2)\label{eq9.9}
\end{eqnarray}

\begin{figure}[t]
 \center
 \resizebox{0.5\textwidth}{!}
 {\includegraphics{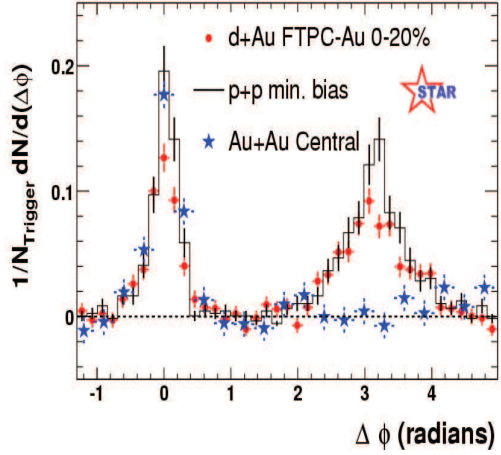}}
\caption{Two particle correlation in the azimuthal plane, in $\sqrt{s}_{NN}$=200 GeV, p+p, d+Au and Au+Au collisions is shown. Note the two peak structure in P=P and d+Au collisions. In Au+Au collisions however, the away side peak is vanished.}
   \label{F36}
\end{figure}

In Eq.\ref{eq9.9}, $f_{a/A}(x,Q^2)$ is the structure function of the parton $a$ in the hadron $A$, $f_{b/B}(x,Q^2)$ is the structure function of the parton $b$ in the hadron $B$. The $Q^2$ is the factorisation scale. $\frac{d\sigma}{dt}$ is the partonic cross section for the reaction $a+b\rightarrow c+d$.   $D_{C/c}(z_c,\mu^2)$ is the fragmentation function for the parton $c$ into hadron $C$, $\mu^2$ being the fragmentation scale. $z_c=\frac{E_C}{E_c}$ is the fraction of the partonic energy carried by the hadron $C$. $K$ in Eq.\ref{eq9.9} takes into account higher order effects. In nucleus-nucleus collisions, if colored medium is formed, the parton $c$ can lose energy in the medium.
If the parton $c$ travels through a medium and lose energy $\Delta E_c$, then $z_c=\frac{E_C}{E_c}\rightarrow z_c(1-\frac{\Delta E_c}{E_c})^{-1}$. The fragmentation function $D_{C/c}(z)$ is a rapidly falling function of $z$ and increase in $z_c$ will lead to reduced production for the hadron  $C$.

After the prediction of jet quenching phenomena, it was discovered in Relativistic Heavy Ion Collider (RHIC) \cite{star1}\cite{star2}.
 In Fig.\ref{F36}, di-hadron correlation in the azimuthal plane, in p+p, d+Au and Au+Au collisions are shown. The data are obtained in the following manner. A high $p_T$ trigger particle is fixed and in coincidence with the trigger particle, associated particles are measured as a function of the azimuthal angle. $\Delta \phi=\phi_{associate}-\phi_{trigger}$ is the difference of azimuthal angle between the trigger and associated particles. In p+p and d+Au collisions, di-hadron correlations shows a double peak structure, which can be understood in terms of two jet events. The peak at $\Delta \phi$=0 is called the near side peak (nearer to the trigger particle) and the peak at $\Delta \phi=\pi$ is called the away side peak (away from the trigger particle). 
In Au+Au collisions however, the away side peak is strongly suppressed. Strong suppression  of the away side peak is the experimental evidence of jet quenching. The understanding is as follows: a di-jet is produced near the surface the medium. One of the jet escapes into the vacuum and fragments. The other enters the medium and loses its energy in the medium before fragmentation. 

Jet quenching lead to high $p_T$ suppression i.e. production of high $p_T$ particles in A+A collision is less than that would have expected in a p+p collision, scaled by the collision number. High-$p_T$
suppression is usually expressed in terms of the nuclear modification factor
($R_{AA}$), 

\begin{figure}[t]
 \center
 \resizebox{0.5\textwidth}{!}
 {\includegraphics{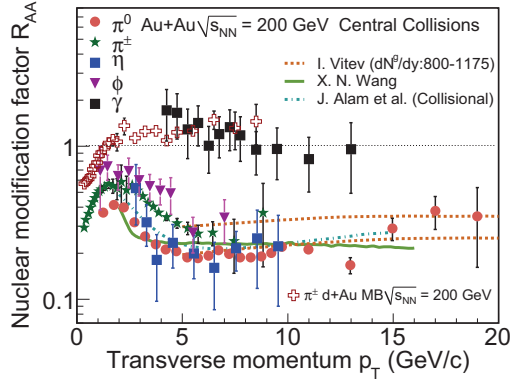}}
\caption{ Nuclear modification factor in $\sqrt{s}_{NN}$=200 GeV Au+Au collisions for $\pi^0$,$\pi^{\pm}$, $\eta$ $\phi$ and $\gamma$. The lines are theoretical calculations.}
  \label{F37}
\end{figure}

\begin{equation}\label{eq9.10}
R_{AA}=\frac{dN^{AA}/dyd^2p_T}{T_{AA}d\sigma^{pp}/dyd^2p_T },
\end{equation}

\noindent $T_{AA}$ in Eq.\ref{eq9.10} is the nuclear thickness function, calculable in a Glauber model. 
If AA collision is a superposition of pp collisions, the expected ratio is unity.  
Fig.\ref{F37} shows the experimental measurements for $R_{AA}$ in $\sqrt{s}_{NN}$=200 GeV Au+Au collisions.
Mesons are largely suppressed at high $p_T$. Photons are however are not suppressed.
The observation is a strong evidence that high $p_T$ suppression is not an initial state effect, but a final state effect. High density colored medium is created in the collision and cause the suppression.  

Parton energy loss $\Delta E$ provides fundamental information on the medium properties. There are several models for the energy loss calculations, e.g. BDMPS \cite{Baier:1996sk}\cite{Baier:1996kr}, GLV \cite{Gyulassy:2000er}\cite{Gyulassy:2000fs}. I will not discuss them here. Just mention that energy loss in the medium is generally characterised by the following variables,

(i) the mean free path $\lambda=1/(\rho\sigma)$, $\rho$ the medium density and sigma the particle medium cross section, (ii) the opacity $N=L/\lambda$ or the number of scattering centres in a medium of thickness $L$, (iii) the Debye mass $m_D(T)\sim g T$, $g$ the coupling parameter, (iv) the transport coefficient $\hat{q}=m_D^2/\lambda$ controls the radiative energy loss, responsible for jet quenching in the
induced gluon bremsstrahlung picture, (v) the diffusion constant $D=\mu T$ characterising the dynamics of heavy, non-relativistic particles, $\mu$ being the mobility of the particles, i.e the ratio of drift velocity and applied force.

The various curves in Fig.\ref{F37} are from different theoretical calculations. The dashed
curve shows a theoretical prediction using the GLV parton energy loss model  \cite{Gyulassy:2000fs}\cite{Vitev:2002pf}.
The model assumes an initial parton density dN/dy = 800 - 1100, which corresponds
to an energy density of approximately 5-15 GeV/fm3. The solid curves are predictions
from reference \cite{Wang:2004yv}. The effect of parton energy loss was implemented
through an effective modified fragmentation function. The modified fragmentation function approximates   the medium effect  in multiple parton scattering formalism. The dot-dashed curve in Fig.\ref{F37} is a theoretical result on $R_{AA}$ by considering only the collisional energy loss  \cite{Alam:2006bu}. Theoretical predictions are approximately consistent with the experimental data.
 
Our discussion on parton energy loss is rather sketchy. Interested reader may look into 
\cite{d'Enterria:2010zz} for more detailed discussion.

\section{Summary}

In this short lecture course, I have discussed some aspects of 
relativistic heavy ion collisions. 
Our centre is actively engaged in experimental and theoretical
studies on Quark-Gluon-Plasma. The emphasis of the course, naturally
was directed to the study of QGP. I have discussed some topics at some length,
some topics briefly and completely left out some topics.
The choice of the topics is personal, which the author felt important enough for 
a student, pursuing his career in theoretical or experimental high energy nuclear 
physics should know. In future, I may extend the scope of the lecture note. I hope 
the students find this lecture course useful. 
I will be  obliged to receive any comments or suggestions. Without any hesitation,
reader may contact me 
by e-mail: akc@vecc.gov.in.

\end{document}